\documentclass[a4paper,11pt]{article}
\usepackage{jcappub} 
\usepackage{lineno}
\linenumbers
\usepackage{geometry}                
\usepackage{float}
\usepackage{amssymb,amsmath,amsopn,bm,dsfont}
\DeclareMathAlphabet{\matheul}{U}{eus}{m}{n}
\usepackage{setspace,enumitem}

\allowdisplaybreaks

\begin{document}
\nolinenumbers

\title{Anisotropic Segre [(11)(1,1)] dark energy following a particular equation of state} 
\author{Philip Beltracchi}
\emailAdd{philipbeltracchi@gmail.com}
\affiliation{Independent Researcher, Parker, Colorado, USA}
\maketitle

\begin{abstract}
\noindent  A generally anisotropic equation of state originally derived in the context of Newman-Janis rotating systems allows for vacuum energy at a specific density. In this paper we examine the possibility of using that equation of state for cosmological dark energy. We treat the case of large scale ordering of the directions of the energy-momentum tensor eigenvectors with a Bianchi cosmological model, and treat the case where the ordering is random on small scales with an effectively isotropic FLRW system. We find particular spacetimes which evolve towards a vacuum energy/ de Sitter like configuration in either case. In the  anisotropic Bianchi case, the system can have behavior reminiscent of big bounce cosmologies, in which the matter content approaches vacuum energy at large scale factor and can behave in a variety of ways at small scale factor. For particular conditions in the effectively isotropic case, we can evolve between true and false vacuum configurations, or between radiation like and vacuum energy configurations. We also show how some simpler equations of state behave under the same assumptions to elucidate the methods for analysis.
\end{abstract}
\section{Introduction}
There are various models for dark energy. The simplest models are vacuum energy or cosmological constant like for which the energy momentum tensor with mixed indices $T^\mu_{~\nu}$ has four identical eigenvalues, and is classified as Segre type [(111,1)].

More generally, one may consider ``perfect fluid" type dark energy, which have isotropic pressures $p$ and an energy density $\rho$, which follows the isotropy principle commonly assumed in cosmology. This sort of system, with degenerate eigenvalues corresponding to all spacelike eigenvectors and a distinct eigenvalue for the timelike eigenvector is classified as Segre type [(111),1]. These ``fluids" may have a simple equation of state such as $p=w \rho$ where the constant $w<-1/3$, the limiting case for the strong energy condition and a requirement for expansion. In the case $w=-1$, perfect fluid dark energy degenerates into vacuum energy and $p=-\rho$, the limiting case for the null energy condition. A variety of quintessence models, involving homogeneous scalar fields $\Phi$ \cite{PhysRevD.37.3406}, follow the equation of state $p=w \rho$ with $w=(\dot{\Phi}^2/2-V(\Phi))/(\dot{\Phi}^2/2+V(\Phi))$ where $V$ is the scalar field potential, see \cite{2013CQGra..30u4003T, quintessence2} for reviews on quintessence.   More complicated scenarios have been considered, such as more general $p=f(\rho)$ equations of state (e.g.\cite{Amendola_2003,2003GReGr..35.2063B,2005PhRvD..71f3004N}), equations of state which depend directly on redshift (e.g.\cite{2001IJMPD..10..213C}) or Hubble parameter (e.g. \cite{2005PhRvD..72b3003N}), viscous models (e.g. \cite{2005GReGr..37.2039B,2005PhRvD..72b3003N}), or phenomenological models (e.g. \cite{2019ApJ...883L...3L}). See \cite{2012ApSS.342..155B,2021Univ....7..163M} for reviews of various dark energy cosmologies.

A third class of models are those with anisotropic dark energy (see e.g. \cite{Dymnikova:2000zi,Koivisto_2008,2010GReGr..42..763A,2021PDU....3200806M}), such as the fully anisotropic [111,1] models considered in \cite{2010GReGr..42..763A}, or models with additional degeneracies such as [(11)(1,1)] which have been used in schemes to unify nonsingular black holes with a background cosmological constant \cite{Dymnikova:2000zi}, and are also the main focus of this paper. Along a space-time plane, this behaves like standard vacuum energy in that the equation of state $p_\parallel=-\rho$, where $p_\parallel$ is a pressure eigenvalue associated with a specific spacelike direction. For the eigenvectors along the other space-space plane, the eigenvalues are degenerate with each other (we call this pressure $p_\perp$), but are not in general degenerate with the eigenvalues from the space-time plane. Such energy momentum tensors arise for static fields in standard \cite{Stephani:2003tm} and nonlinear electrodynamics theories \cite{Gibbons:1995cv,AyonBeato:1998ub,Lobo:2006xt,2019arXiv191008166B}, cosmic strings \cite{1977SPhD...22..312S,PhysRevD.23.852,PhysRevD.31.3288,Letelier:1979ej,gursesnambu}, and nonsingular black hole models \cite{Bardeen,Dymnikova1992,AyonBeato:1998ub,Hayward:2005gi}. Segre type [(11)(1,1)] systems must follow the equations of state
\begin{align}
    p_\parallel=-\rho,~p_2=p_3=p_\perp, \label{unimatdef}
\end{align}
and may follow an equation of state
\begin{align}
    p_\perp=f(\rho). \label{unieos}
\end{align}

The standard version of the Newman-Janis algorithm \cite{Newman:1965tw,Newman:1965my,Kerr:2007dk,Gurses:1975vu} maps a spherically symmetric Segre type [(11)(1,1)] to a rotating axisymetric Segre type [(11)(1,1)] system. While the equation of state for pure vaccuum $p_\perp=\rho=0$ and an electric field $p_\perp=\rho$ are preserved by the Newman-Janis algorithm (resulting in the mappings of Schwarzschild to Kerr and Reissner-Nordstrom to Kerr-Newman black holes), general equations of state of the type Eq.~\ref{unieos} are not preserved. For example, when non rotating vacuum energy/ de Sitter space is fed in to the Newman-Janis algorithm (see e.g. \cite{Ibohal:2004kk,2006PhLB..639..368D,Azreg-Ainou:2014nra,deUrreta:2015nla,Beltracchi2021a} ), the corresponding ``rotating de Sitter space" has inhomogeneous and anisotropic stress in general and no longer follows $-\rho=p_\perp=const$. However, the more general equation of state 
\begin{align}
    (\rho-p_\perp)^2=4 \rho \rho_\Lambda. \label{NJeos}
\end{align}
is preserved by the Newman-Janis algorithm \cite{Beltracchi2021a}. The standard electric field (Reissner-Nordstrom to Kerr-Newman) and true vacuum cases (Schwarzschild to Kerr) can be described with an equation of state of the form (\ref{NJeos} ). Interestingly, ``rotating de Sitter space" can also be interpreted as being a substance following the equation of state (\ref{NJeos}) in general, that happens to be at the point $\rho=\rho_\lambda=-p_\perp$ when the rotation disappears. This makes Eq.~(\ref{NJeos}) is possibly relevant for cosmological dark energy, as it can describe the generalized behavior of vacuum energy which has been made to rotate with the Newman-Janis algorithm.

The question remained as to what sort of properties a universe filled with a Segre type [(11)(1,1)] substance, which behaves like standard dark energy only along a single axis, would have. To answer this question, we first in Section \ref{derivation section}  examine a particular metric for a homogeneous but anisotropic cosmology, and derive the Einstein equations and other quantities in the general case. We then examine some simpler equations of state of the form Eq.~(\ref{unieos}) for which exact solutions to the Einstein equation may be found in Subsections \ref{stringsec}, \ref{EMsec}, and \ref{vacsec}. To our knowledge these exact solutions have not yet been presented in the literature. In Section \ref{NJsec} we use covariant energy conservation to determine the scale dependence of the matter functions, then use perturbation theory and numerical methods to examine time evolution for systems obeying Eq.~(\ref{NJeos}). In Section \ref{isoavg} we examine the possibility of isotropic averaging for systems which are fundamentally [(11)(1,1)] on small scales but have disorder in the direction of the spatial axes on large scales, and show how isotropically averaged systems consisting of the previously considered matter types would behave. We then summarize and give possible avenues for future work in Section \ref{conclusion}.

 In  appendix \ref{geoapdx}, we examine Killing vectors, conserved quantities, and geodesics for metrics of the form (\ref{met}). In appendix \ref{bianchi}, we explain the Bianchi classification process and classify our general metric (\ref{met}) and its partially specified metric (\ref{unimet}).  Appendix \ref{constB} shows that an alternative way of restricting (\ref{met}) to describe a Segre [(11)(1,1)] spacetime is not conducive to studying evolution of a system with equation of state (\ref{NJeos}).
\section{Derivation and Basic Examples}
\label{derivation section}
A metric inspired by the FLRW metric, modified to having two spatial axes the same but another different, can be conveniently written in cylindrical type coordinates $t,z,r,\theta$, with the line element 
\begin{align}
    ds^2=-dt^2+a(t)^2dz^2+b(t)^2\Big[\frac{dr^2}{1-k r^2/2 } +r^2 d\theta^2\Big]. \label{met}
\end{align}
This four dimensional spacetime is composed of a two dimensional space of constant curvature\footnote{Specifically $k$ is the Ricci scalar of the two dimensional metric formed by the term in square brackets of Eq.(\ref{met}).} $k$ and scale factor $b(t)$, a perpendicular spacelike axis with scale factor $a(t)$, and a time axis. Appendix \ref{bianchi} shows that metric (\ref{met}) corresponds to a Bianchi type $III$ when $k\ne0$ and a Bianchi type $I$ when $k=0$.

One can construct a tetrad for metric (\ref{met}) which allows for the examination of components in an orthonormal frame,
 \begin{align}
     e^\alpha_{~\hat{\alpha}}=\left(
\begin{array}{cccc}
 1 & 0 & 0 & 0 \\
 0 & 1/a & 0 & 0 \\
 0 & 0 & \frac{\sqrt{1-k r^2/2 }}{b} & 0 \\
0 & 0 & 0 &
   \frac{1}{r b } \\
\end{array}
\right) .
\label{tet}
 \end{align}
Here the spacetime index $\alpha$ labels the rows and the orthonormal index $\hat{\alpha}$ labels the columns.  The metric in this orthornormal frame can be computed as
\begin{align}
g_{\hat{\alpha}\hat{\beta}} = g_{\alpha\beta} e^\alpha_{~\hat{\alpha}} e^\beta_{~\hat{\beta}} = diag( - 1,  1,  1,  1).
\end{align}
With the metric (\ref{met}) and tetrad (\ref{tet}) we can give compute relevant curvature and physical quantities.
The nonzero principle components of the Riemann tensor in the orthonormal frame are
\begin{subequations}
\begin{align}
    R_{\hat{0}\hat{1}\hat{0}\hat{1}}=\frac{-\ddot{a}}{a},\\
    R_{\hat{0}\hat{2}\hat{0}\hat{2}}=R_{\hat{0}\hat{3}\hat{0}\hat{3}}=\frac{-\ddot{b}}{b},\\
    R_{\hat{2}\hat{3}\hat{2}\hat{3}}=\frac{k+2\dot{b}^2}{2b^2},\\
     R_{\hat{3}\hat{1}\hat{3}\hat{1}}= R_{\hat{1}\hat{2}\hat{1}\hat{2}}=\frac{\dot{a}\dot{b}}{ab}.
\end{align}
\label{ReimannA}
\end{subequations}
Dots indicate time derivatives throughout the paper. Notice this has six total components out of a possible twenty and there are two sets of degenerate components, so there are effectively four functions specifying the orthonormal Riemann components. The Weyl tensor simplifies as well, and can be written very compactly in terms of the Q matrix \cite{Stephani:2003tm}
\begin{subequations}
\begin{align}
    Q_{11}=E_{11}=W_{\hat{2}\hat{3}\hat{2}\hat{3}}=\frac{a \left(k-2 b \ddot{b}+2 \dot{b}^2\right)+2 b \left(b \ddot{a}-\dot{a} \dot{b}\right)}{6 a b^2},\\
    Q_{22}=Q_{33}=-Q_{11}/2,
\end{align}
\label{QA}
\end{subequations}
with other $Q$ matrix components vanishing, such that it is effectively dependent on a single function.
The Ricci and Kretchman scalars are
\begin{align}
    \mathcal{R}=\frac{a \left(4 b \ddot{b}+2 \dot{b}^2+k\right)+2 b \left(b \ddot{a}+2 \dot{a} \dot{b}\right)}{a b^2},\\
    \mathcal{K}=\frac{8 \left(\frac{\dot{a}^2 \dot{b}^2}{a^2}+\ddot{b}^2\right)}{b^2}+\frac{4 \ddot{a}^2}{a^2}+\frac{\left(2 \dot{b}^2+k\right)^2}{b^4}.
\end{align}
The energy-momentum tensor/Einstein tensor components are given by
\begin{subequations}
\begin{align}
    8\pi T^t_{~t}=-\frac{ 2 \dot{b}^2+k}{2  b^2}-\frac{2 \dot{a} \dot{b}}{ a b},\label{rhoA}\\
    8\pi T^z_{~z}=-\frac{2\ddot{b}}{ b}-\frac{2 \dot{b}^2+k}{2 b^2},\label{pzA}\\
    8\pi T^r_{~r}=8\pi T^\theta_{~\theta}=-\frac{b \ddot{a}+a \ddot{b}+\dot{a} \dot{b}}{a b}\label{ptA}.
\end{align}
\label{TA}
\end{subequations}
As we can see from the energy-momentum tensor (\ref{TA}), this metric generally describes a Segre [(11)1,1] spacetime with three independent functions specifying the Ricci sector.  It degenerates to [(11)(1,1)] when $T^t_{~t}=T^z_{~z}$, which implies that either: 1) $b=B$ is not a function of time, which we briefly examine in the third appendix, or 2) the  scale functions $a$ and $b$ follow the relationship
\begin{align}
\frac{\dot{a}}{a}=\frac{\ddot{b}}{\dot{b}},\\
    a=c \dot{b},\label{unienforcer}
\end{align}
 here $c$ is an arbitrary integration constant. The metric becomes
 \begin{align}
    ds^2=-dt^2+c^2\dot{b}^2dz^2+b^2(\frac{dr^2}{1-k r^2/2 } +r^2 d\theta^2). \label{unimet}
\end{align}
Since the Bianchi classification is dependent on $k$ rather than the relationship between $a$ and $b$ in metric (\ref{met}), the classification for the partially specified case (\ref{unimet}) is the same.

With the specification (\ref{unienforcer}), the geometric items simplify, the principle orthornormal Riemann components become
\begin{subequations}
\begin{align}
    R_{\hat{0}\hat{1}\hat{0}\hat{1}}=\frac{-\overset{\therefore }{b}}{\dot{b}},\\
    R_{\hat{0}\hat{2}\hat{0}\hat{2}}=R_{\hat{0}\hat{3}\hat{0}\hat{3}}=-R_{\hat{3}\hat{1}\hat{3}\hat{1}}=-R_{\hat{1}\hat{2}\hat{1}\hat{2}}=\frac{-\ddot{b}}{b},\\
    R_{\hat{2}\hat{3}\hat{2}\hat{3}}=\frac{k+2\dot{b}^2}{2b^2}.
\end{align}
\label{ReimannB}
\end{subequations}
Notice here we have an additional degeneration, so there are now three effective Riemann functions. We still have the same basic form of the Weyl tensor/Q matrix, but the independent component becomes
\begin{align}
    Q_{11}=\frac{2 b^2 ~\overset{\therefore }{b}+\dot{b} \left(k-4 b \ddot{b}\right)+2 \dot{b}^3}{6 b^2 \dot{b}}.\label{QB}
\end{align}
The Ricci and Kretchman scalars become
\begin{align}
    \mathcal{R}=\frac{2 \dot{b}^2+k}{b^2}+\frac{2 \overset{\therefore }{b}}{\dot{b}}+\frac{8 \ddot{b}}{b},\label{RicciB}\\
    \mathcal{K}=\frac{\left(2 \dot{b}^2+k\right)^2}{b^4}+\frac{16 \ddot{b}^2}{b^2}+\frac{4~\overset{\therefore }{b}~^2}{\dot{b}^2}\label{KretchmanB}.
\end{align}
Finally, the energy-momentum tensor obeys
\begin{align}
    8\pi T^t_{~t}=8\pi T^z_{~z}=-\frac{4 b \ddot{b}+2 \dot{b}^2+k}{2 b^2},\label{rhoB}\\
     8\pi T^r_{~r}=8\pi T^\theta_{~\theta}=-\frac{\overset{\therefore }{b}}{\dot{b}}-\frac{2 \ddot{b}}{b},\label{pTB}
\end{align}
which is properly Segre type [(11)(1,1)]. Notice that the constant $c$ from Eq.~(\ref{unienforcer}) has not appeared in any of these quantities. This is because the constant c is a ``gauge" constant simply related to the scaling of the $z$ coordinate and can be trivially removed with the coordinate change $z->z'/c$.

The covariant conservation of energy equation $\nabla_\mu T^\mu_{~\nu}$ gives one nontrivial equation, being
\begin{align}
    2(\rho+p_\perp)\frac{\dot{b}}{b}=-\dot{\rho}.\label{TConservation}
\end{align}

With these quantities calculated, we can now examine what happens when an equation of state between the energy-momentum tensor eigenvalues $-\rho$ and $p_\perp$ is enforced. We find that for simple examples of a ``stringy" equation of state 
\begin{align}
    p_\perp=0,\label{stringeos}
\end{align}
 an ``electromagnetic" equation of state 
\begin{align}
    p_\perp=\rho,\label{EMeos}
\end{align}
and a ``vacuum" equation of state
\begin{align}
    p_\perp=-\rho,\label{vaceos}
\end{align}
there can exist closed form relations between $b$ and $t$. We use these names for the equations of state because the stringy equation of state shows up in systems with cosmic strings, the electromagnetic equation of state shows up for static electric fields, and for the vacuum equation of state the energy-momentum tensor has the structure of vacuum energy. These closed form solutions to the Einstein equations may be new, as they have important differences from the similar spacetimes we have found in the literature.

While we do not find a closed form solution in the case of the Newman-Janis derived equation of state Eq.~(\ref{NJeos}), perturbative analysis shows that in certain circumstances it naturally approaches the standard isotropic vacuum energy equation of state, and we can numerically retrodict from the perturbative analysis to earlier stages in the evolution of the cosmology.
\subsection{Example: Stringy Equation of State}
\label{stringsec}
Our first basic example is the using Eq.~(\ref{stringeos}). In \cite{Letelier:1979ej}, cosmic string spacetimes following this equation of state for various symmetry systems were considered. One of these cosmic string spacetimes had similar symmetry to which we consider here, which was called ``planar" in that work, but did not include the transverse curvature $k$.
 The stringy equation of state $p_\perp=0$, combined with Eq.~(\ref{pTB}), gives the condition
\begin{align}
    -\frac{\overset{\therefore }{b}}{\dot{b}}=\frac{2 \ddot{b}}{b}.
    \label{stringeosc1}
\end{align}
As it turns out, if we invert this equation to find $t(b)$ rather than $b(t)$ the relationship is expressible in terms of elementary functions. We require the derivative replacement rules
\begin{subequations}
\begin{align}
  \overset{\therefore }{b}=-t'''\dot{b}^4+3\frac{ \ddot{b}^2}{\dot{b}},\\
  \ddot{b}=-t''\dot{b}^3,\\
  \dot{b}=\frac{1}{t'},
\end{align}
\label{Dinverter}
\end{subequations}
where a prime denotes a derivative with respect to $b$.
Using the derivative replacement rules Eq.~(\ref{Dinverter}) and the condition Eq.~(\ref{stringeosc1}), we obtain the differential equation
\begin{align}
    2 t'' t'+t''' b t'=3 b (t'')^2.
    \label{stringeosc2}
\end{align}
One way to express $t(b)$ satisfying the differential equation (\ref{stringeosc2}) is
\begin{align}
    t(b)=Y\Bigg(\frac{\sqrt{b}\sqrt{2+b X^2}}{X^2}-\frac{2~ArcSinh(\frac{\sqrt{b}X}{\sqrt{2}})}{X^3}\Bigg)+Z,
    \label{tbstring}
\end{align}
where $X,Y,Z$ are integration constants. The other scale factor $a$ then follows
\begin{align}
    a=c\frac{\sqrt{2 +bX^2}}{Y\sqrt{b}}
\end{align}
from Eqs. (\ref{tbstring},\ref{Dinverter},\ref{unienforcer}).
We can also obtain expressions for the important geometric and physical quantities, such as the energy-momentum tensor components
\begin{align}
    -\rho= T^t_{~t}=T^z_{~z}=p_z=-\frac{2X^2/Y^2+k}{16 \pi b^2},\qquad T^r_{~r}=T^\theta_{~\theta}=p_\perp=0,\label{stringTmv}
\end{align}
which show the equation of state (\ref{stringeos}) is indeed satisfied. The result Eq~(\ref{stringTmv}) also agrees with the analysis of the energy conservation equation (\ref{TConservation}) given the equation of state (\ref{stringeos}), because upon separation and integration it yields $\rho\propto b^{-2}$.
The orthonormal Riemann tensor components are
\begin{subequations}
\begin{align}
    R_{\hat{0}\hat{1}\hat{0}\hat{1}}=\frac{-2}{b^3Y^2},\\
    R_{\hat{0}\hat{2}\hat{0}\hat{2}}=R_{\hat{0}\hat{3}\hat{0}\hat{3}}=-R_{\hat{3}\hat{1}\hat{3}\hat{1}}=-R_{\hat{1}\hat{2}\hat{1}\hat{2}}=\frac{1}{b^3Y^2},\\
    R_{\hat{2}\hat{3}\hat{2}\hat{3}}=\frac{2}{b^3Y^2}+\frac{2 X^2/Y^2+ k }{2b^2}.
\end{align}
\label{Reimannstring}
\end{subequations}
The Weyl sector is specified by the Q matrix component
\begin{align}
    Q_{11}=\frac{2}{ b^3 Y^2}+\frac{2 X^2/Y^2+ k }{6 b^2 }.\label{Qstring}
\end{align}
Finally, the  Ricci and Kretchman scalars are
\begin{align}
    \mathcal{R}=\frac{1}{b^2}\Big(\frac{2 X^2}{Y^2}+k\Big),\label{Riccistring}\\
    \mathcal{K}=\frac{b^2 \left(k Y^2+2 X^2\right)^2+8 b \left(k Y^2+2 X^2\right)+48}{b^6
   Y^4}\label{Kretchmanstring}.
\end{align}

Because the stringy equation of state with metric (\ref{unimet}) leads to a simple solution, it is possible to do some analysis which will illustrate techniques and patterns which will be useful in the later cases.

As stated previously, the constant $c$ is a gauge constant, not showing up in any of the physical or geometric quantities, because it is simply a scaling of the $z$ coordinate. Likewise, the constant $Z$ from Eq.~(\ref{tbstring}) does not show up in any of the physical or geometric quantities, on inspection of (\ref{tbstring}) we see that it is simply a shift of the time coordinate, so it is also a gauge constant. This means that we have two integration constants $X,Y$ and the transverse spatial curvature constant $k$ which dictate the behavior. 

\begin{figure}
    \centering
    \includegraphics{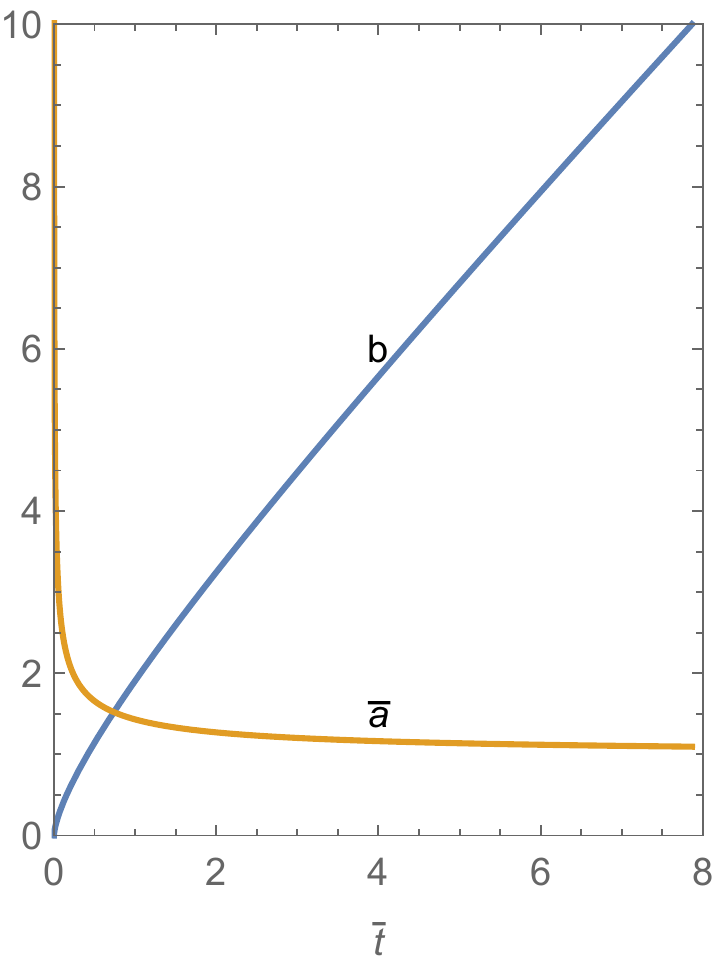}
    \caption{Figure detailing the evolution of the scale functions $b$ and $\bar{a}$ with respect to $\bar{t}$ for $|X|=1$.}
    \label{stringab}
\end{figure}

We can analyze the behavior more easily by looking at the quantities
\begin{align}
 \bar{t}=\frac{t-Z}{Y},\qquad \bar{a}=\frac{a Y}{c}.   
\end{align}
For small $b\rightarrow0$, we have 
\begin{align}
    \bar{t}\approx \frac{\sqrt{2}}{3} b^{3/2},\qquad \bar{a}\approx\sqrt{\frac{2}{b}}.
\end{align}
For large $b\rightarrow\infty$, we have
\begin{align}
    \bar{t}\approx \frac{b}{|X|},\qquad \bar{a}\approx|X|.
\end{align}
Notice that these asymptotic behaviors are simple enough to be inverted, such that 
\begin{align}
    b(\bar{t}\approx0)\propto \bar{t}^{~2/3},\qquad b(\bar{t}\approx\infty)\propto \bar{t}.\label{btasymptotestring}
\end{align}
Loosely, these behaviors describe a configuration that is long $a\rightarrow\infty$ and thin $b\rightarrow0$, which evolves to a constant length parameter $a=c|X|/Y$ but ever increasing width parameter $b\propto \bar{t}$. It is important to note that if $Y$ is negative, then the evolution of the system in coordinate time $t$ will be reversed. We give a plot of $\bar{a}$ and $b$ with respect to $\bar{t}$ for $|X|=1$ in Figure \ref{stringab}. 
 From the Ricci scalar Eq.~(\ref{Riccistring}) and these asymptotic behaviors Eq.~(\ref{btasymptotestring}) we see that this spacetime has a singularity when $\bar{t}=0$. 
 There are two separate quantities which show up repeatedly throughout the geometric and physical items, being 
 \begin{align}
     \frac{1}{b^3Y^2},\qquad \frac{1}{b^2}\Big(\frac{2 X^2}{Y^2}+k\Big)
 \end{align}
 The former combination does not show up in the Ricci sector but does enter into the Weyl sector, the latter combination is the Ricci scalar itself but also features in the Weyl sector.
 \subsection{Example: Electromagnetic Equation of State}
 \label{EMsec}
 
 An example of a Segre type [(11)(1,1)] system which obeys the equation of state $p_\perp=\rho$ is a non null electric or magnetic field \cite{Stephani:2003tm}, which is why we use the terminology electromagnetic equation of state. Using Eq.~(\ref{EMeos}) and Eqs.~(\ref{rhoB} ,\ref{pTB}) we determine that we require
 \begin{align}
     \frac{4\ddot{b}}{b}+\frac{\dot{b}^2}{b^2}+\frac{k}{2b^2}=-\frac{\overset{\therefore }{b}}{\dot{b}}.
     \end{align}
Once again, a solution in terms of elementary functions for $t(b)$ is possible if we invert using Eq.~(\ref{Dinverter}); we obtain
\begin{align}
    \frac{-4 t''}{ (t')^3b}+\frac{1}{(t')^2b^2}+\frac{k}{2b^2}=\frac{-t'''}{(t')^3}+\frac{3 (t'')^2}{(t')^4},
\end{align}
which is solved by
\begin{align}
    t=\sqrt{2} \left(\frac{Y \tan ^{-1}\left(\pm\frac{b k-Y}{\sqrt{k} \sqrt{2 b Y-4 X-k b^2}}\right)}{k^{3/2}}\mp\frac{\sqrt{2 b Y-4
   X-kb^2}}{k}\right)+Z.\label{tEM}
\end{align}
This leads to 
\begin{align}
    a=\mp c\frac{\sqrt{bY-2X-\frac{k b^2}{2}}}{b}.\label{aEM}
\end{align}
The form of Eq.~(\ref{tEM}) makes the $k=0$ case somewhat difficult to see, but we can easily set $k=0$ in Eq.~(\ref{aEM}), then integrate Eq.~(\ref{unienforcer}) to obtain
\begin{align}
    t(k=0)=\mp\frac{2(bY+4X)}{3Y^2}\sqrt{bY-2X}+Z.
\end{align}

With Eq.~(\ref{tEM}) and the derivative replacement rules (\ref{Dinverter}), we can specify our general forms of the physical and geometric quantities to this electromagnetic system. The energy-momentum tensor obeys
\begin{align}
    T^t_{~t}=T^z_{~z}=-T^r_{~r}=-T^\theta_{~\theta}=-\frac{X}{4\pi b^4}.\label{TEM}
\end{align}
Again we could have guessed the $b$ dependence of this from the equation of state (\ref{EMeos}) and energy conservation equation (\ref{TConservation}), because upon separation and integration it yields $\rho\propto b^{-4}$.
For the Riessner-Nordstrom black hole, which is the most famous standard spacetime 
\footnote{ There are other known but less famous spacetimes which contain a non-null Maxwell field (and positive density), which have symmetries more similar to what we consider here than the Riessner-Nordstrom \cite{Stephani:2003tm}.  The ``Robinson-Bertotti" universe \cite{2011GReGr..43.2307L,Robinson:1959ev,PhysRev.116.1331} is one example, as is a spacetime originating in \cite{1975JMP....16.2306M}. Neither of these spacetimes is equivalent to ours, as the former has vanishing $Q$ matrix and the latter has ``magnetic" contributions to the Weyl sector (imaginary contributions to the $Q$ matrix). }  with a non null electromagnetic field , the sign of the density $\rho$ is positive and the weak energy condition is satisfied. If the weak energy condition is to be satisfied for our spacetime, we require $X\ge0$.

The principle orthonormal Riemann components are
\begin{subequations}
\begin{align}
    R_{\hat{0}\hat{1}\hat{0}\hat{1}}=\frac{6X-bY}{b^4},\\
    R_{\hat{0}\hat{2}\hat{0}\hat{2}}=R_{\hat{0}\hat{3}\hat{0}\hat{3}}=-R_{\hat{3}\hat{1}\hat{3}\hat{1}}=-R_{\hat{1}\hat{2}\hat{1}\hat{2}}=\frac{bY-4X}{2b^4},\\
    R_{\hat{2}\hat{3}\hat{2}\hat{3}}=\frac{bY-2X}{b^4}.
\end{align}
\label{ReimannEM}
\end{subequations}
The Q matrix is specified by the component
\begin{align}
    Q_{11}=\frac{bY-4X}{b^4},\label{QEM}
\end{align}
and the Kretchman scalar is
\begin{align}
    \mathcal{K}=\frac{4 \left(3 b^2 Y^2-24 b X Y+56 X^2\right)}{b^8}.
\end{align}
The Ricci scalar vanishes
\begin{align}
    \mathcal{R}=0\label {ricciEM},
\end{align}
which follows from taking the trace of the Einstein equation for any Segre[(11)(1,1)] system obeying the equation of state (\ref{EMeos}).

As in the stringy case, we can now perform further analysis to obtain further insight into techniques and patterns for the behavior of metrics of the form (\ref{unimet}).

Notice how the integration constant $Z$ is analogous in the electromagnetic and stringy examples, amounting to a shift of the time and not entering into any curvature quantities. Interestingly, the constant transverse spatial constant $k$ from the metric did not show up in any curvature quantities for the  electromagnetic spacetime due to cancellations, at least when they are written directly in terms of $b$. It is also worth mentioning that the sign choice in the solutions for $a$ Eq.~(\ref{aEM}) and $t$ Eq.~(\ref{tEM}) does not enter in to any curvature quantities Eqs. (\ref{TEM}-\ref{ricciEM}). Once again, we have two quantities from which the curvature and physical quantities are built, being 
\begin{align}
    \frac{X}{b^4},\qquad\frac{Y}{b^3},
\end{align}
where one of these combinations (in this case, $Y/b^3$) does not show up in the Ricci sector.

While the physical and curvature quantities take a fairly simple form when written in terms of $b$, the relationships between the scale factors $a$, $b$, and time $t$ are more complicated here than they were in the stringy case, in part because of the appearance of complex numbers for certain values of $b,~X,$ and $Y$ . The combination
\begin{align}
  Arg=  2bY-4X-k b^2 \label{EMarg}
\end{align}
must be nonnegative for the solution to make sense (in that the metric function $a^2$ stays positive and $t$ is real),  which means the $a$ solution may be bounded by one or more scale factors $b$, namely
\begin{align}
    b_1=\frac{Y+\sqrt{Y^2-4kX}}{k},\qquad  b_2=\frac{Y-\sqrt{Y^2-4kX}}{k},\qquad b_{k=0}=\frac{2X}{Y}.
\end{align}
Additionally, the solution may be bounded by the value
\begin{align}
    b_z=0,
\end{align}
not due to the appearance of complex numbers, but because that leads to curvature singularities for nonzero $X,Y$.

 Based on the values of the integration constants $X,~Y$ and $k$ there are several cases for time evolution based on the existence or not of the various boundaries. We give a brief account here of some possibilities.  One factor that is worth mentioning is that restricting to positive $X$ is equivalent to looking at positive energy density in the framework of the weak energy condition, and is one criteria that can be used to reduce the possibilities.

In the $k=0$ case, which we can call Case 0, we have the solution exists for $bY\ge2X$. Positive energy density requires $X>0$. In this case we have that $b$ and $Y$ must have the same sign and that $|b|\ge|2X/Y|$, so in the $k=0$ case the spacetime is bounded by an instant where $a=0$, $b=2X/Y=b_{k=0}$ and never encounters $b=b_z=0$. Note this is not in general true for $X<0,k=0$ spacetimes which violate the WEC, which may be bounded by $b=b_z=0$. 

Turning our attention now to $k\ne0$, we find that the condition
\begin{align}
    Y^2-4kX\ge0
\end{align}
dictates the existence of $b_1$ and $b_2$ as boundary points.
The global extremum of $Arg$ is at
\begin{align}
    b=Y/k,\qquad ArgEX=Y^2/k-4X.
\end{align}

For $ Y^2-4kX<0$ the $b_1$ and $b_2$ are complex and do not restrict the values $b$ actually takes. In the case that $ArgEX$ is negative, there are no values of $b$ for which $a^2>0$ , one example of this is $X=1,Y=1,k=1$. We name situations like this Case 1.

It is also possible that $ Y^2-4kX<0$ and $ArgEX>0$ if $k<0$, which we note as Case 2, in which case all real $b$ give the correct sign for $a^2$ and the boundary point on the spacetime is $b=0$. This situation requires $X<0$, or violation of the WEC. One example of this is $X=-1,Y=1,k=-1$.

Turning our attention now to $Y^2-4kX\ge0$ and $b_1,b_2$ being actual boundary points, we have the Case 3 where $ArgEX<0$ if $k<0$. The values between $b_1$ and $b_2$ are forbidden, an example of this is $Y=1,X=1,k=-1$. It is possible for $b_z=0$  outside of this interval, which can lead to the $b_z$ type boundary. For the given example the boundary points are $b_1=-1-\sqrt{5}$, $b_2=-1+\sqrt{5}$

Finally we can look at $Y^2-4kX\ge0$, with $ArgEX<0$ and $k>0$ for Case 4. In this case the allowed values for $b$ are between $b_1$ and $b_2$ (excluding the point $b_z$ if it is in that interval). An example of this is $X=1/3,Y=1,k=1/3$, for which the boundary points are $b_1=3+\sqrt{5}$, $b_2=3-\sqrt{5}$.

\begin{figure}
    \centering
    \includegraphics{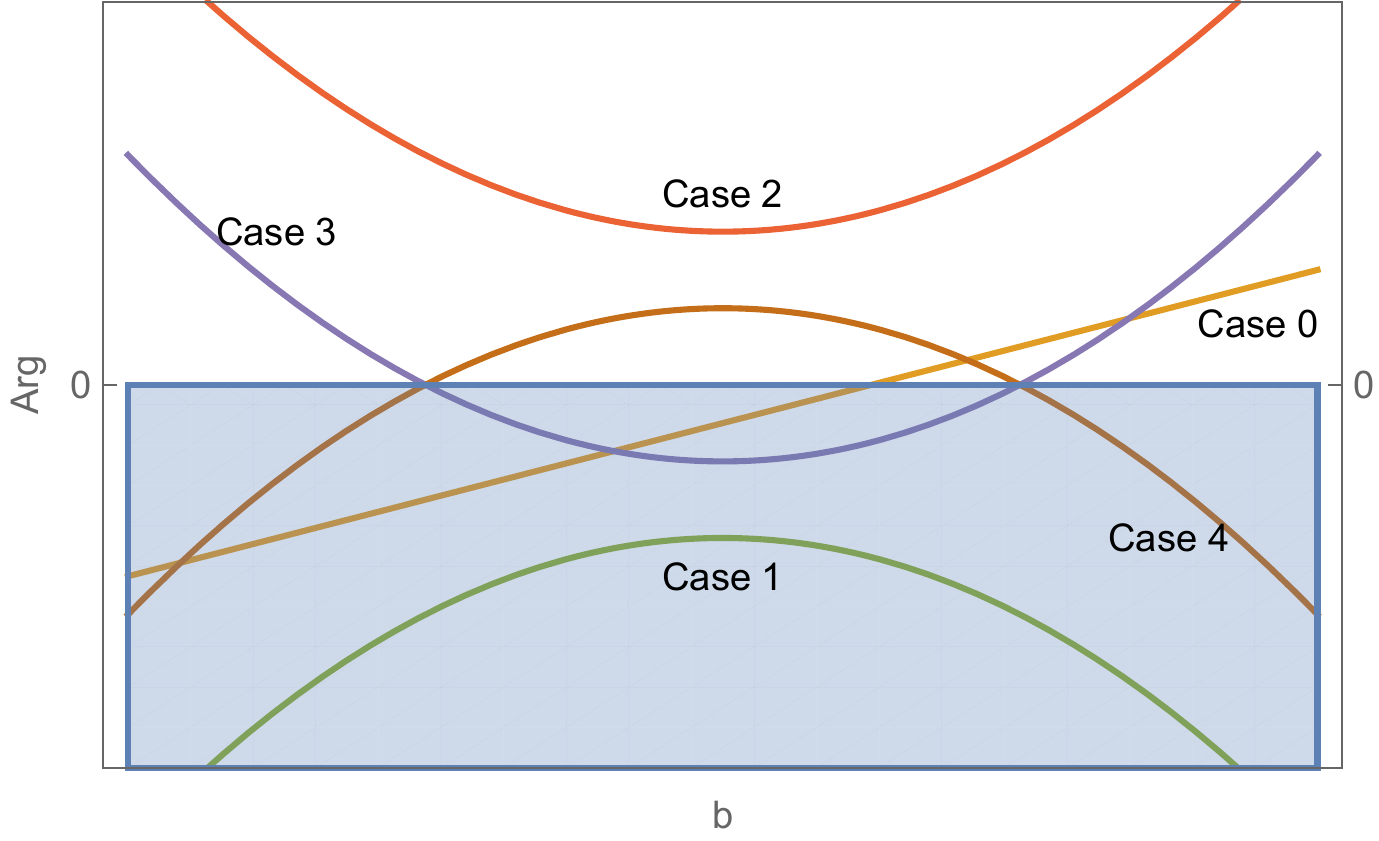}
    \caption{Schematic graph showing the different cases for $Arg$. For Case 0, $k=0$, $Arg$ is linear and we have a single point at which it transfers between negative and positive, there may be a point at which $Arg$ is positive and $b=0$.  For Case 1, $Arg$ is always negative so there are no $b$ for which $a^2$ is positive z coordinate remains spacelike. For Case 2, $Arg$ is always positive, and there will be a point where $b=0$. For Case 3, $Arg$ is positive outside a given range between $b_1$ and $b_2$, there may be a point at which $Arg$ is positive and $b=0$. For Case 4, $Arg$ is positive inside a given range between $b_1$ and $b_2$, there may be a point at which $Arg$ is positive and $b=0$. }
    \label{fig:emcase}
\end{figure}

A schematic drawing of the cases is shown in Figure \ref{fig:emcase}, which may elucidate the classification scheme.

\subsection{Example: Vacuum Energy EOS}
\label{vacsec}
One final simple example of an equation of state is that of vacuum energy $p_\perp=-\rho$. Of all the simple examples, this is the most physically relevant in that it can be compared to standard de Sitter cosmologies. Additionally, it can act as a useful precursor to the perturbatively/ numerically defined solutions for our main equation of state.

Ultimately the vacuum equation of state and our metric (\ref{unimet}) leads to a differential equation
\begin{align}
    \frac{2 b^2 \overset{\therefore }{b}}{\dot{b}}-2\dot{b}^2=k \label{vacBdeq}
\end{align}
Note this equation is set up as an inhomogeneous differential equation for arbitrary $k$. The general solution $b$ 
for arbitrary $k$ involves inverses of elliptic functions, but
 $a(b)$ can be written in terms of elementary functions. One can invert the Eq.~(\ref{vacBdeq}) to obtain 
\begin{align}
    \frac{3(t'')^2-t' t'''}{(t')^4}=\frac{k/2+1/(t')^2}{b^2},
\end{align}
from which it is possible to determine
\begin{align}
    t'=\pm\sqrt{\frac{6b}{6 b^3 V-3 b k-4 W}},\label{avack}
\end{align}
where $W$ and $V$ are the integration constants. Recall that from Eqs.~(\ref{unienforcer},\ref{Dinverter}) we have $t'\propto1/a$, from which we can determine expressions for physical and curvature quantities
\begin{align}
     T^t_{~t}=T^z_{~z}=T^r_{~r}=T^\theta_{~\theta}=-\frac{3V}{8\pi}.\label{vacTk}
\end{align}
Notice that the energy momentum tensor eigenvalues are constant in space and time (the energy conservation equation (\ref{TConservation}) gives $\dot{\rho}=0$ for this equation of state) and fully degenerate. However, this is not simply an unusual coordinate system for de Sitter space, as there is a nonzero Q matrix component, therefore the Weyl tensor is nonvanishing
\begin{align}
      Q_{11}=-\frac{2W}{3b^3}.\label{vacQk}
\end{align}
Interestingly, the $R_{\hat{2}\hat{3}\hat{2}\hat{3}}$ is the negative of $ R_{\hat{0}\hat{1}\hat{0}\hat{1}}$ for this particular spacetime
\begin{align}     
      R_{\hat{0}\hat{1}\hat{0}\hat{1}}= -R_{\hat{2}\hat{3}\hat{2}\hat{3}}=\frac{2W}{3b^3}-V,\\
      R_{\hat{0}\hat{2}\hat{0}\hat{2}}=R_{\hat{0}\hat{3}\hat{0}\hat{3}}=-R_{\hat{3}\hat{1}\hat{3}\hat{1}}=-R_{\hat{1}\hat{2}\hat{1}\hat{2}}=-\frac{W}{3b^3}-V,
\end{align}
and as in the electromagnetic case, we see the $k$ does not show up directly in the curvature quantities when they are written in terms of $b$. The Ricci and Kretchman scalars can be written
\begin{align}
      \mathcal{R}=12V,\label{riccivacK}\\
      \mathcal{K}=24 V^2+\frac{16W^2}{3b^6}.
\end{align}
  
As was the case for the stringy and electromagnetic equations of state, there are two quantities from which the curvature quantities are built, being in this case
\begin{align}
   V,\qquad\frac{W}{b^3},
\end{align}
which are $\mathcal{R}$ and $Q_{11}$ up to constant factors.

We can directly solve the homogeneous case $k=0$ of Eq.~(\ref{vacBdeq}) to obtain  a reasonably simple solution for the time dependence, being
\begin{align}
   b(k=0)=X \sqrt[3]{e^{-2 Y (t+Z)}+e^{2 Y (t+Z)}-2},\label{bvac0}
\end{align}
from which we obtain
\begin{align}
    a=c\frac{2 X Y \left(e^{2 Y (t+Z)}-e^{-2 Y (t+Z)}\right)}{3 \big(e^{-2 Y
   (t+Z)}+e^{2 Y (t+Z)}-2\big)^{3/2}}.
\end{align}
In this case, the energy-momentum tensor is
\begin{align}
    T^t_{~t}=T^z_{~z}=T^r_{~r}=T^\theta_{~\theta}=-\frac{Y^2}{6\pi}.\label{Tvac}
\end{align}
The Q matrix component is
\begin{align}
    Q_{11}=\frac{16}{9}Y^2\frac{ e^{2 Y (t+Z)}}{\left(e^{2 Y (t+Z)}-1\right)^2},\label{Qvac}
\end{align}
and the principle orthonormal Riemann components are 
\begin{subequations}
\begin{align}
   R_{\hat{0}\hat{1}\hat{0}\hat{1}}=- R_{\hat{2}\hat{3}\hat{2}\hat{3}}= -\frac{4}{9} Y^2 \left(\frac{4 e^{2 Y (t+Z)}}{\left(e^{2 Y
   (t+Z)}-1\right)^2}+1\right),\\
    R_{\hat{0}\hat{2}\hat{0}\hat{2}}=R_{\hat{0}\hat{3}\hat{0}\hat{3}}=-R_{\hat{3}\hat{1}\hat{3}\hat{1}}=-R_{\hat{1}\hat{2}\hat{1}\hat{2}}=-\frac{4}{9} Y^2 \left(1-\frac{2 e^{2 Y (t+Z)}}{\left(e^{2 Y
   (t+Z)}-1\right)^2}\right).
\end{align}
\label{Reimannvac}
\end{subequations}
 The Kretchman and Ricci scalars are
\begin{align}
    \mathcal{K}&=\frac{128 Y^4 \left(-4 e^{2 Y (t+Z)}+14 e^{4 Y (t+Z)}-4 e^{6 Y (t+Z)}+e^{8
   Y (t+Z)}+1\right)}{27 \left(e^{2 Y (t+Z)}-1\right)^4},\label{vacK}\\
   \mathcal{R}&=\frac{16 Y^2}{3}.\label{riccivac}
\end{align}
By comparing the curvature quantities in the arbitrary $k$ case with the $k=0$ case, particularly Eqs. (\ref{riccivacK},\ref{riccivac}) and (\ref{vacQk},\ref{Qvac}), we can infer a correspondence of the integration constants when $k=0$
\begin{align}
  k\rightarrow0,\qquad  V\rightarrow\frac{4Y^2}{9},\qquad W\rightarrow\frac{-8X^3 Y^2}{3}.
\end{align}

Because it has an elementary form for $b(t)$ and because it corresponds more closely with the standard de Sitter space, we examine the time dependence of the $k=0$ vacuum energy solution in more detail.

\begin{figure}
    \centering
    \includegraphics{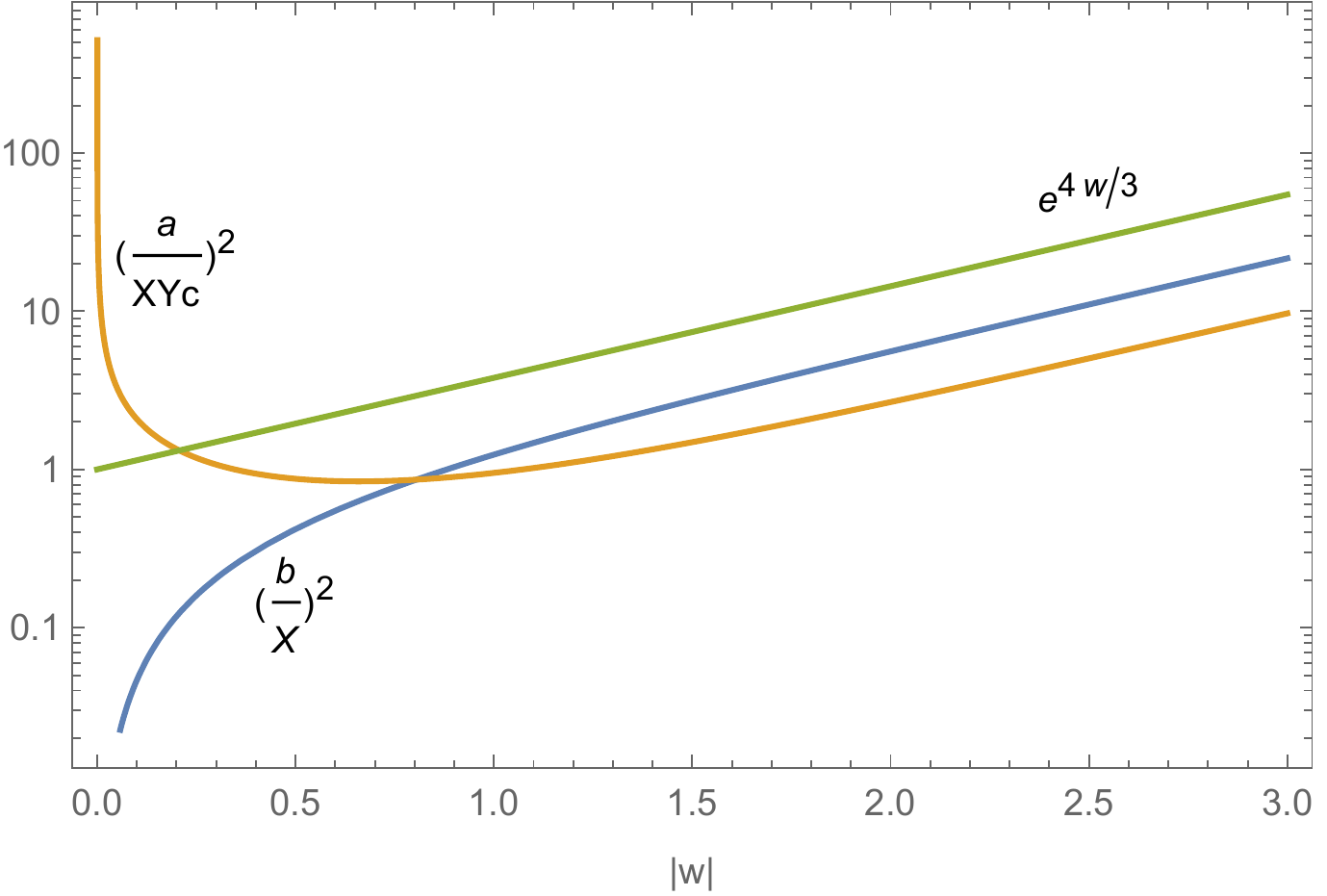}
    \caption{Plots of the scale factors from the metric $(b/X)^2$ and $(a/YXc)^2$ with their large $w$ behavior indicated. It is important to remember that $w=0$ can either be a point we evolve towards or away from with increasing $t$ depending on the values of $Y,Z,t$.  }
    \label{fig:vacab}
\end{figure}

Notice that the constant $X$ does not show up in the curvature quantities, suggesting it is a gauge constant. Indeed, upon examination one can see that the constant $X$ behaves much like the constant $c$ in the definition of $a$ as a removable scale factor, so it is in fact a gauge constant. The integration constant $Y$ shows up both as an overall factor and in the argument to all of the exponential functions from Eq.(\ref{bvac0}) to Eq. (\ref{vacK})
\begin{align}
    w=Y(t+Z),
\end{align} 
$w$ is also the only place in which $Z$ shows up. Usage of the combination $w$ simplifies the analysis.
We can begin with the scale factors. For large $|w|$, the scale functions behave much like standard de Sitter space in that there is exponential behavior, specifically
\begin{align}
    a\propto b\propto e^{2|w|/3}.
\end{align}
At small $|w|$, we have different behavior in that the scale factor $b$ tends toward zero and $a$ tends toward an infinite value. The magnitude of $a$ takes a minimum value at $w\approx0.6585$. In Figure \ref{fig:vacab} we show how the scale factors change with $w$.

The main difference between this spacetime and standard de Sitter space is the presence of a nontrivial Weyl sector as is evident by the nonzero $Q_{11}$ from Eq.~(\ref{Qvac}). At large $|w|$, we have
\begin{align}
    Q_{11}\propto e^{-2|w|},
\end{align}
which shows that at large $|w|$ this space becomes more like standard de Sitter as the $Q$ matrix tends toward zero. When $|w|\rightarrow0$, we have $Q_{11}$ and the Kretchman scalar $\mathcal{K}$ from Eq.~(\ref{vacK}) diverging, indicating the presence of a curvature singularity.

It is possible to consider changes in $w$ as being due to time evolution, for instance one may approach $w=0$ or move away from $w=0$ as $t$ increases depending on the values of $Y,Z,t$. Alternatively, one may be interested in seeing how the normal inflationary de Sitter solution shows up as a specific choice of integration constants. In order to see this, we redefine the constant $X\rightarrow A e^{-2YZ/3}$, then move items inside the cube root to obtain
\begin{align}
    b=A\sqrt[3]{e^{2Yt}-2e^{-2YZ}+e^{-2Yt-4YZ}}.
\end{align}
Finally, we can set $Y=\sqrt{3\Lambda}/2$ and $Z\rightarrow \infty$ to obtain
\begin{align}
    b=A e^{t\sqrt{\Lambda/3}}, \label{dsB}
\end{align}
which is the standard de Sitter scale factor.

\section{Equation of State from the Newman-Janis Algorithm}
\label{NJsec}
\begin{figure}
    \centering
    \includegraphics{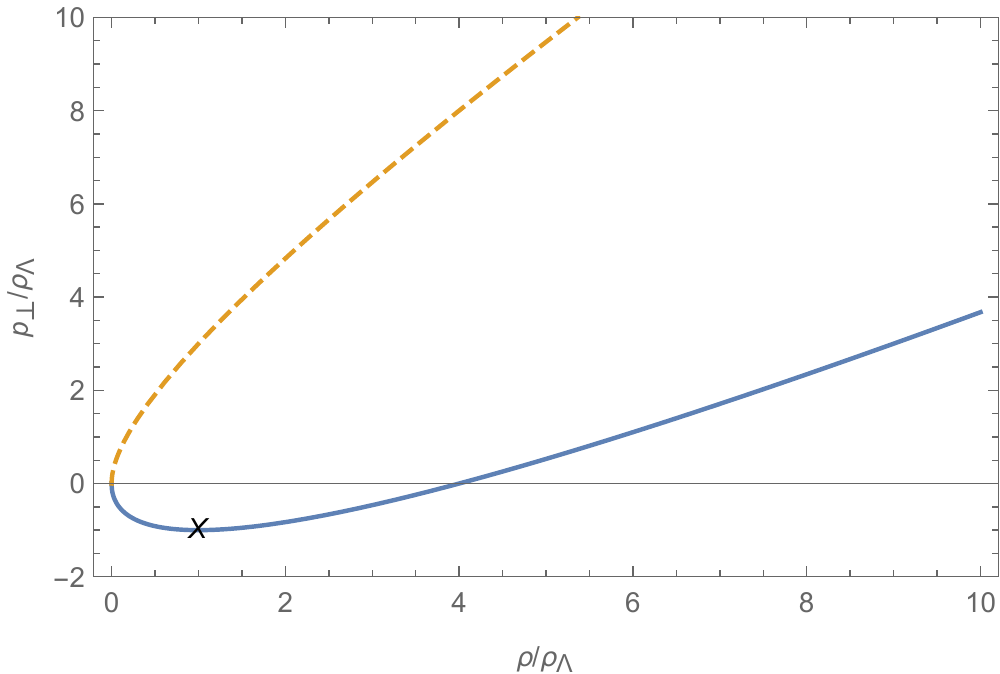}
    \caption{Plot of the equation of state \ref{NJeos}. Isolating $p_\perp$ requires taking a square root and therefore results in two branches which are shown, the lower branch Eq.~(\ref{branchD}) which contains the vacuum energy point (marked with an X) is solid and the upper branch Eq.~(\ref{branchU}) is dashed.}
    \label{NJeosplot}
\end{figure}

The equation of state from the Newman-Janis algorithm Eq.~(\ref{NJeos}) allows vacuum energy $p_\perp=-\rho$ like behavior at $\rho=\rho_\Lambda$,  string like $p_\perp=0$ behavior at $\rho=4\rho_\Lambda$, and electromagnetic like $p_\perp=\rho$ behavior as $\rho/\rho_\Lambda \rightarrow\infty$. As stated in the introduction, vacuum energy which has been made to rotate with the Newman-Janis algorithm does not follow $p_\perp=-\rho$ in general, but it is one of the systems that follows Eq.~(\ref{NJeos}). It is important to realize that in order to solve for $p_\perp$ one obtains two branches, namely
\begin{align}
    p_\perp=\rho+2\sqrt{\rho\rho_\Lambda},\label{branchU}\\
    p_\perp=\rho-2\sqrt{\rho\rho_\Lambda}.\label{branchD}
\end{align}
Notice that it is the lower branch Eq.~(\ref{branchD}) which contains the minus sign that contains the vacuum energy point (marked with an X on Figure \ref{NJeosplot}). On the graphs in this subsection, results obtained from the upper branch Eq.~(\ref{branchU}) are dashed and results from the lower branch Eq.~(\ref{branchD}) are solid.
Usage of the equation of state (\ref{NJeos}) with Eqs.~(\ref{rhoB},\ref{pTB}) results in a differential equation which could be solved for $b$, namely
\begin{align}
    \frac{\left(2 b^2 \overset{\therefore }{b}+\dot{b} \left(8 b \ddot{b}+k\right)+2 \dot{b}^3\right)^2}{4 b^4 \dot{b}^2}=4\Lambda\frac{4 b \ddot{b}+2 \dot{b}^2+k}{2 b^2},\label{NJdifEQ}
\end{align}
where $\Lambda=8\pi \rho_\Lambda$. We find this equation may be treated perturbatively near the de Sitter $b$ from Eq.~(\ref{dsB}), which we show later in Subsection \ref{pert1}, and could be solved numerically given suitable initial conditions, but do not find a general solution to quantify the relationship between $b$ and $t$.

However, it is possible to use the covariant energy conservation equation (\ref{TConservation}) and Eqs.~(\ref{branchD}, \ref{branchU}) to perform analysis 
of how $\rho$ and $p_\perp$ change with $b$. Rearranging the covariant energy conservation equation and substituting for $p_\perp$ from the equation of state gives
\begin{align}
    -\frac{\dot{b}}{b}=\frac{\dot{\rho}}{4(\rho \pm \sqrt{\rho \rho_\Lambda})}
\end{align}
Integration here leads to
\begin{align}
    c-\log(b)=\frac{1}{2}\log(\sqrt{\rho}\pm \sqrt{\rho_\Lambda})
\end{align} where c is an integration constant. Finally we may solve this for $\rho$ to obtain
\begin{align}
    \rho=\Big(\frac{C}{b^2}\mp\sqrt{\rho_\Lambda}\Big)^2.
\end{align}
For plotting purposes, we introduce the variable $\bar{b}=b(\rho_\lambda/C^2)^{1/4}$.
For both branches, as $\bar{b}\rightarrow0$ the density goes to infinity as $b^{-4}$ and the transverse pressure follows, such that $p_\perp/\rho\rightarrow1$. At $\bar{b}=1$,  the lower branch is at $\rho=4\rho_\lambda$, $p_\perp=0$ and the  upper branch is at $\rho=p_\perp=0$. At large $\bar{b}$ the density of both branches is $\rho_\lambda$, but the transverse pressure on the lower branch is $-\rho_\lambda$ and on the upper branch is $3\rho_\lambda$. It is important to remember that without a solution for $b(t)$, one can not say how these configurations can be passed through in a time evolving system.
\begin{figure}
    \centering
    \includegraphics[width=5.5cm]{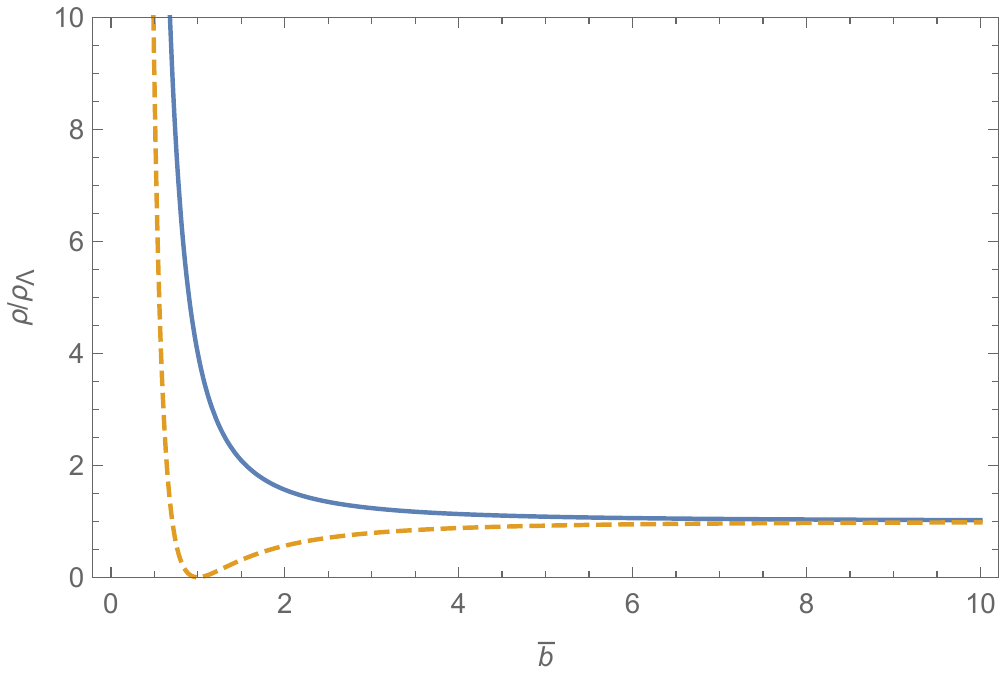}
    \includegraphics[width=5.5cm]{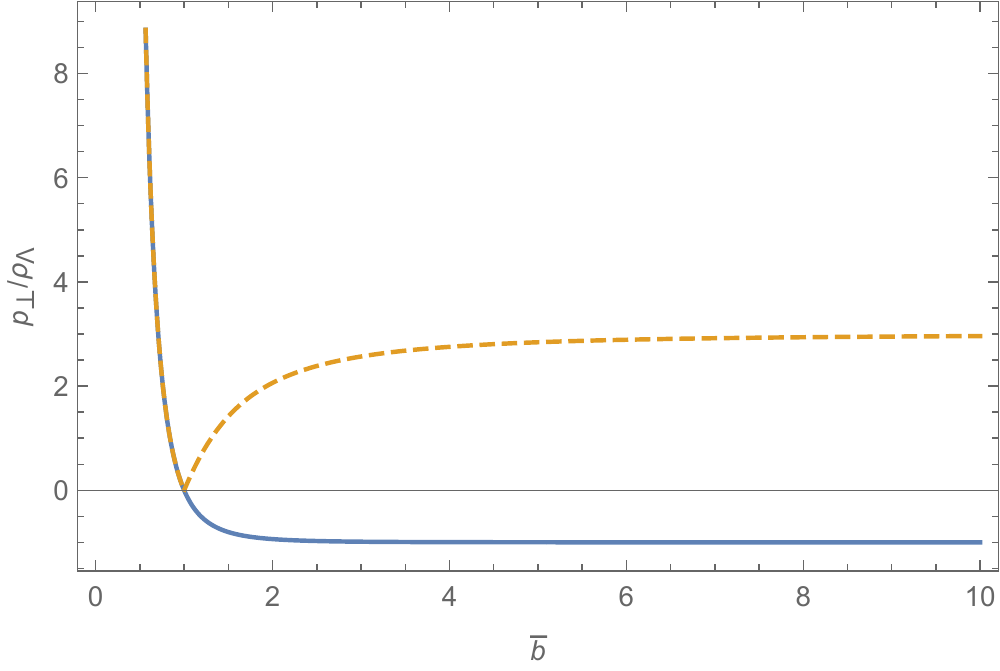}
    \includegraphics[width=5.5cm]{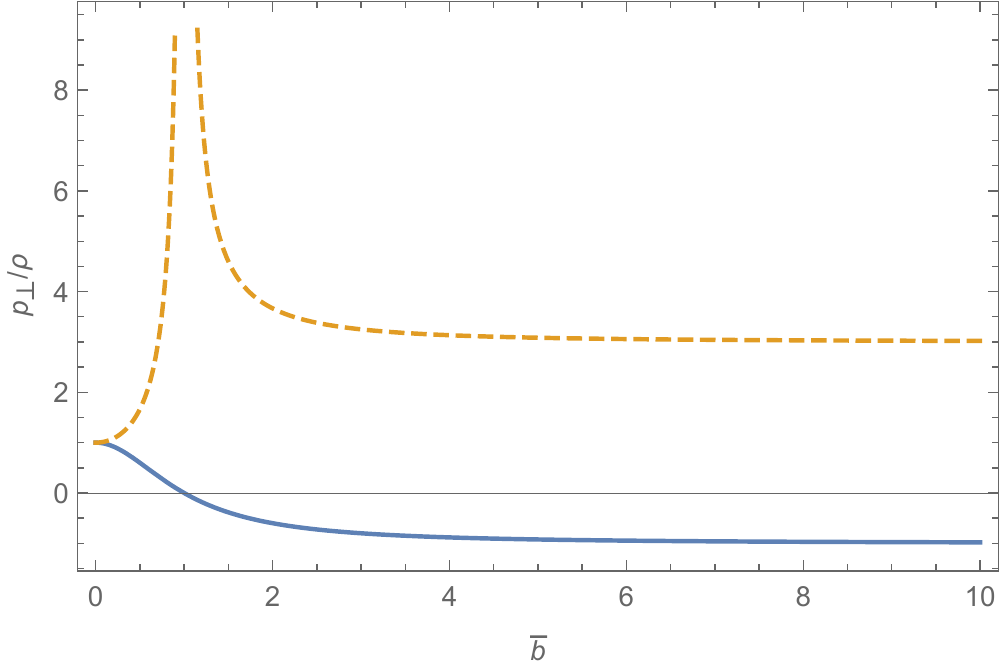}
    \caption{Plots of $\rho/\rho_\Lambda$, $p_\perp/\rho_\Lambda$, and $p_\perp/\rho$ versus $\bar{b}$ for the branches (\ref{branchU}) (dashed) and (\ref{branchD}) (solid). Both branches have high density and roughly electromagnetic $p_\perp=\rho$ behavior as $\bar{b}\rightarrow0$ and approach $\rho=\rho_\Lambda$ at high $\bar{b}$. Below $\bar{b}=1$, both branches have the same $p_\perp$, whereas at large $\bar{b}$ the solid branch goes to $p_\perp=-\rho_\Lambda$ and the dashed branch goes to $p_\perp=3\rho_\Lambda$. The solid branch therefore approximates standard vacuum energy at high $\bar{b}$, while the dashed branch approaches a different configuration which violates the dominant energy condition. At $\bar{b}=1$, the solid branch has $p_\perp=0$ and finite $\rho$, therefore approximating a stringy configuration, while the dashed branch has both $\rho$ and $p_\perp$ approaching zero, but in such a way $p_\perp/\rho\rightarrow\infty$.}
    \label{NJbvs}
\end{figure}
\subsection{Perturbations about de Sitter space}
\label{pert1}
In spite of a lack of solution to Eq. (\ref{NJdifEQ}), we find that the system should evolve toward a standard de Sitter configuration if starting sufficiently close to it. We demonstrate this by defining
\begin{align}
     b=A e^{t\sqrt{\Lambda/3}}\Big(1+\epsilon f(t)\Big), \label{pertdef}
\end{align}
where $\epsilon$ is some small parameter and the background solution is the same as Eq.~(\ref{dsB}), and setting $k\rightarrow0$ as is appropriate for that background solution. With this definition, Eq.~(\ref{NJdifEQ}) becomes
 \begin{align}
    4\epsilon\Big(2\sqrt{3}\Lambda^{3/2}\dot{f}+5\Lambda\ddot{f}+\sqrt{3\Lambda}~\overset{\therefore}{f}\Big)+O(\epsilon^2)=0.
 \end{align}
 We can thus cancel the first order in $\epsilon$ terms with the appropriate $f$, namely
 \begin{align}
     f=Z +Y e^{-2t\sqrt{\Lambda/3}}+X e^{-t\sqrt{3\Lambda}}\label{pertsol}
 \end{align}
 With $f$, we can compute the physical and curvature quantities to first order in $\epsilon$, such as the energy-momentum tensor components

 \begin{align}
     -\rho=T^t_{~t}=T^z_{~z}=-\frac{\Lambda}{8\pi}+\epsilon\frac{Y\Lambda}{6\pi }e^{-2t\sqrt{\Lambda/3}},\label{pertden}\\
     p_\perp=T^r_{~r}=T^\theta_{~\theta}=-\frac{\Lambda}{8\pi}.\label{pertp}
\end{align}
 The energy-momentum elements (\ref{pertden}) and (\ref{pertp}) satisfy the equation of state (\ref{NJeos}) to first order in $\epsilon$. The Null Energy condition, and therefore all the other standard energy conditions will be violated if $Y>0$.
The orthonormal Riemann components are
\begin{subequations}
\begin{align}
      R_{\hat{0}\hat{1}\hat{0}\hat{1}}=-\frac{\Lambda }{3}+\epsilon  2\Lambda X e^{-t \sqrt{3 \Lambda }},   \\
    R_{\hat{0}\hat{2}\hat{0}\hat{2}}=R_{\hat{0}\hat{3}\hat{0}\hat{3}}=-R_{\hat{3}\hat{1}\hat{3}\hat{1}}=-R_{\hat{1}\hat{2}\hat{1}\hat{2}}=-\frac{\Lambda }{3}-\epsilon \Lambda X e^{-t \sqrt{3 \Lambda }},  \\
    R_{\hat{2}\hat{3}\hat{2}\hat{3}}=\frac{\Lambda}{3}-\epsilon\Lambda\Big(2X e^{-t \sqrt{3 \Lambda }}   +\frac{4 Y e^{-2 t \sqrt{\Lambda /3}}  }{3 }\Big),
\end{align}\label{pertRiemann}
\end{subequations}
 and the curvature scalars and $Q$ matrix component are
\begin{align}
    \mathcal{R}=4\Lambda-\epsilon\Big(\frac{8 \Lambda Y e^{-2 t \sqrt{\Lambda /3}}  }{3}\Big),\label{pertR}\\
    \mathcal{K}=\frac{8 \Lambda ^2}{3}-\epsilon\frac{32 \Lambda ^2 Y  e^{-2 t\sqrt{\Lambda/3 } }}{9 },\label{pertK}\\
    Q_{11}=-\epsilon\Big(2\Lambda X e^{-t \sqrt{3 \Lambda }} +\frac{4\Lambda Y e^{-2 t \sqrt{\Lambda /3}}  }{9}\Big)\label{pertQ}.
 \end{align}
Once again, we see that there are a few items which appear repeatedly in the curvature quantities, namely
\begin{align}
    \Lambda,\qquad \epsilon Y \Lambda e^{-2t\sqrt{\Lambda/3}},\qquad\epsilon X \Lambda e^{-t\sqrt{3\Lambda}}.
\end{align}
 Notice that $Z$ does not show up in these terms, this is because it can be absorbed into the definition of the background gauge constant $A$. The $Y$ shows up in the Ricci sector/ perturbations to the energy-momentum tensor eigenvalues, while the $X$ term does not to first order in $\epsilon$.
 It is also important to realize that the perturbations fall off exponentially with time, indicating that for configurations close to vacuum energy-de Sitter space, systems following the equation of state (\ref{NJeos}) evolve toward vacuum energy-de Sitter space. 
 \subsection{Numerical Retrodiction}
\begin{figure}[H]
    \centering
    \includegraphics[width=7cm]{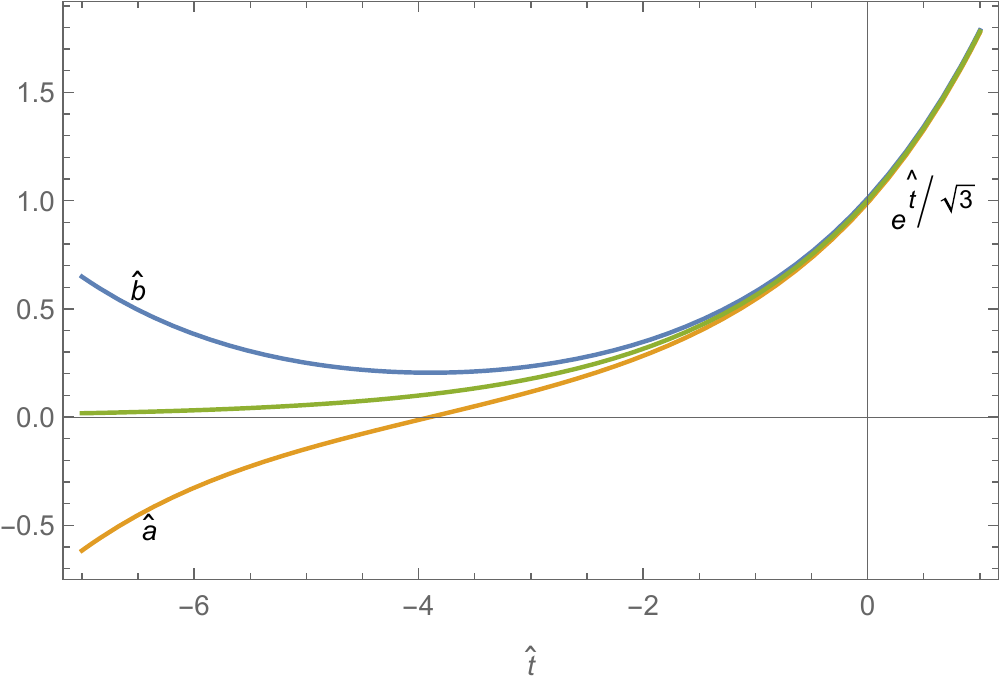}
    \includegraphics[width=7cm]{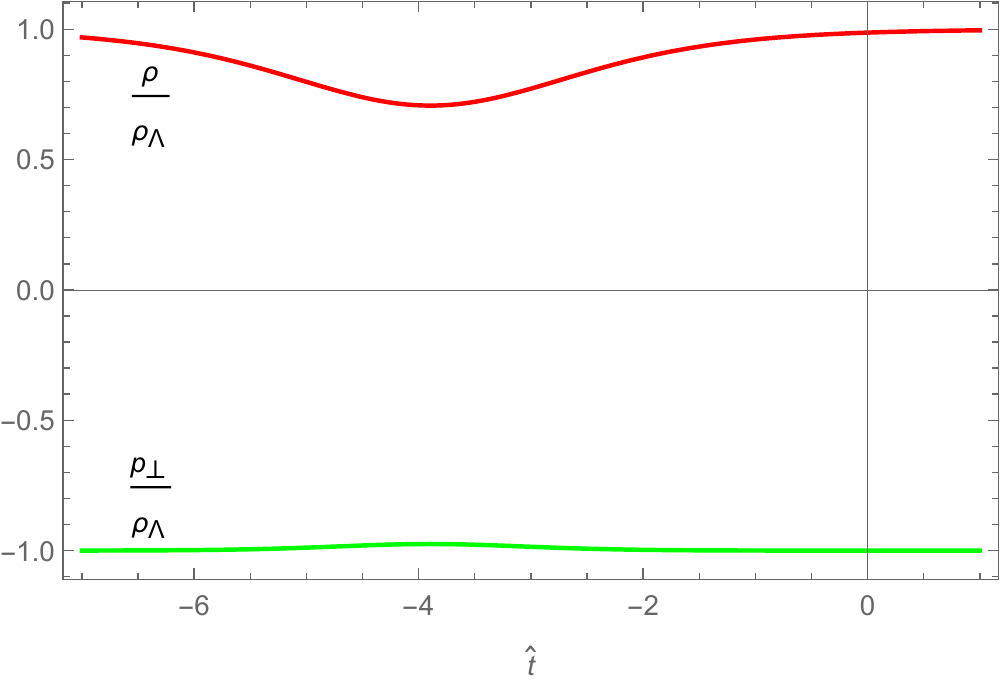}
     \includegraphics[width=7cm]{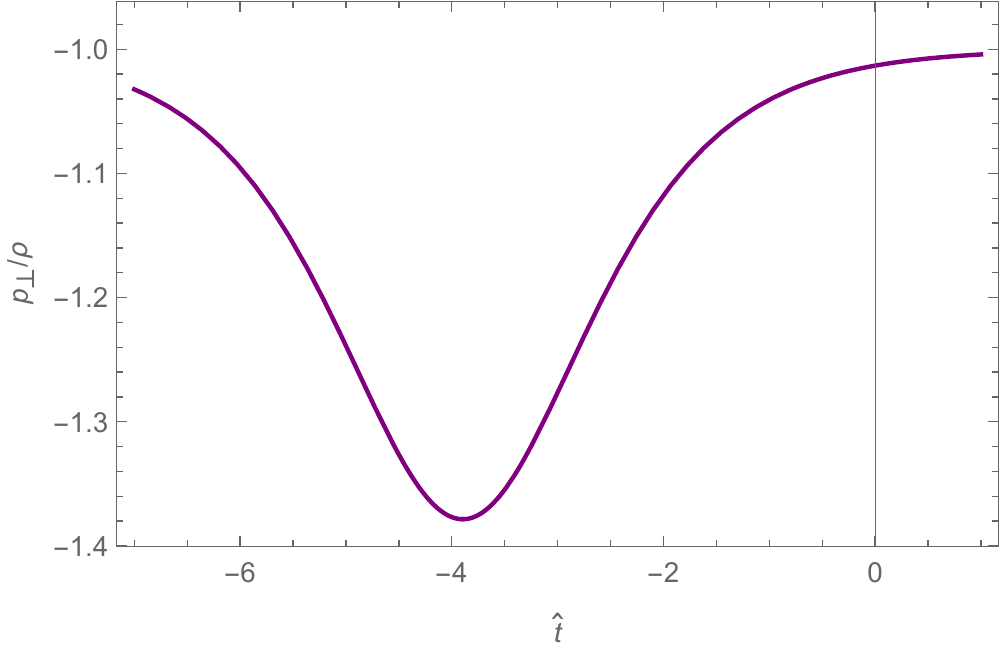}
       \includegraphics[width=7cm]{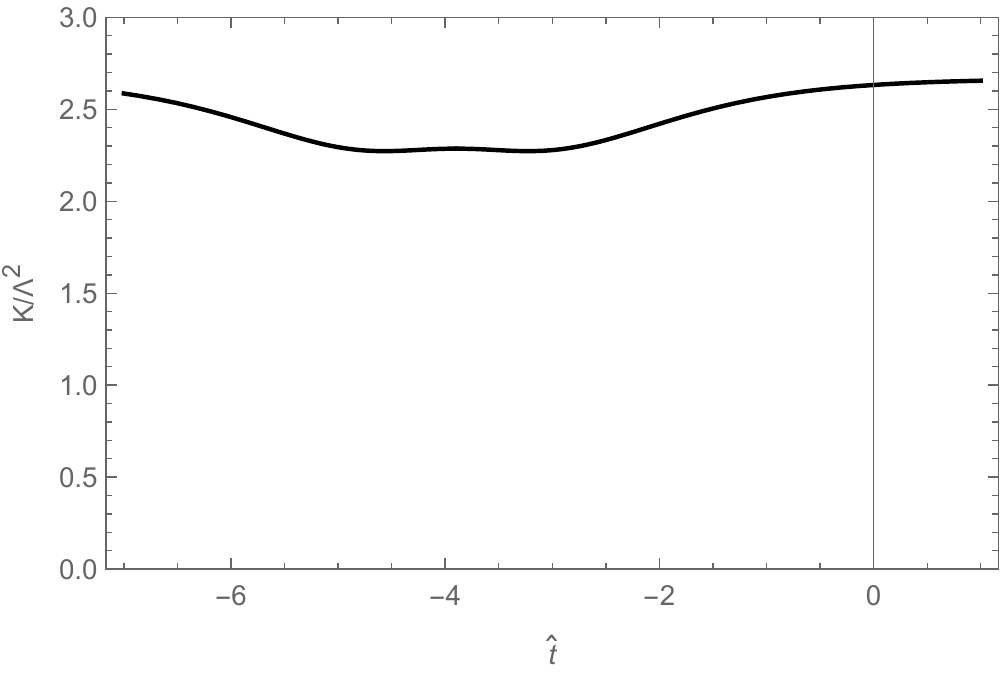}
       \includegraphics[width=7cm]{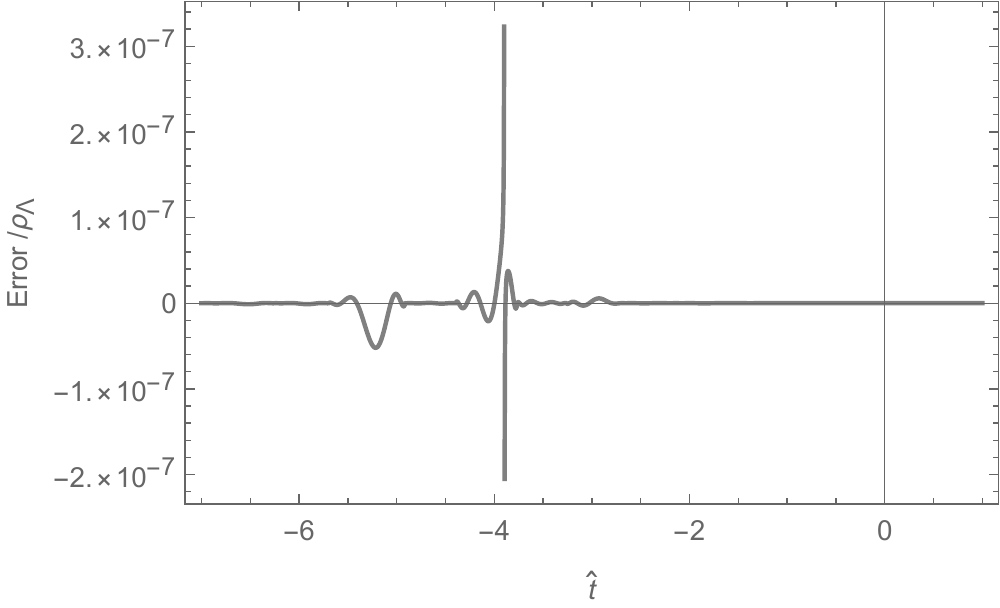}
    \caption{ Plots of the scale factors, energy-momentum tensor eigenvalues, ratio of the energy momentum tensor eigenvalues, Kretchman Scalar, and residual error in Mathematica's interpolating functions with regard to satisfying Eq.~(\ref{branchD}) for $X=0,Y=1$. Notice that near $\hat{t}=-3.896$, $\hat{a}$ goes through 0, $\hat{b}$ changes from collapsing to expanding, the density reaches its minimum, the transverse pressure reaches its maximum, and the error is largest (although still quite small). Notice that although $\hat{a}=0$, the Kretchman scalar is perfectly regular at that point. Since $p_\perp/\rho\le-1$ the null energy condition is violated, which agrees with our analysis for perturbative analysis for positive $\epsilon Y$. Agreeing with the perturbative analysis, at positive times the quantities approach their background configurations $\bar{a}=\bar{b}=e^{\hat{t}/\sqrt{3}}$, $\rho=\rho_\Lambda=-p_\perp$, $K/\Lambda^2=8/3$.}
    \label{NJzp}
\end{figure}

\begin{figure}[H]
    \centering
    \includegraphics[width=7cm]{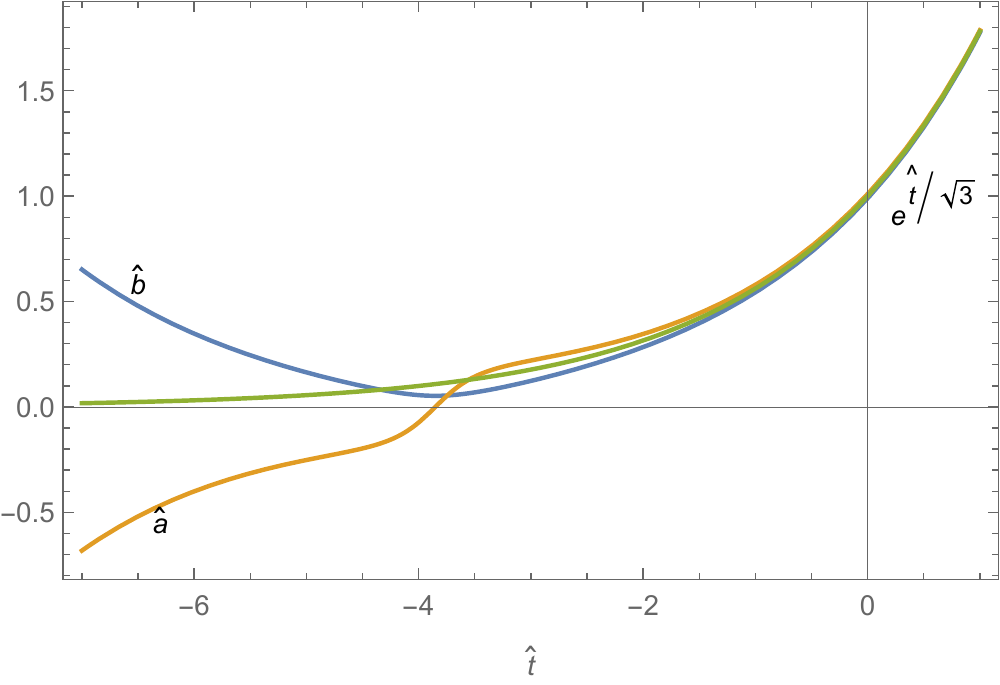}
    \includegraphics[width=7cm]{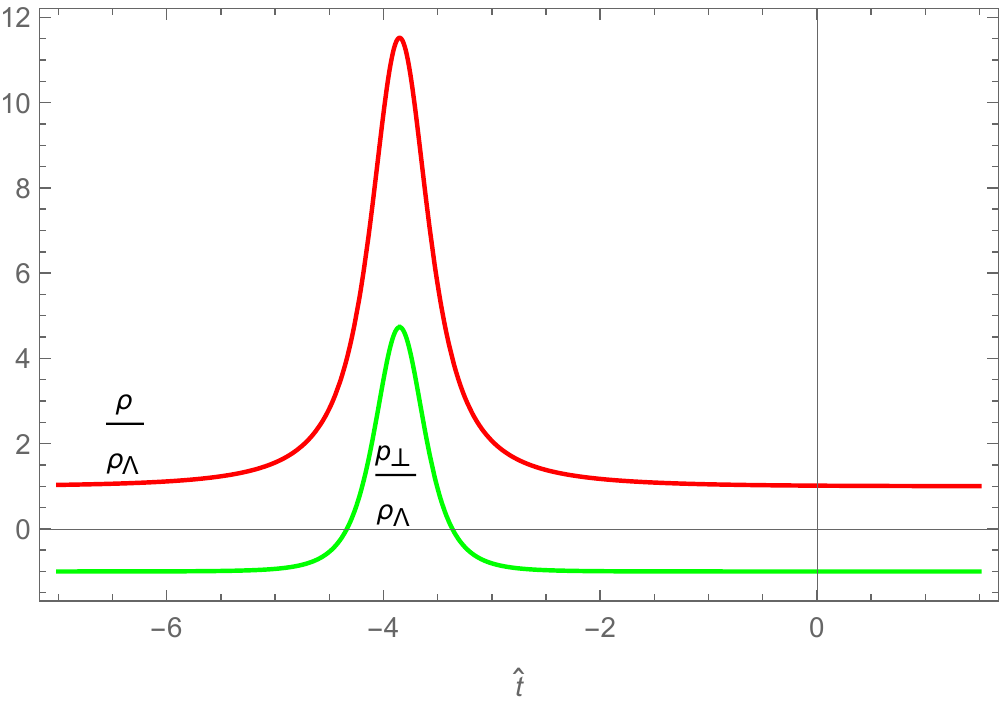}
     \includegraphics[width=7cm]{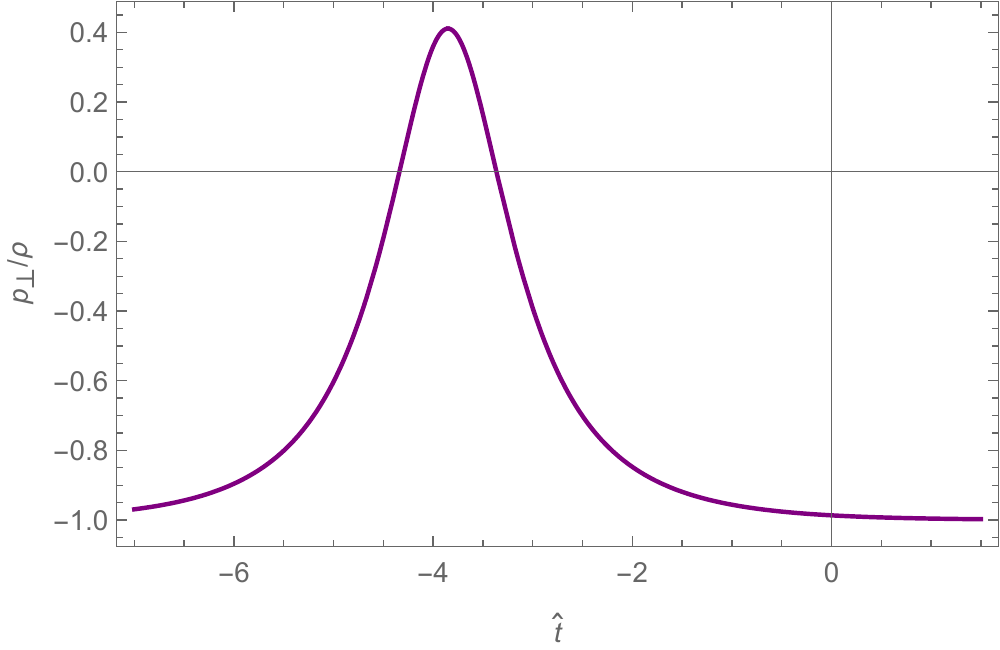}
       \includegraphics[width=7cm]{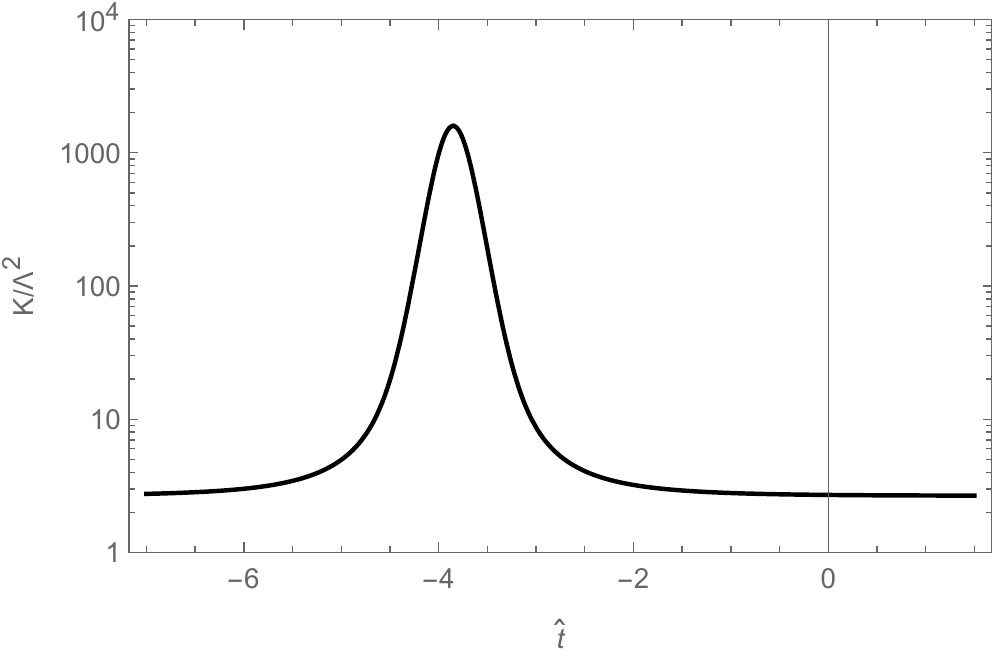}
       \includegraphics[width=7cm]{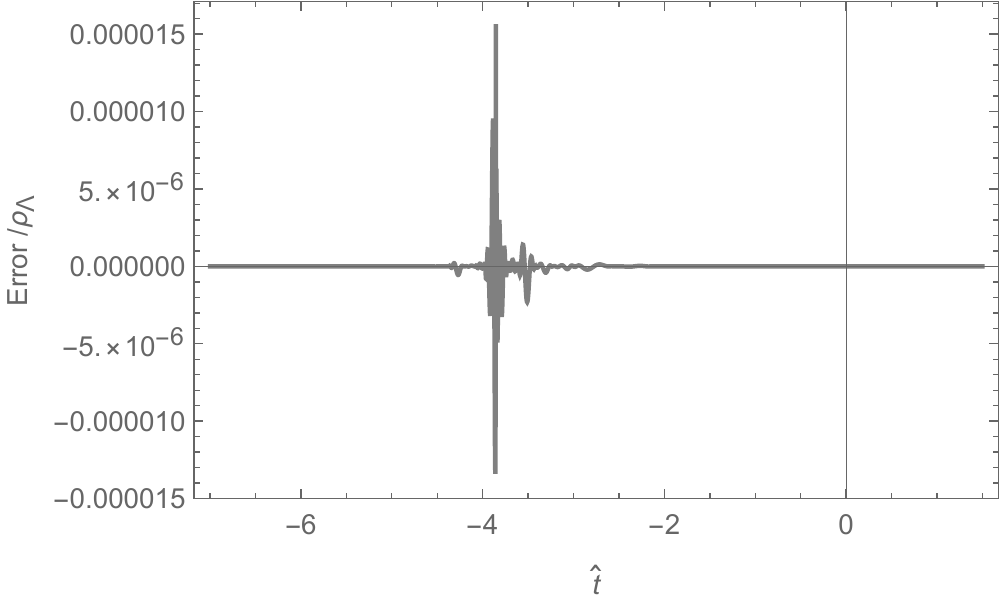}
    \caption{ Plots of the scale factors, energy-momentum tensor eigenvalues, ratio of the energy momentum tensor eigenvalues, Kretchman Scalar, and residual error in Mathematica's interpolating functions with regard to satisfying Eq.~(\ref{branchD}) for $X=0,Y=-1$. Notice that near $\hat{t}=-3.851$, $\hat{a}$ goes through 0, $\hat{b}$ changes from collapsing to expanding. In this $X=0,Y=-1$, case, both the density and the transverse pressure are maximum at this point. Again we have $\hat{a}=0$, and the Kretchman scalar is large but still finite at that point. Since $p_\perp/\rho\ge-1$ the null energy condition is satisfied in this case, which agrees with our analysis for perturbative analysis for negative $\epsilon Y$. At positive times the functions approach the vacuum-de Sitter background configurations $\bar{a}=\bar{b}=e^{\hat{t}/\sqrt{3}}$, $\rho=\rho_\Lambda=-p_\perp$, $K/\Lambda^2=8/3$.}
    \label{NJzm}
\end{figure}

\begin{figure}[H]
    \centering
    \includegraphics[width=7cm]{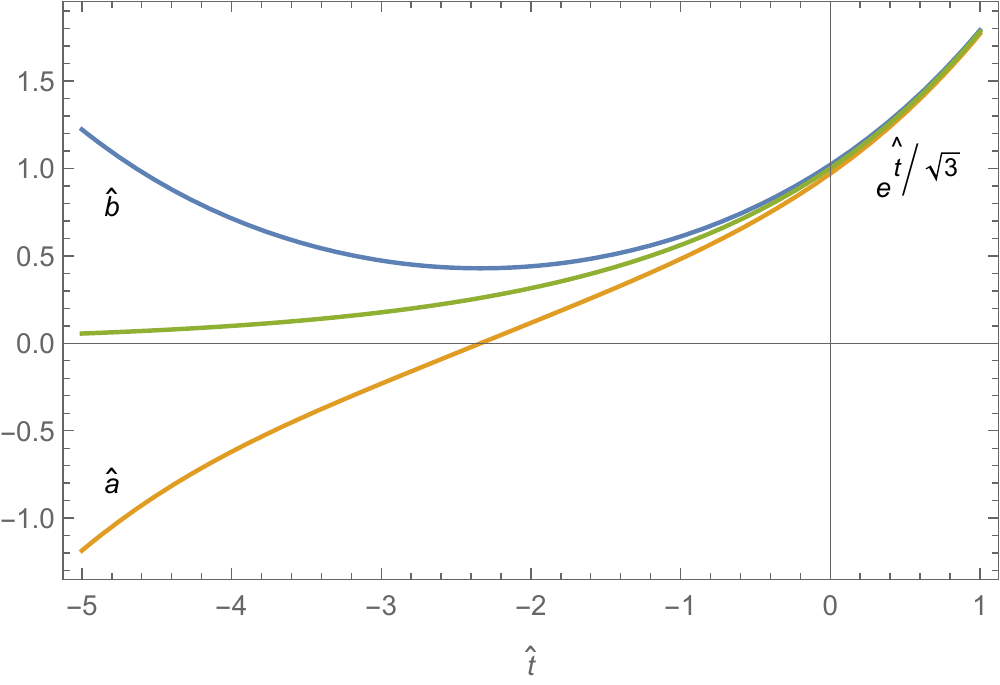}
    \includegraphics[width=7cm]{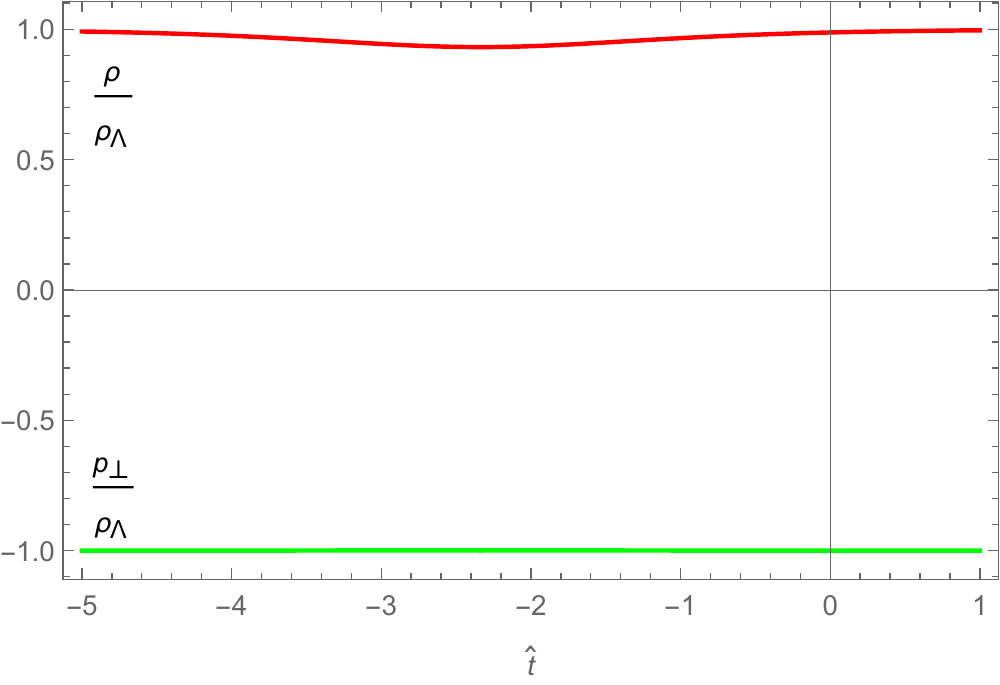}
     \includegraphics[width=7cm]{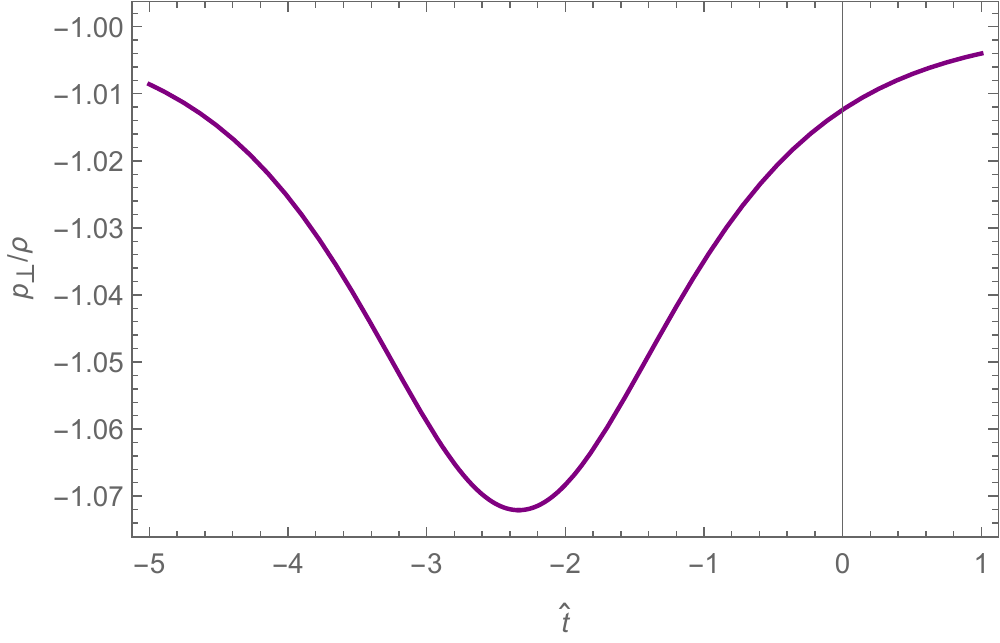}
       \includegraphics[width=7cm]{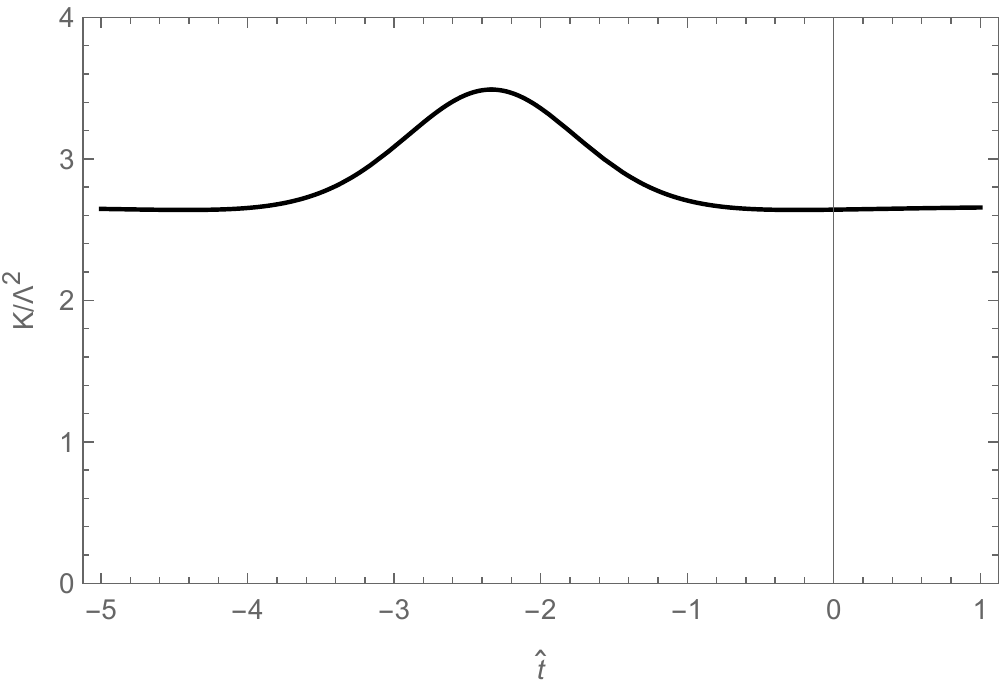}
       \includegraphics[width=7cm]{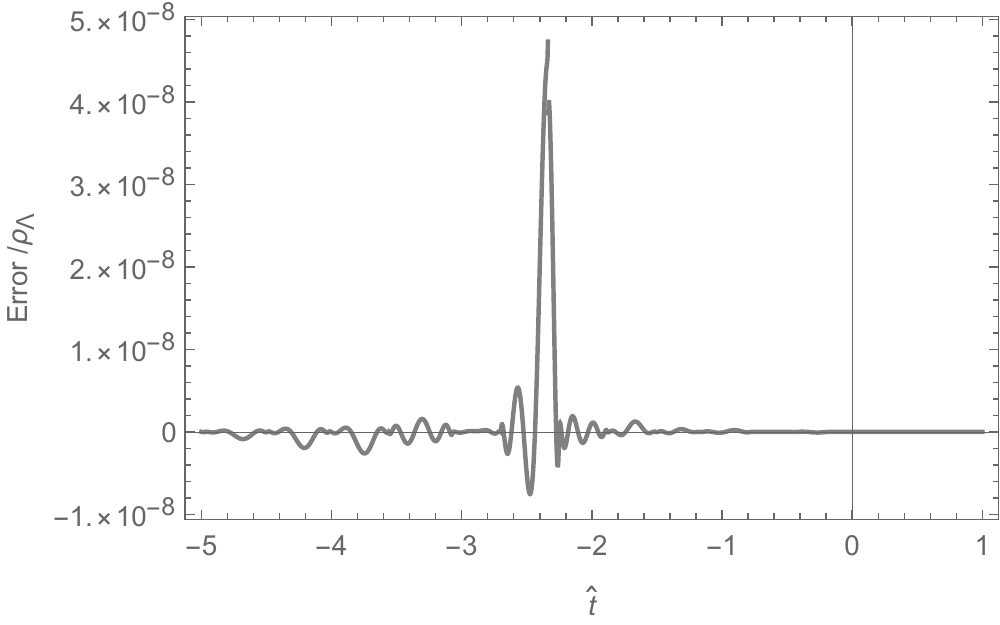}
    \caption{ Plots of the scale factors, energy-momentum tensor eigenvalues, ratio of the energy momentum tensor eigenvalues, Kretchman Scalar, and residual error in Mathematica's interpolating functions with regard to satisfying Eq.~(\ref{branchD}) for $X=1,Y=1$. In this case $\hat{a}$ goes through 0 at $\hat{t}=-2.336$. The energy-momentum tensor functions are similar to the $X=0,Y=1$ case in that at this point $\rho$ is minimum $p_\perp$ is maximum, although this is difficult to see as $p_\perp$ does not vary much, and the Null energy condition is violated. It differs in that the Kretchman scalar at the $\hat{a}=0$ point is above its background configuration value, while in the $X=0,Y=1$ case the Kretchman scalar was below its background value.}
    \label{NJpp}
\end{figure}

\begin{figure}[H]
    \centering
    \includegraphics[width=7cm]{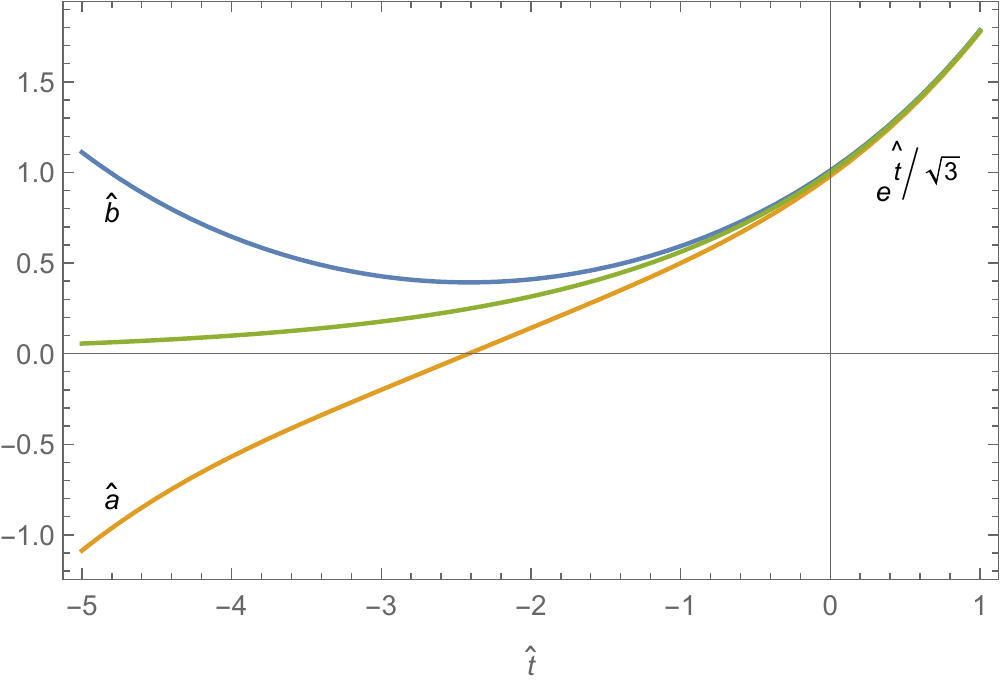}
    \includegraphics[width=7cm]{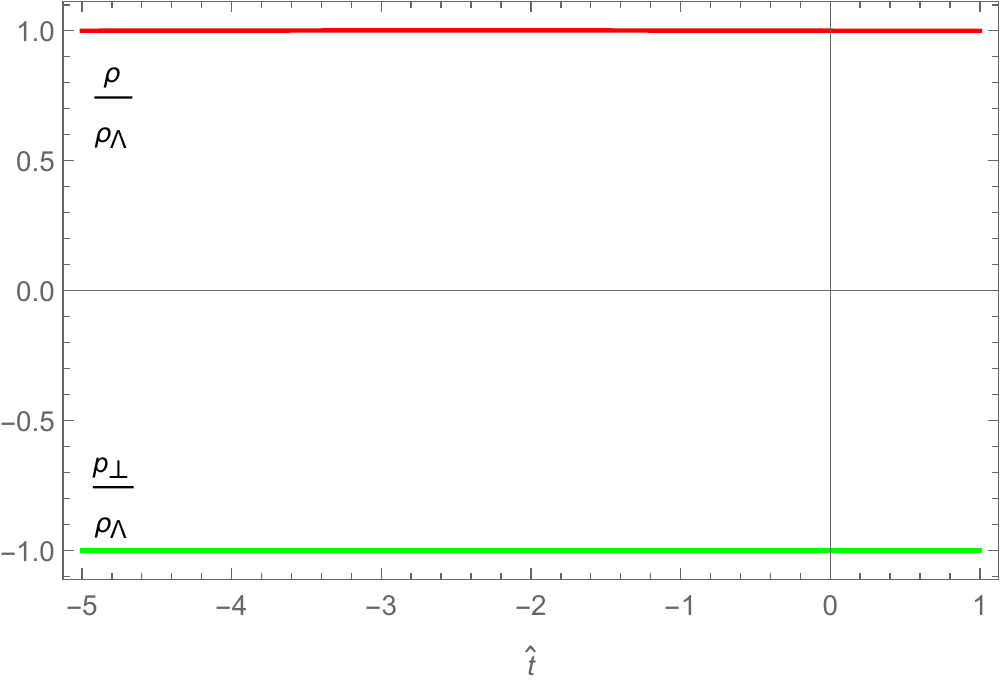}
     \includegraphics[width=7cm]{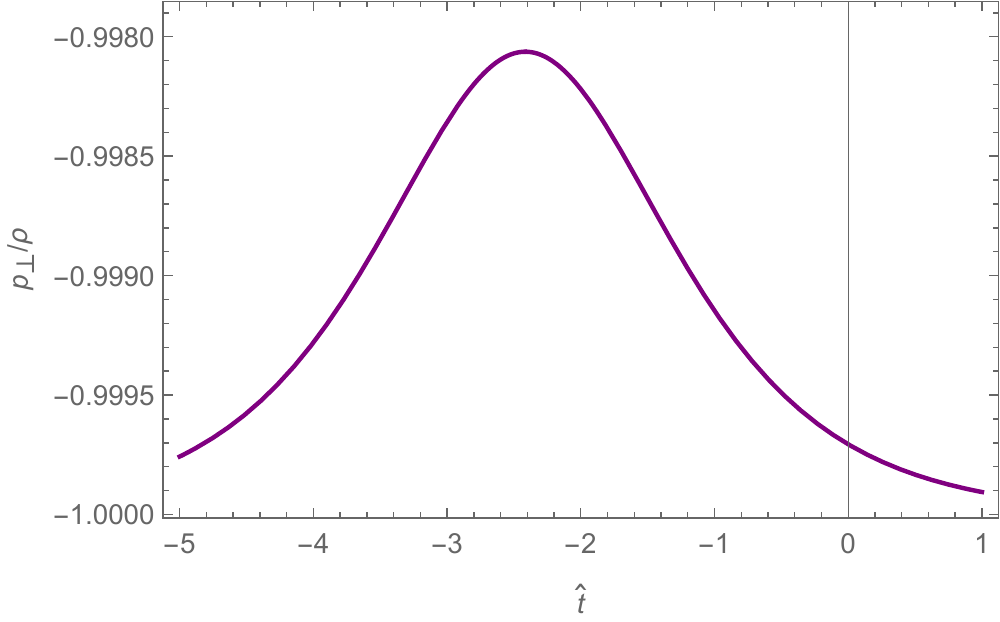}
       \includegraphics[width=7cm]{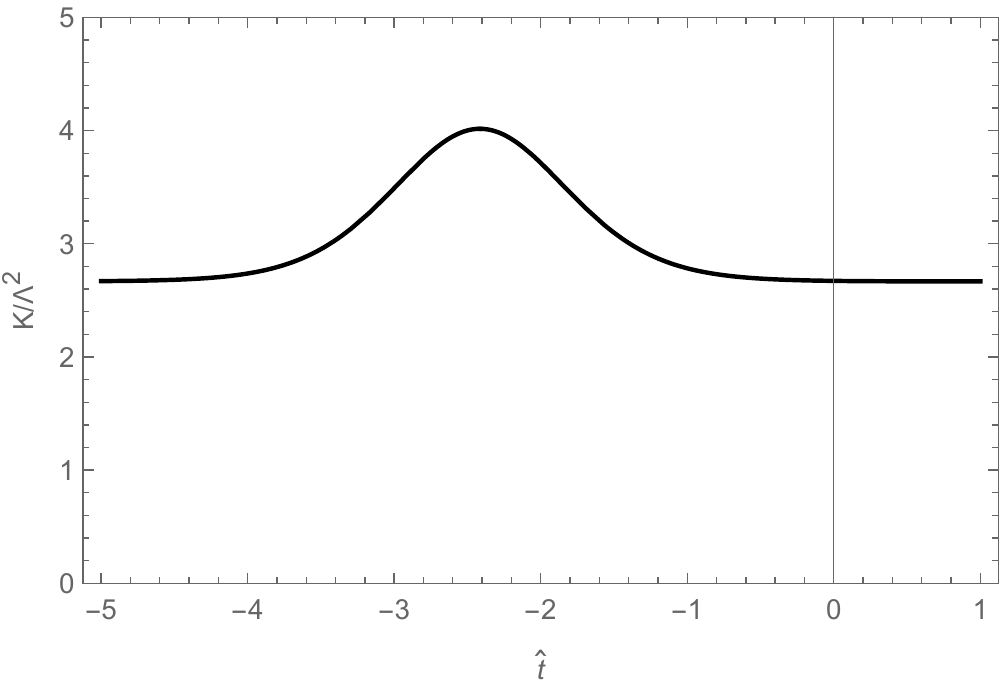}
       \includegraphics[width=7cm]{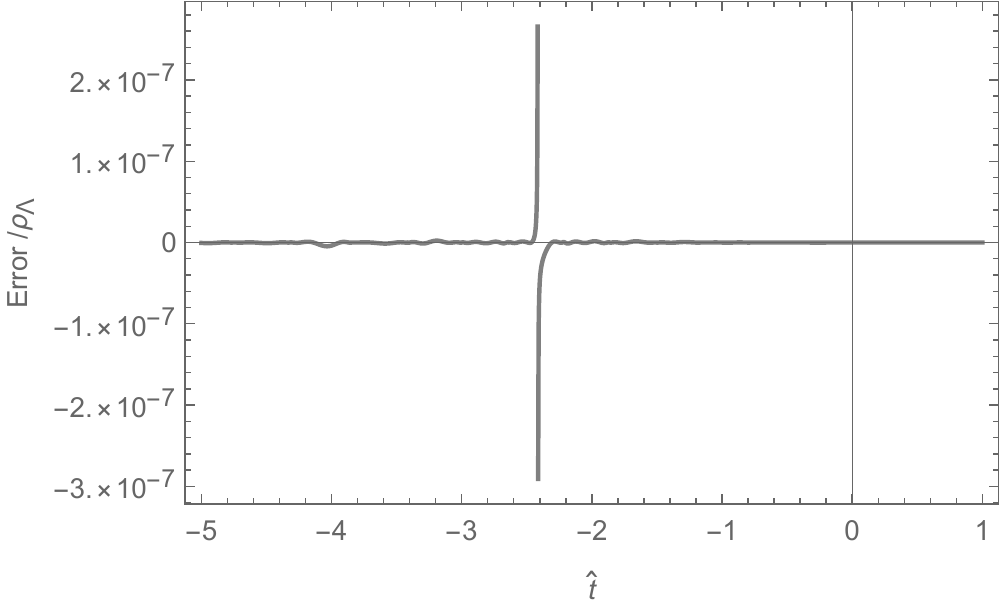}
    \caption{ Plots of the scale factors, energy-momentum tensor eigenvalues, ratio of the energy momentum tensor eigenvalues, Kretchman Scalar, and residual error in Mathematica's interpolating functions with regard to satisfying Eq.~(\ref{branchD}) for $X=1,Y=0$. In this case $\hat{a}$ goes through 0 at $\hat{t}=-2.415$. The scale factors and Kretchman scalar have very similar behavior to the $X=1,Y=1$ case. This case differs in that the energy-momentum tensor functions barely change, although they seem to go through a maximum at the $\hat{a}=0$ point and since $p_\perp/\rho\ge-1$ the Null energy condition is satisfied. This in important because to first order in $\epsilon$, $X$ did not show up to first order in $\epsilon$ in the energy-momentum tensor Eqs. (\ref{pertden},\ref{pertp}) so we did not know this from perturbative analysis.}
    \label{NJpz}
\end{figure}

\begin{figure}[H]
    \centering
    \includegraphics[width=7cm]{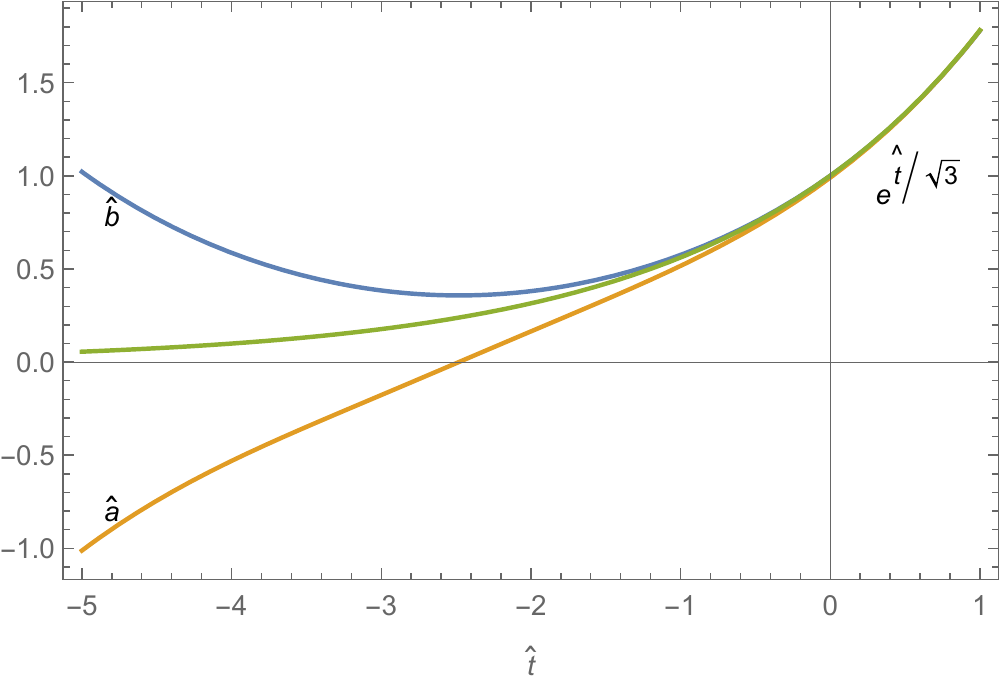}
    \includegraphics[width=7cm]{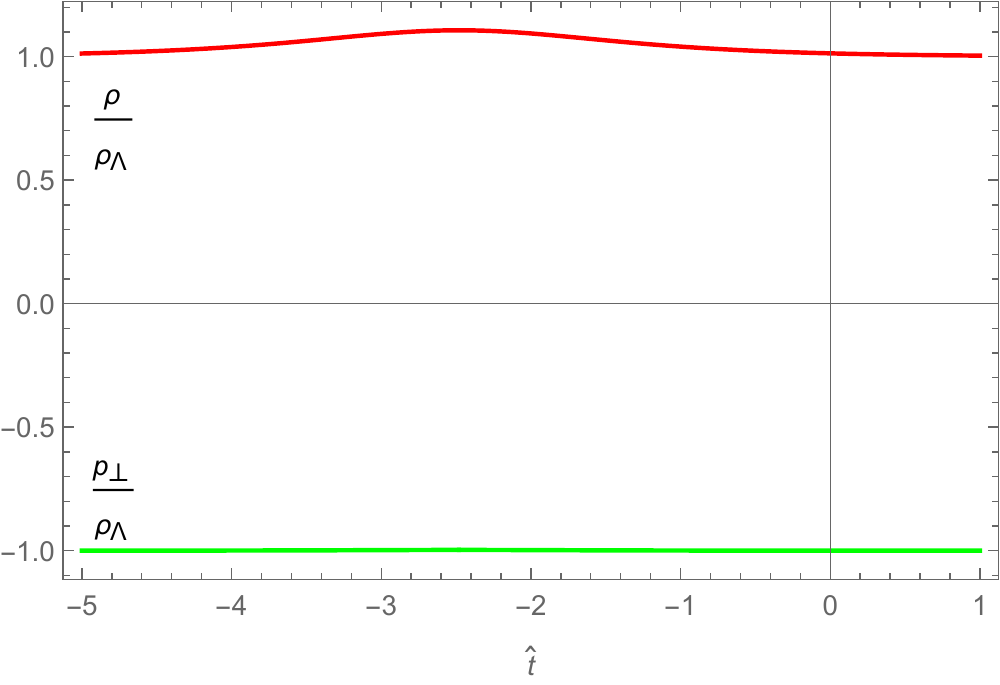}
     \includegraphics[width=7cm]{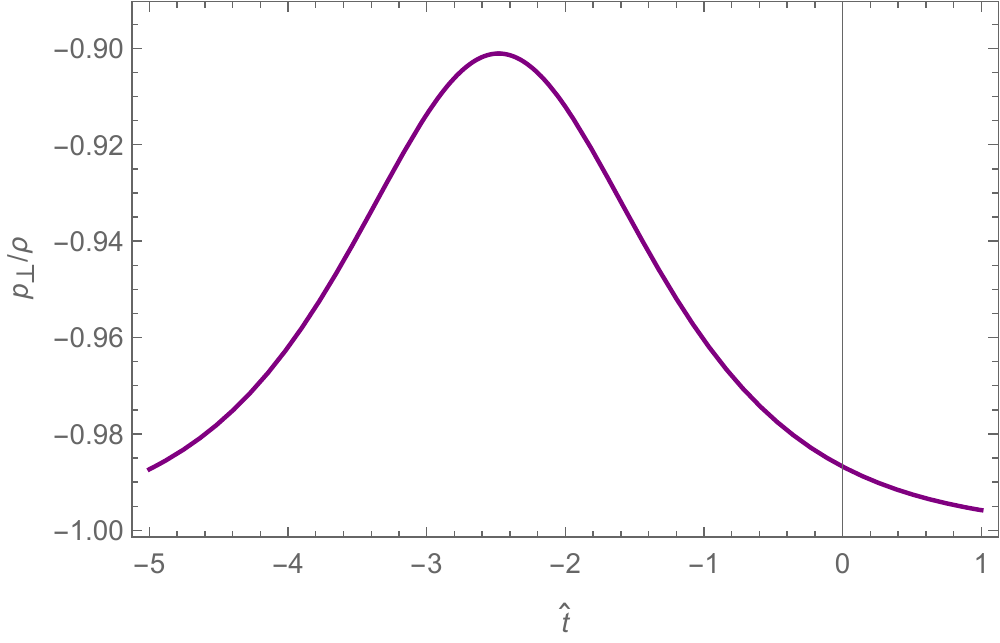}
       \includegraphics[width=7cm]{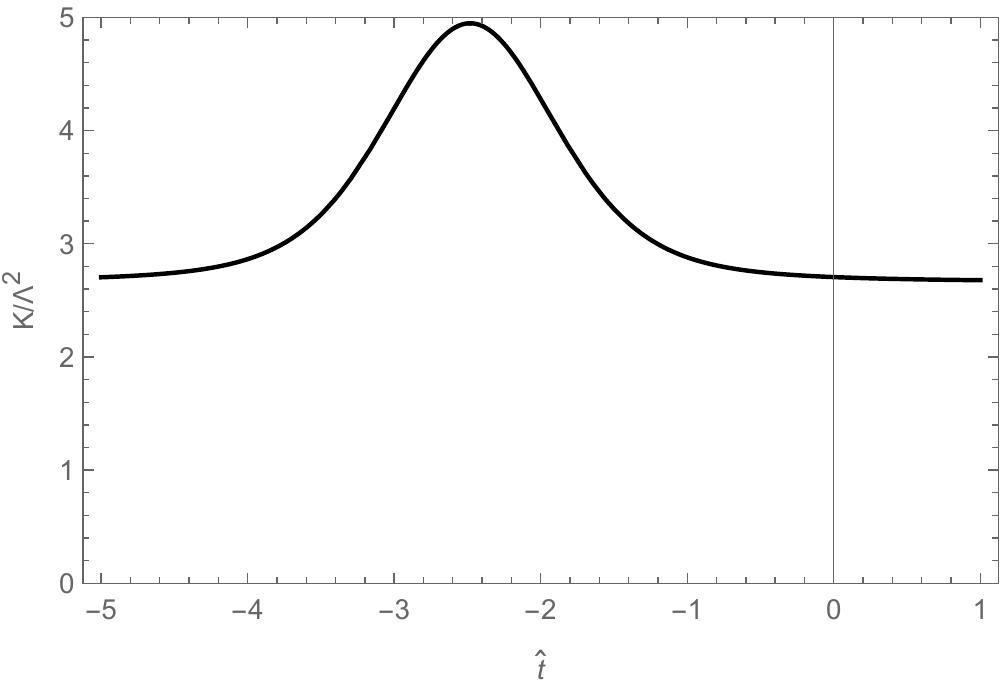}
       \includegraphics[width=7cm]{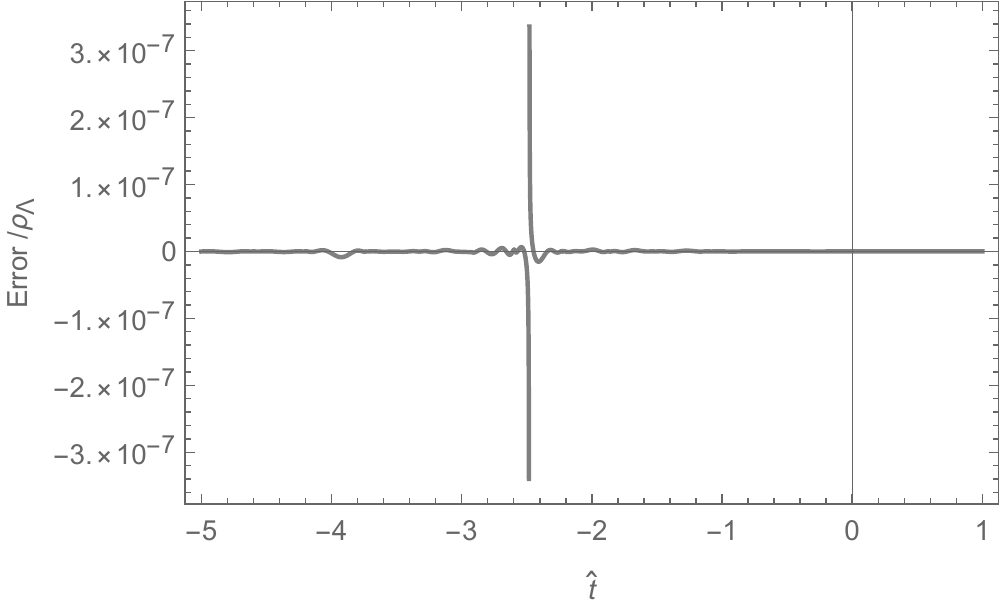}
    \caption{ Plots of the scale factors, energy-momentum tensor eigenvalues, ratio of the energy momentum tensor eigenvalues, Kretchman Scalar, and residual error in Mathematica's interpolating functions with regard to satisfying Eq.~(\ref{branchD}) for $X=1,Y=-1$. In this case the critical point is at $\hat{t}=-2.481$. The behavior of this case is qualitatively similar to $X=1,Y=0$, except that the variation in density is higher more obvious on the graph.}
    \label{NJpm}
\end{figure}

\begin{figure}[H]
    \centering
    \includegraphics[width=7cm]{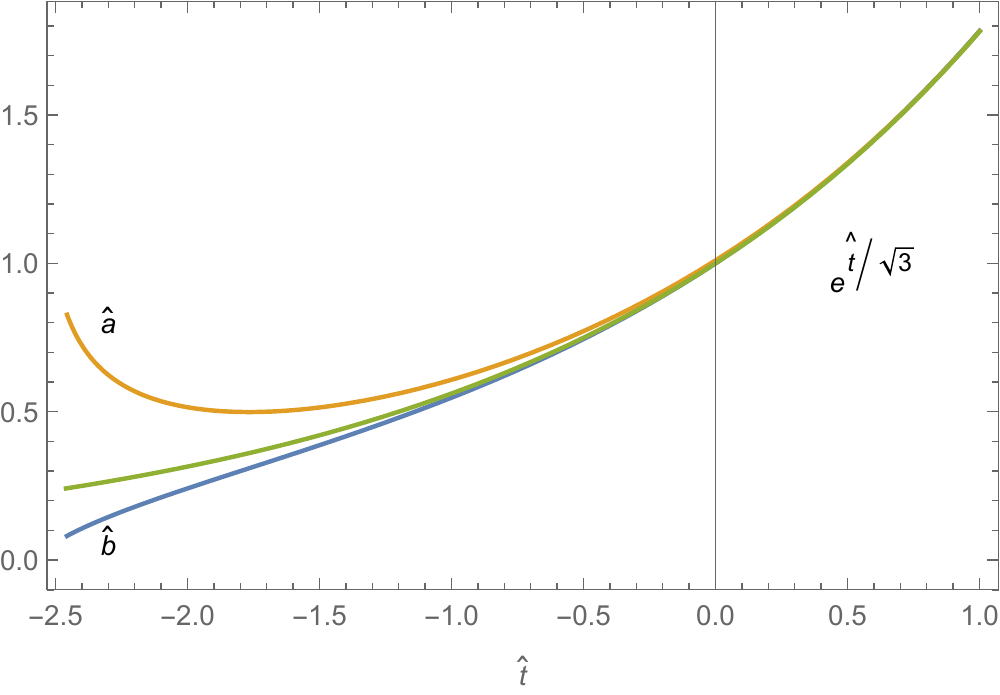}
    \includegraphics[width=7cm]{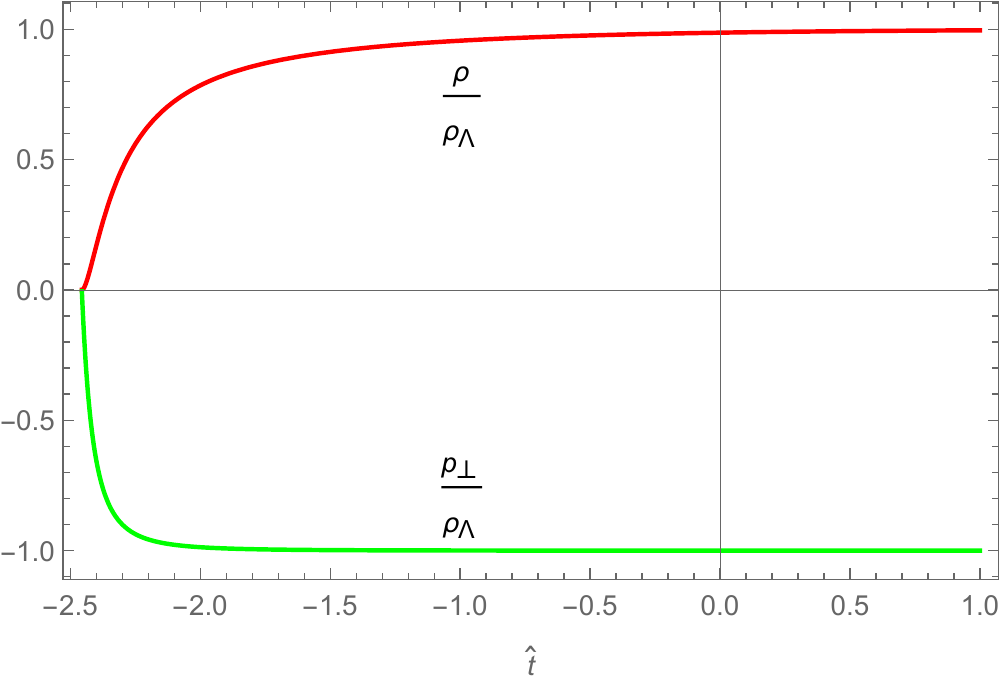}
     \includegraphics[width=7cm]{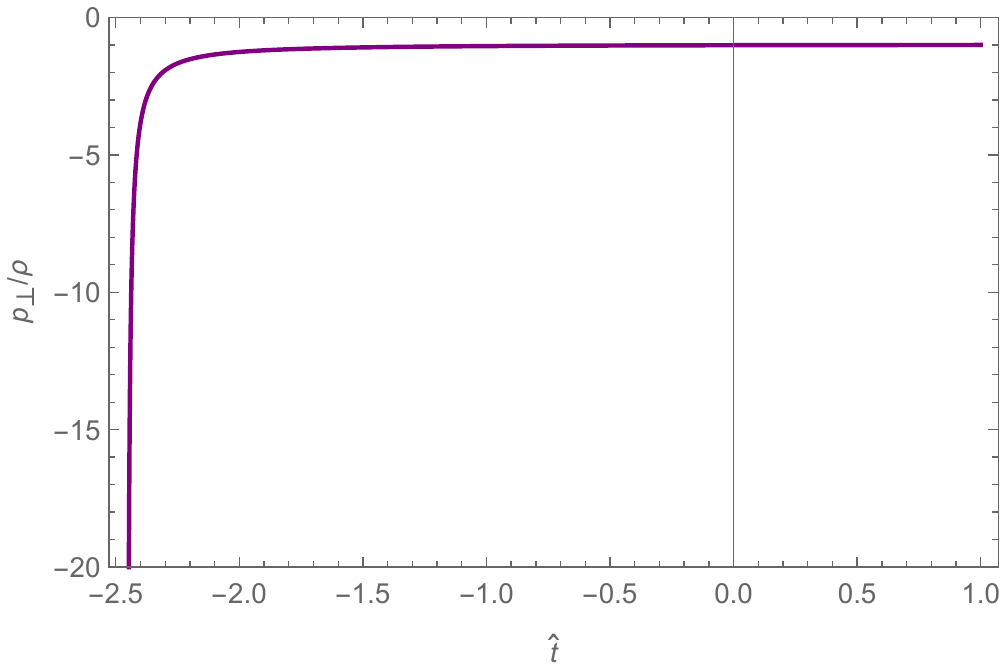}
       \includegraphics[width=7cm]{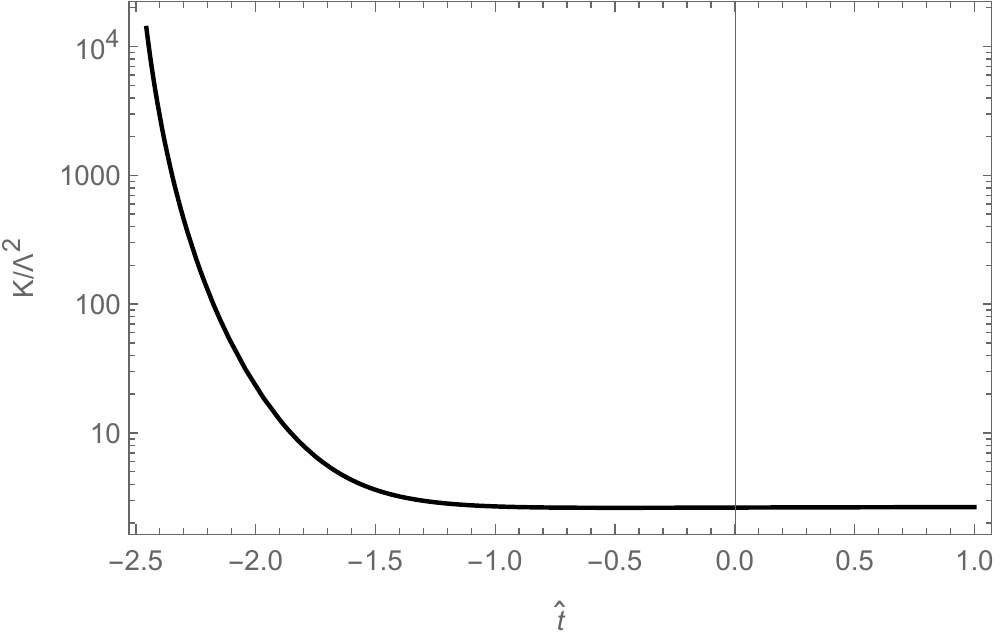}
       \includegraphics[width=7cm]{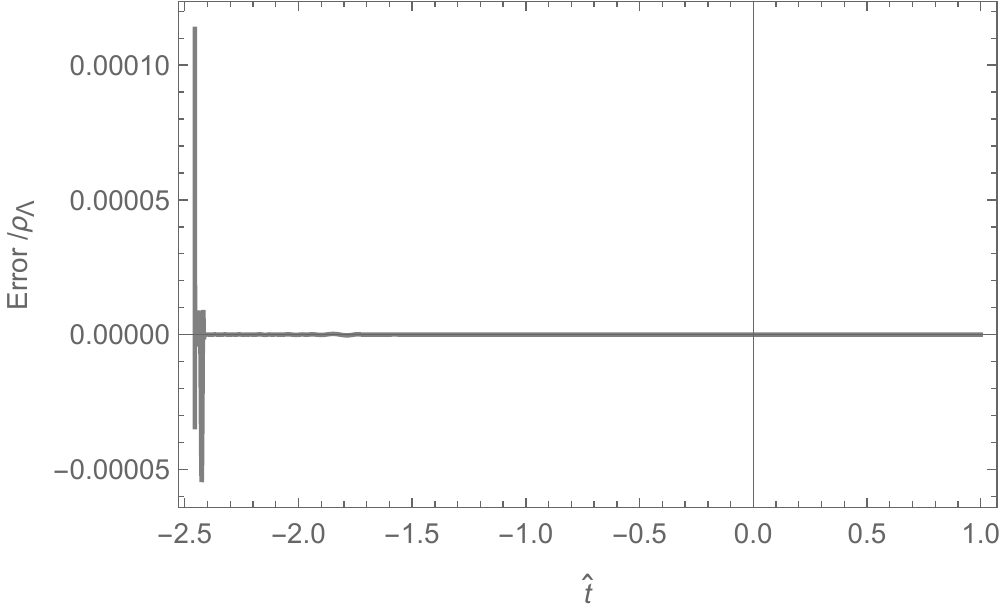}
    \caption{ Plots of the scale factors, energy-momentum tensor eigenvalues, ratio of the energy momentum tensor eigenvalues, Kretchman Scalar, and residual error in Mathematica's interpolating functions with regard to satisfying Eq.~(\ref{branchD}) for $X=-1,Y=1$. This case is noteworthy that both the density and transverse pressure reach zero at $\hat{t}\approx-2.456$. As we approach this point, the density tends toward zero faster than the transverse pressure, leading to a divergence in $p_\perp/\rho$. The energy-momentum terms, and hence Ricci sector of the curvature, go to zero.  It appears the Kretchman scalar may diverge at that point, indicating a curvature singularity in the Weyl sector. Since this solution goes to the boundary of the equation of state branches (\ref{branchD},\ref{branchU}), it is conceivable that this could be joined to a solution which is on branch (\ref{branchU}) at $\hat{t}<-2.456$. }
    \label{NJmp}
\end{figure}

\begin{figure}[H]
    \centering
    \includegraphics[width=7cm]{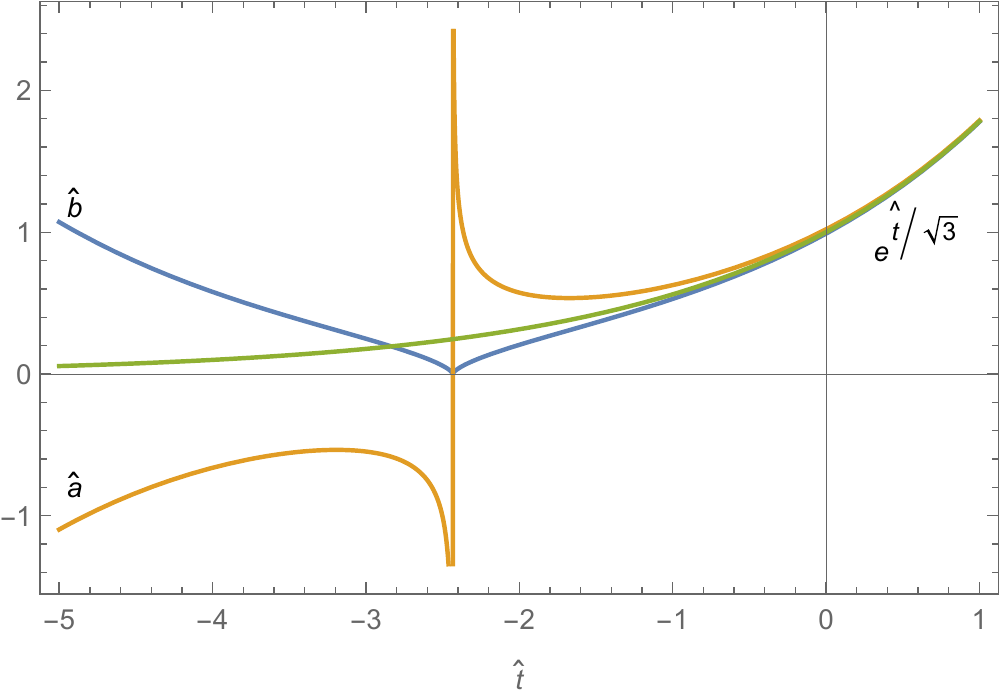}
    \includegraphics[width=7cm]{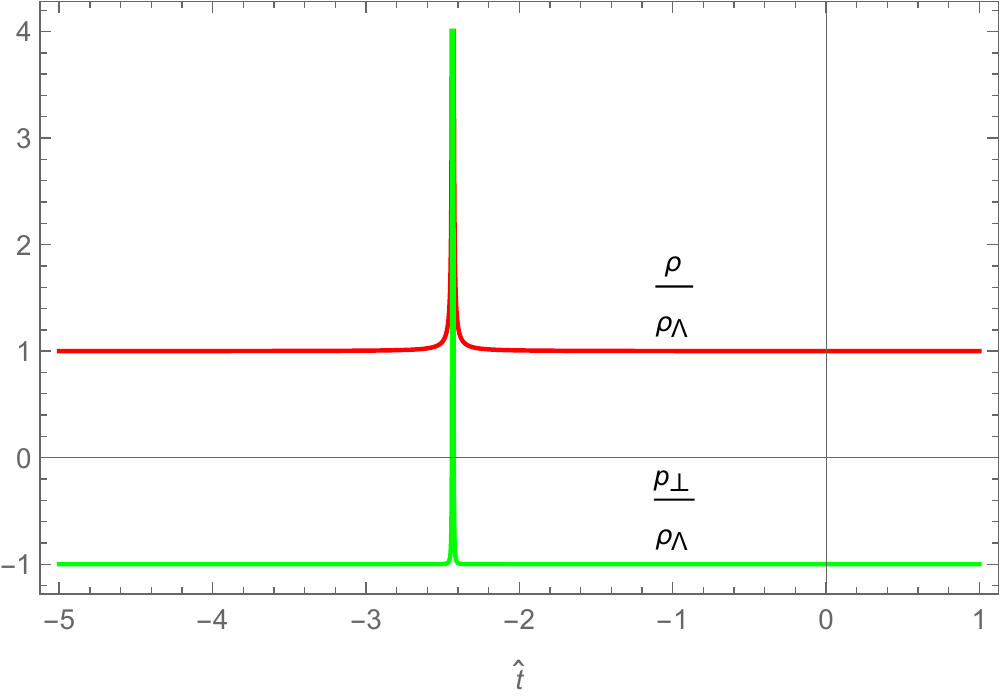}
     \includegraphics[width=7cm]{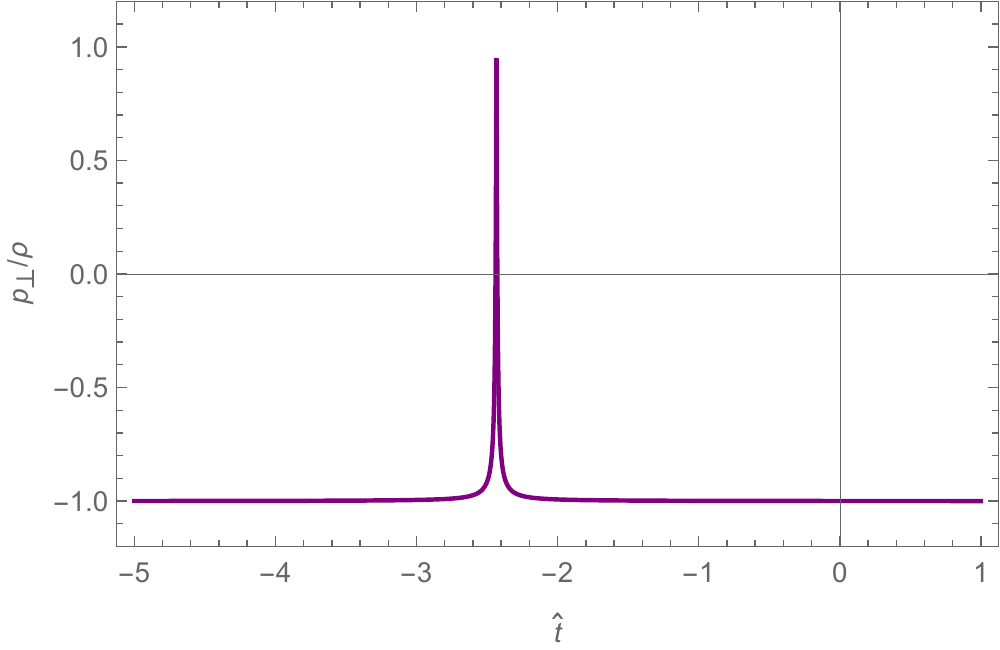}
       \includegraphics[width=7cm]{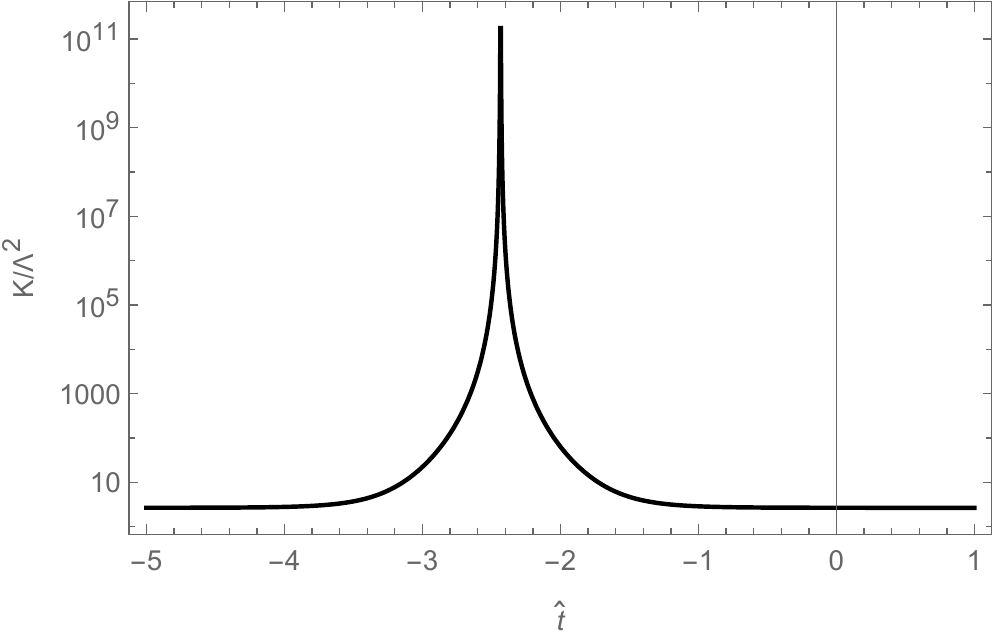}
       \includegraphics[width=7cm]{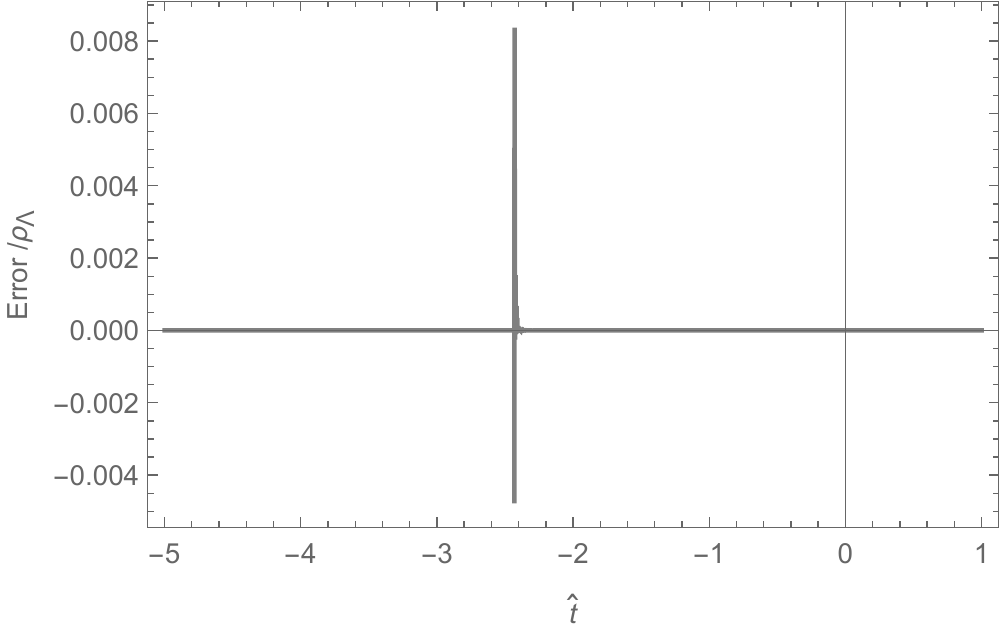}
    \caption{ Plots of the scale factors, energy-momentum tensor eigenvalues, ratio of the energy momentum tensor eigenvalues, Kretchman Scalar, and residual error in Mathematica's interpolating functions with regard to satisfying Eq.~(\ref{branchD}) for $X=-1,Y=0$. In this case, at $\hat{t}=-2.433$ we have $\hat{b}$ seeming to touch zero in a cusp like manner and $\hat{a}$ diverging asymmetrically. The energy density and transverse pressure also seem to diverge, and their ratio $p_\perp/\rho\rightarrow1$ which indicates electromagnetic like behavior. Notice that the error is somewhat higher in this case than some of the other cases, however this occurs when the density and pressure themselves are extremely high so the relative error with regard to the equation of state is still small. The Kretchman scalar also appears to diverge at this point. }
    \label{NJmz}
\end{figure}

\begin{figure}[H]
    \centering
    \includegraphics[width=7cm]{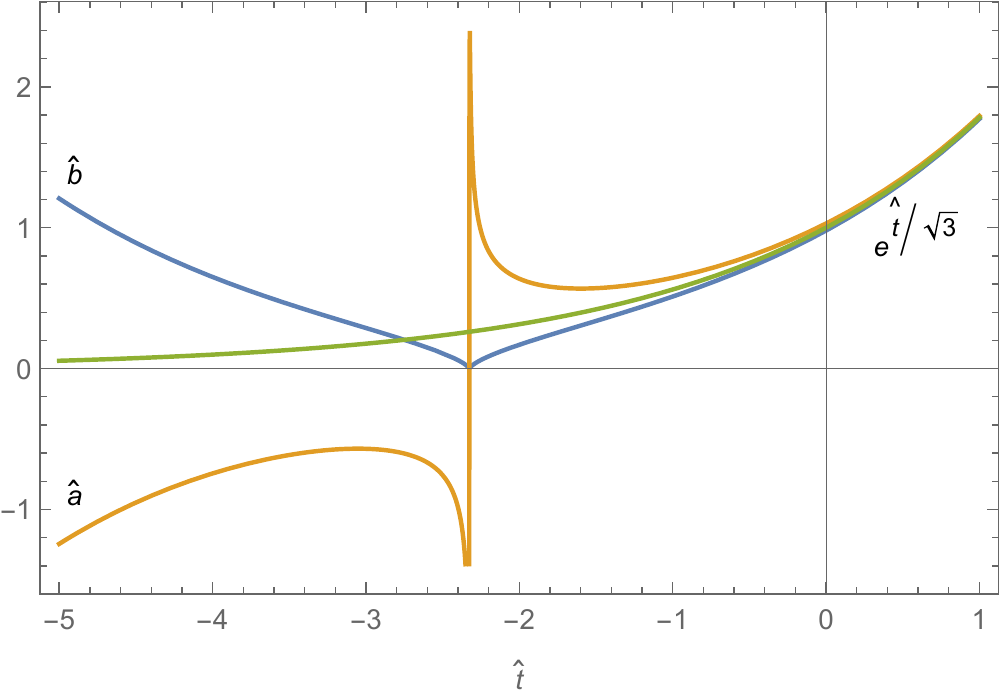}
    \includegraphics[width=7cm]{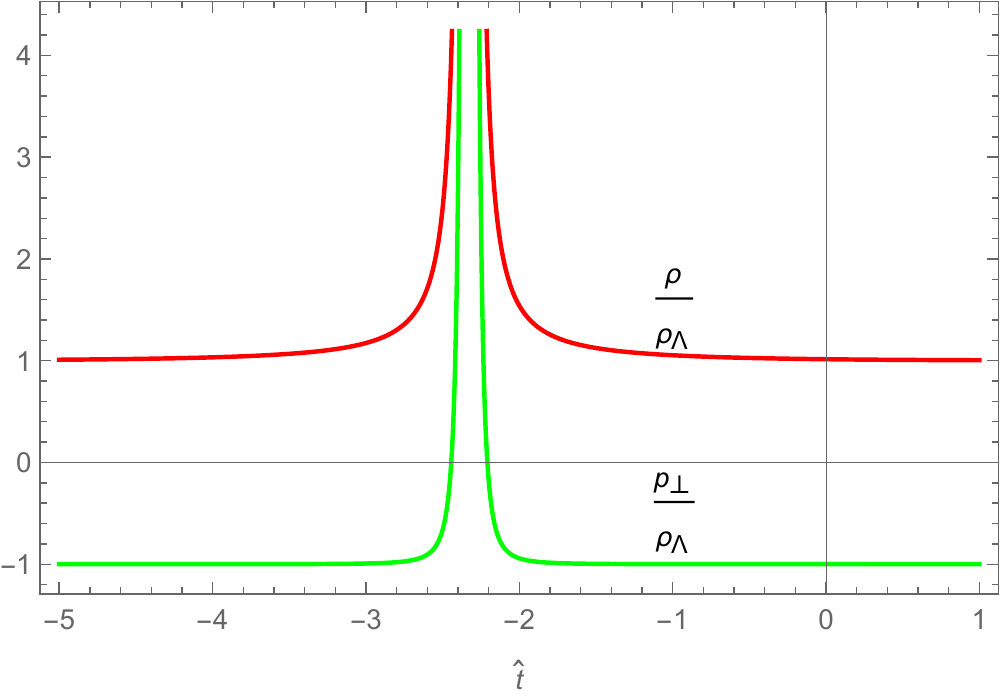}
     \includegraphics[width=7cm]{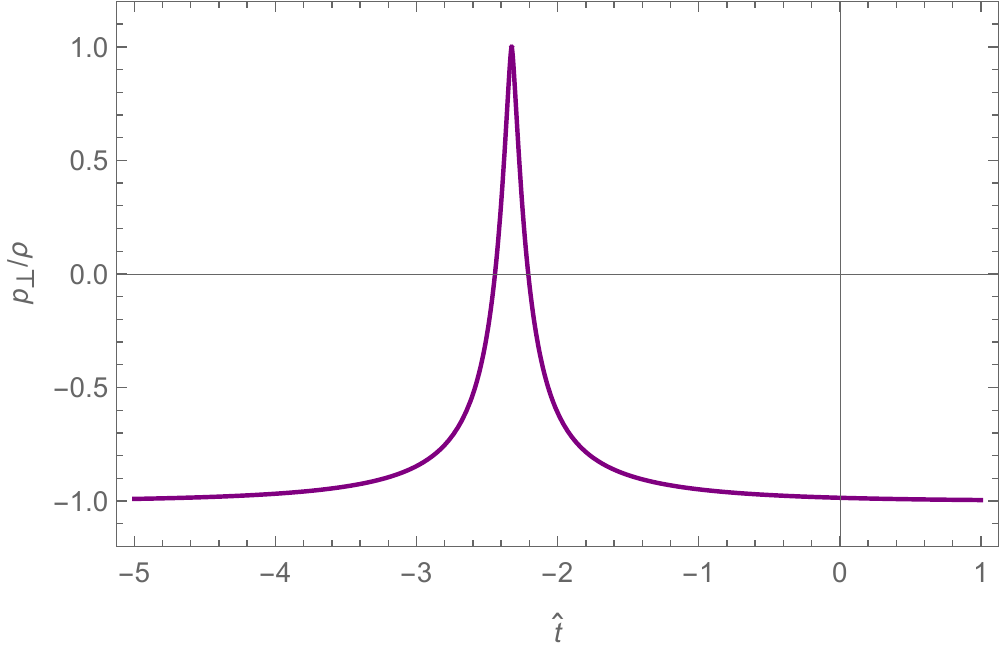}
       \includegraphics[width=7cm]{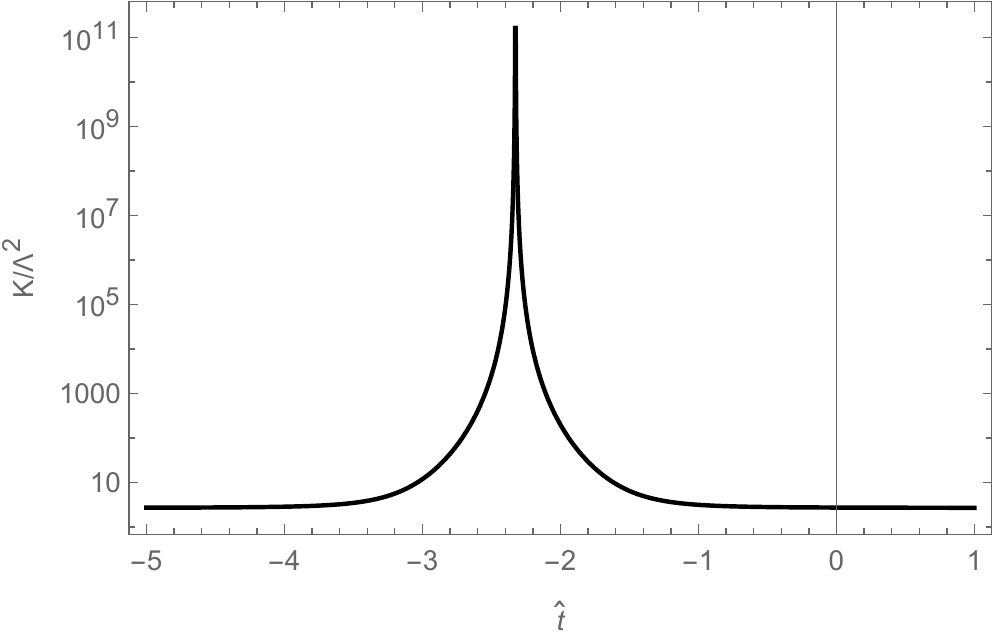}
       \includegraphics[width=7cm]{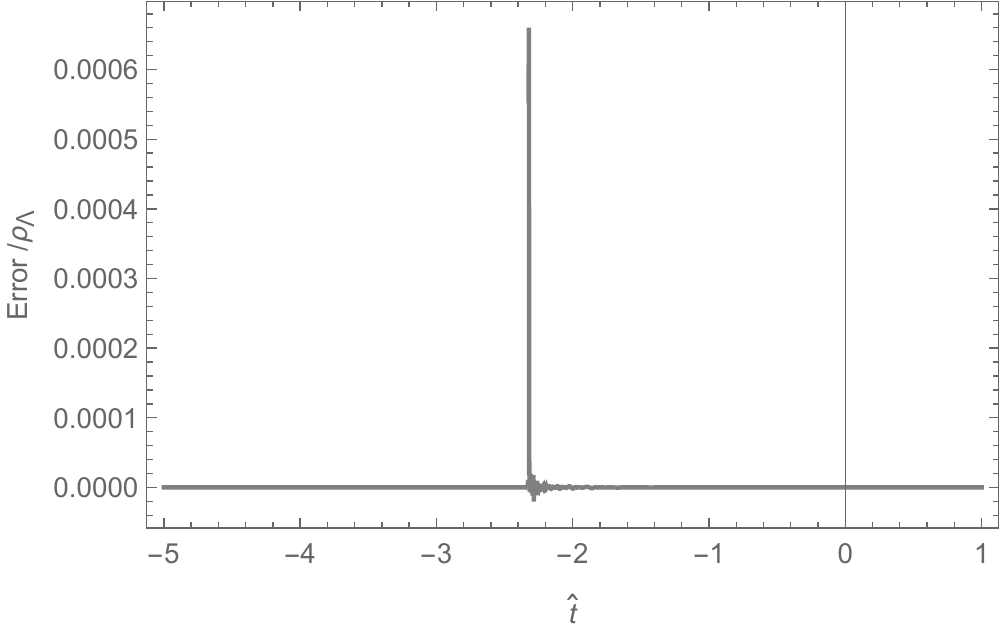}
    \caption{ Plots of the scale factors, energy-momentum tensor eigenvalues, ratio of the energy momentum tensor eigenvalues, Kretchman Scalar, and residual error in Mathematica's interpolating functions with regard to satisfying Eq.~(\ref{branchD}) for $X=-1,Y=-1$. In this case the critical point is at $\hat{t}=-2.326$. The behavior is extremely similar to the $X=-1,Y=0$ case except that the peaks in density and transverse pressure are wider.}
    \label{NJmm}
\end{figure}

 With the perturbative solution in hand, we can use it to find appropriate initial conditions and then use numerical methods to retrodict the behavior further in the past.
 For the numerical work, the plots are shown in variables of 
 \begin{align}
     \hat{b}=\frac{b}{A},\qquad \hat{a}=\frac{a}{cA\sqrt{\Lambda/3}},\qquad \hat{t}=\sqrt{\Lambda}t\label{Nvars}.
 \end{align}
 and we specifically use the lower branch Eq.~(\ref{branchD}) and $k=0$ which is appropriate near the vacuum energy-de Sitter background case, such that the differential equation becomes
 \begin{align}
     -\frac{2\ddot{b}}{b}-\frac{\overset{\therefore}{b}}{\dot{b}}=\frac{\dot{b}^2+2 b\ddot{b}}{b^2}-2\sqrt{\Lambda\frac{\dot{b}^2+2 b\ddot{b}}{b^2}}.\label{NDEq1}
 \end{align}
In terms of the rescaled $\hat{t}$ and $\hat{b}$ variables, this becomes
\begin{align}
   -\frac{2\frac{d^2 \hat{b}}{d\hat{t}^2}}{\hat{b}}-\frac{\frac{d^3 \hat{b}}{d\hat{t}^3}}{\frac{d \hat{b}}{d\hat{t}}}=\frac{\frac{d \hat{b}}{d\hat{t}}^2+2 \hat{b}\frac{d^2 \hat{b}}{d\hat{t}^2}}{\hat{b}^2}-2\sqrt{\frac{\frac{d \hat{b}}{d\hat{t}}^2+2 \hat{b}\frac{d^2 \hat{b}}{d\hat{t}^2}}{\hat{b}^2}}.\label{NDeq2}
\end{align}
 To examine the possible behaviors, we examine the cases when at $t=0$, the boundary conditions are given by the perturbative solution (with $Z=0$ since it is gauge), or
 \begin{align}
     \hat{b}\rightarrow e^{\hat{t}/\sqrt{3}}\Big(1+\epsilon \big(Y e^{-2\hat{t}/\sqrt{3}}+X e^{-\sqrt{3}\hat{t}}\big)\Big) 
 \end{align}
 with $\epsilon=1/100$, $X=-1,0,1$ and $Y=-1,0,1$, discounting the $X=0,~Y=0$ case because we know the analytic solution will be the unperturbed vacuum energy-de Sitter background. We plot the scale factors, energy momentum tensor eigenvalues, the ratio of energy momentum tensor eigenvalues, Kretchman scalar, and error parameter for these cases in Figures \ref{NJzp}-\ref{NJmm}. The error parameter is computed by using  Mathematica's \cite{Mathematica} interpolating function\footnote{The default settings led to a large error parameter near the critical point and $\hat{t}=0$, it is important to set InterpolationOrder$\rightarrow$All. The graphs were generated using the ExplicitRungeKutta, DifferenceOrder$\rightarrow8$ settings, although these settings have lesser effect on the error near the critical point. } for $b$, taking derivatives of the interpolating function to obtain $p_\perp$ and $\rho$ via Eqs.~(\ref{rhoB}, \ref{pTB}), then computing
 \begin{align}
     Error=p_\perp-\rho+2\sqrt{\rho_\Lambda \rho},
 \end{align}
 which would be zero for a perfect solution as a consequence of the equation of state.
 
 There are several noteworthy patterns in the various cases. In all cases, as we move forward in time, we approach the equilibrium solution $\hat{a}=\hat{b}=e^{\hat{t}/\sqrt{3}}$, with $\rho=\rho_\lambda=-p_\perp$ and $\mathcal{K}=8\Lambda^2/3$ which we expected from the perturbative analysis.  Also in all cases, there appears to be a critical point in time when $\hat{b}$ reaches a minimum value , which at least for these examples seems to be related largely to the value of $X$.
 
 For $X=0$, Figures (\ref{NJzp}, \ref{NJzm}), we have $\hat{b}$ reaching a local minimum and $\hat{a}=0$ at roughly $\hat{t}\approx-3.9$. For $X=1$ Figures (\ref{NJpm}, \ref{NJpz}, \ref{NJpp}), we also have a local minimum in $\hat{b}$ and $\hat{a}$ going through zero at roughly $\hat{t}\approx-2.5$. In both $X=0,1$ cases, despite the axial scale function $a$ going through zero, the Kretchman scalar remains regular, indicating the zero in axial scale factor may be due to a coordinate singularity. The energy-momentum tensor eigenvalues also remain regular in these cases, with violations of the Null energy condition occurring at the critical point only in the $Y=1$ cases. In these cases, since the real scale factors $a^2$ and $b^2$ transition between a shrinking and growing universe at the critical point, this could be described as a big bounce type of universe.

 For $X=-1$ with $Y=-1,0$ Figures (\ref{NJmm},\ref{NJmz}), the critical point appears to be a cusp like minimum in $\hat{b}$ and a pole like behavior in $\hat{a}$ at roughly $\hat{t}=-2.5$, the Kretchman scalar appears\footnote{Given the absence of an analytic solution we can not be certain if these are actual divergences or extremely high values.} to diverge indicating a curvature singularity. The energy-momentum tensor eigenvalues also appear to diverge and approach the ``electromagnetic" configuration $p_\perp=\rho$ at the critical point. These universes still have the transverse scale factor $b^2$ behaving in a big bounce sort of matter, but the axial scale factor $a^2$ has a more complicated behavior featuring local minima at either side of the critical point and a maximum or divergence at the critical point.
 
 The case $X=-1,Y=1$ Figure (\ref{NJmp}) is unusual in that the solution does not pass though the critical point, although it looks similar to the cusp/pole behavior in the other $X=-1$ cases. This is likely because, uniquely among all the considered cases, this case goes through the zero density point in the equation of state. It is possible that this solution could be joined to a solution on the upper branch Eq.~(\ref{branchU}). Because this $X=-1,Y=1$ solution approaches the zero density point on the equation of state from below, the null energy condition is violated. Despite the energy-momentum tensor eigenvalues going to zero, the Kretchman scalar is very high and increasing rapidly, possibly indicating the onset of a singularity in the Weyl sector of the curvature. 
 
 \subsubsection{$X=0,~Y=1$}

\subsubsection{$X=0,~Y=-1$}
\subsubsection{$X=1,~Y=1$}
\subsubsection{$X=1,~Y=0$}
\subsubsection{$X=1,~Y=-1$}
\subsubsection{$X=-1,~Y=1$}
\subsubsection{$X=-1,~Y=0$}
\subsubsection{$X=-1,~Y=-1$}
\subsubsection{Summary of Critical Point Properties }
 We summarize the properties of the critical points in this subsection and Table \ref{crittable}. Notice that for $X=0$, the critical points are at $\hat{t}\approx-3.9$, while for $|X|=1$, the critical points are at $\hat{t}\approx-2.4$. For these cases, the null energy condition is violated if and only if $Y=1$. Despite the singular behavior of the metric functions at the critical point, it is only the cases where $X=-1$ where the Kretchman scalar seems to diverge. Finally, the cases $X=1,Y=0$ and $X=1,Y=-1$ have qualitatively the same behavior as each other at their critical points, as do the cases $X=-1,Y=0$ and $X=-1,Y=-1$.

\begin{table}[]
\begin{center}
\begin{tabular}{ |c|c|c|c|c| }
 X,Y&$\hat{t}$ & Type & Matter& $\mathcal{K}$ \\
 \hline
0,1& -3.896 & Local min, zero crossing & NEC violation & Below background value\\
0,-1& -3.851 & Local min, zero crossing & positive $p_\perp$ & Significantly above background value\\
1,1& -2.336 & Local min, zero crossing & NEC violation & Above background value\\
1,0 & -2.415 & Local min, zero crossing & NEC satisfied & Above background value\\
1,-1 & -2.481 & Local min, zero crossing & NEC satisfied & Above background value\\
-1,1& -2.456 & One sided & $p_\perp,\rho\rightarrow0$ & Possibly divergent\\
-1,0 &-2.433 & Cusp, pole & Approaches EM limit & Probably divergent\\
-1,-1 &-2.326 & Cusp, pole & Approaches EM limit & Probably divergent \\
\hline
\end{tabular}
\caption{Table summarizing the critical points of the different $X,Y$ cases.}
\end{center}
\label{crittable}
\end{table}

\section{Disordered Systems with Isotropic Averaging}
\label{isoavg}
In the previous sections \ref{derivation section} and \ref{NJsec}, we considered systems for which there was one particular spatial axis $z$ throughout all of space along which $p_\parallel=-\rho$. Another possibility that can be considered is that different regions of space have different directions of the preferred spatial axis, like crystal domains in a solid. If the domains are much larger than the region we are trying to describe, then the models involving metric (\ref{unimet}) and equations of state (\ref{unimatdef}), (\ref{unieos})
may be appropriate. If the domains are much smaller than the region we are trying to describe  and randomly oriented, then it may be appropriate to treat the system as a perfect fluid FLRW system with the effective pressure and equation of state
\begin{align}
    p_{iso}=\frac{p_\parallel+2p_\perp}{3}=\frac{-\rho+2f(\rho)}{3}=g(\rho),\label{piso}
\end{align}
if the contribution to the energy-momentum tensor from topological defects can be neglected. A common metric parameterization for FLRW spacetime that has a similar structure to Eq.~(\ref{met}) is 
\begin{align}
   ds^2= -dt^2+a(t)^2\Big(\frac{dr^2}{1-kr^2}+r^2d\theta^2+r^2\sin^2\theta d\phi^2\Big).\label{Fmet}
\end{align}
Because FLRW systems in general are well known we will not perform as thorough of an analysis featuring other quantities, but we still use components of the energy-momentum tensor
\begin{align}
    T^t_{~t}=-\rho=\frac{-3 (k+\dot{a}^2)}{8\pi a^2},\qquad T^r_{~r}=T^{\theta}_{~\theta}=T^\phi_{~\phi}=p=\frac{-(k+\dot{a}^2)}{8\pi a^2}-\frac{\ddot{a}}{4\pi a}.\label{FTmv}
\end{align}
 The covariant energy conservation equation $\nabla_\mu T^\mu_{~\nu}$ gives 
\begin{align}
   \dot{\rho}=-3\frac{\dot{a}}{a}\Big(\rho+p\Big).\label{FTConservation}
\end{align}
\subsection{Isotropic averaging of simple examples}
The isotropicly averaged versions of the simple example stringy, electromagnetic, and vacuum equations of state (\ref{stringeos}) are
\begin{align}
    p_{iso}=\frac{-\rho+2(0)}{3}=\frac{-\rho}{3},\label{isostring}\\
    p_{iso}=\frac{-\rho+2(\rho)}{3}=\frac{\rho}{3},\label{isoEM}\\
    p_{iso}=\frac{-\rho+2(-\rho)}{3}=-\rho. \label{isovac}
\end{align}

Usage of Eqs.~(\ref{FTmv}) and (\ref{isostring})  leads to
\begin{align}
    \ddot{a}=0,\qquad a=X t+Y,\qquad \rho=\frac{3(X^2+k)}{8\pi(Y+Xt)^2}=-3p. \label{Fstringans}
\end{align}
This is a coasting cosmology, and has been previously analyzed in \cite{1989ApJ...344..543K}, in which the $p=-\rho/3$ matter was given the name ``K matter", and the interpretation as cosmic stings was briefly mentioned. For this universe, we have $\rho\propto a^{-2}$.

Usage of Eqs.~(\ref{FTmv}) and (\ref{isoEM}) leads to
\begin{align}
   k+ \dot{a}^2+a\ddot{a}=0,\qquad a=\sqrt{Y-kt^2+Xt},\qquad \rho=\frac{3(X^2+4kY)}{32\pi(Y-kt^2+Xt)^2}=3p. \label{FEMans}
\end{align}
Notice that $\rho\propto a^{-4}$, this can also be derived from the energy conservation Eq.~(\ref{FTConservation}), which is effectively the same as a radiation system. The scale factor has the same functional form, although with different constants, to the radiation plus curvature example from \cite{2006FoPh...36.1736A}. A version of the radiation plus curvature solution with the condition $a(t=0)=0$ is  also mentioned in \cite{Wald:1984rg}.

Usage of Eqs.~(\ref{FTmv}) and (\ref{isovac}) leads to
\begin{align}
  k+\dot{a}^2=a \ddot{a},\qquad a=X e^{\frac{\sqrt{\Lambda }
   t}{\sqrt{3}}}+\frac{3k e^{-\frac{\sqrt{\Lambda } t}{\sqrt{3}}}}{4\Lambda X},\qquad \rho=\Lambda/8\pi=-p.\label{DSK}
\end{align}
Notice that taking the $k\rightarrow0$ limit typically results in an inflationary scenario for this convention of the constants, unless one defines $X\propto k$ before taking the limit in which case it is a deflationary scenario.  Solutions to the Friedman equations for $\Lambda$ plus $k$ can also be found in \cite{2006FoPh...36.1736A}, again having a similar form to what is presented in Eq.~\ref{DSK}.

\subsection{Isotropic Averaging of EOS \ref{NJeos}}

\begin{figure}
    \centering
    \includegraphics{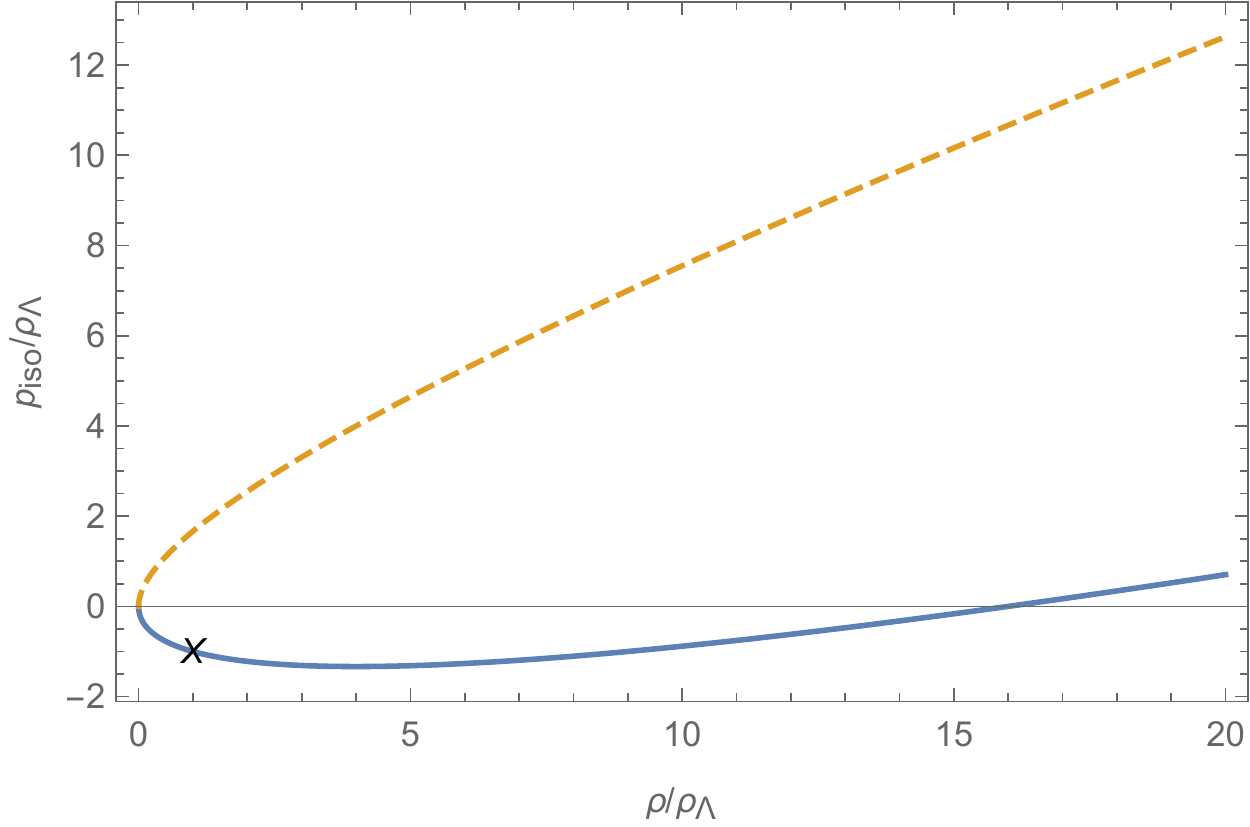}
    \caption{Plot of the equation of state branches for the isotropic version Eq. (\ref{NJisobranch}). This is analogous to Figure \ref{NJeosplot}. Notice the vacuum energy point (marked with X) is not the minimum of $p_{iso}$, while it was the minimum value for $p_\perp$ . }
    \label{EOSbranchiso}
\end{figure}

The isotropically averaged version of the equation of state \ref{NJeos} is
\begin{align}
    (\rho-3p_{iso})^2=16 \rho \rho_\Lambda,\label{NJeosiso}
\end{align}
where solving for $p_{iso}$ we obtain
\begin{align}
    p_{iso}=\frac{\rho \pm 4\sqrt{\rho \rho_\Lambda}}{3}.\label{NJisobranch}
\end{align}
We plot this in Figure \ref{EOSbranchiso}.

Putting the isotropically averaged equation of state in to the energy conservation Eq.~(\ref{FTConservation}) we obtain
\begin{align}
    \frac{\dot{\rho}}{4(\rho \pm \sqrt{\rho \rho_\Lambda})}=-\frac{\dot{a}}{a},
\end{align}
which leads to the following proportionality relation upon integration
\begin{align}
    \sqrt{\rho} \pm \sqrt{\rho_\Lambda} \propto a^{-2}.\label{NJisoprop}
\end{align}
Unlike the anisotropic case, here it is possible to find a closed form solution for $a(t)$. Equations  ~(\ref{FTmv}) and (\ref{NJeosiso}) give
\begin{align}
    4 \Lambda (k+\dot{a}^2)=\frac{3(k+\dot{a}^2+a \ddot{a})^2}{a^2}.\label{NJisoDEQ}
\end{align}
\subsubsection{Nonzero k cases}
 For nonzero $k$, a convenient form for $a$ satisfying Eq.~(\ref{NJisoDEQ}) is
\begin{align}
    a=\sqrt{\frac{3 k-Y\pm 2  \sqrt{3 k Y} \sinh \left(\frac{2 \sqrt{\Lambda }
   (t+Z)}{\sqrt{3}}\right)}{4\Lambda }}. \label{NJaiso}
\end{align}
 Notice that we have to be very careful with this solution because $a^2$ may not be positive for particular values of the parameters, for instance $k$ and $Y$ must not have the opposite sign \footnote{ Most of this subsection concerns arbitrary $k,Y$ of the same sign. Instead setting $k$ or $Y$ to zero in Eq.~(\ref{NJaiso}) gives a static universe with
 \begin{align}
     (Y\rightarrow0),~a=\sqrt{\frac{3k}{4\Lambda}},\qquad \rho=\frac{\Lambda}{2\pi},\qquad p_{iso}=-\frac{\Lambda}{6\pi},
 \end{align}
or empty universe with
 \begin{align}
     (k\rightarrow0),~a=\sqrt{\frac{-Y}{4\Lambda}},\qquad \rho=0,\qquad p_{iso}=0.
 \end{align}
However, it is more useful to set $k\rightarrow0$ in Eq.~(\ref{NJisoDEQ}) and solve again, which we do in Eq.~(\ref{NJaisokz}).}. The proportionality relation Eq.~(\ref{NJisoprop}) implies $\rho=(\sqrt{\rho_\Lambda}+c/a^2)^2$ for some constant $c$, using Eq.~(\ref{NJaiso}) and Eq.(\ref{FTmv}) we can find 
 \begin{align}
     \rho=\Big(\sqrt{\rho_\Lambda}+\frac{3k+Y}{a^2\sqrt{128\pi\Lambda}}\Big)^2.\label{NJisoden}
 \end{align}
 The isotropic average pressure is
 \begin{align}
    p_{iso}=\frac{3 k (7 Y-6 k)-9 k Y \cosh \left(\frac{4 \sqrt{\Lambda }
   (t+Z)}{\sqrt{3}}\right) \pm 4  \sqrt{3 k Y} (Y-6 k) \sinh \left(\frac{2 \sqrt{\Lambda
   } (t+Z)}{\sqrt{3}}\right)}{192\pi \Lambda a^4}.
 \end{align}
 For analysis of this spacetime, we can introduce the shorthand 
 \begin{align}
     T=\frac{\sqrt{\Lambda}(t+Z)}{\sqrt{3}},
 \end{align}
such that $a$ from Eq.~(\ref{NJaiso}) is zero when 
 \begin{align}
     T=\frac{1}{2}\sinh^{-1}\big(\frac{Y-3k}{\pm 2\sqrt{3kY}}\big). \label{Tzk}
 \end{align}
 The minus branch in Eq.~(\ref{NJaiso}) and here corresponds to a deflating universe approaching $a=0$ and ending at the given $T$, whereas the plus branch is an inflating universe starting at the given $T$. Regardless of whether it is inflating or deflating, at $T$ far away from the given value (\ref{Tzk}) the configuration is nearly vacuum energy while near the given value of $T$ the density approaches infinity and the equation of state approaches $p_{iso}=\rho/3$. This indicates that a universe of this nature evolves between a radiation like configuration and a vacuum energy like configuration.
 
\begin{figure}
     \centering
     \includegraphics[width=5.5cm]{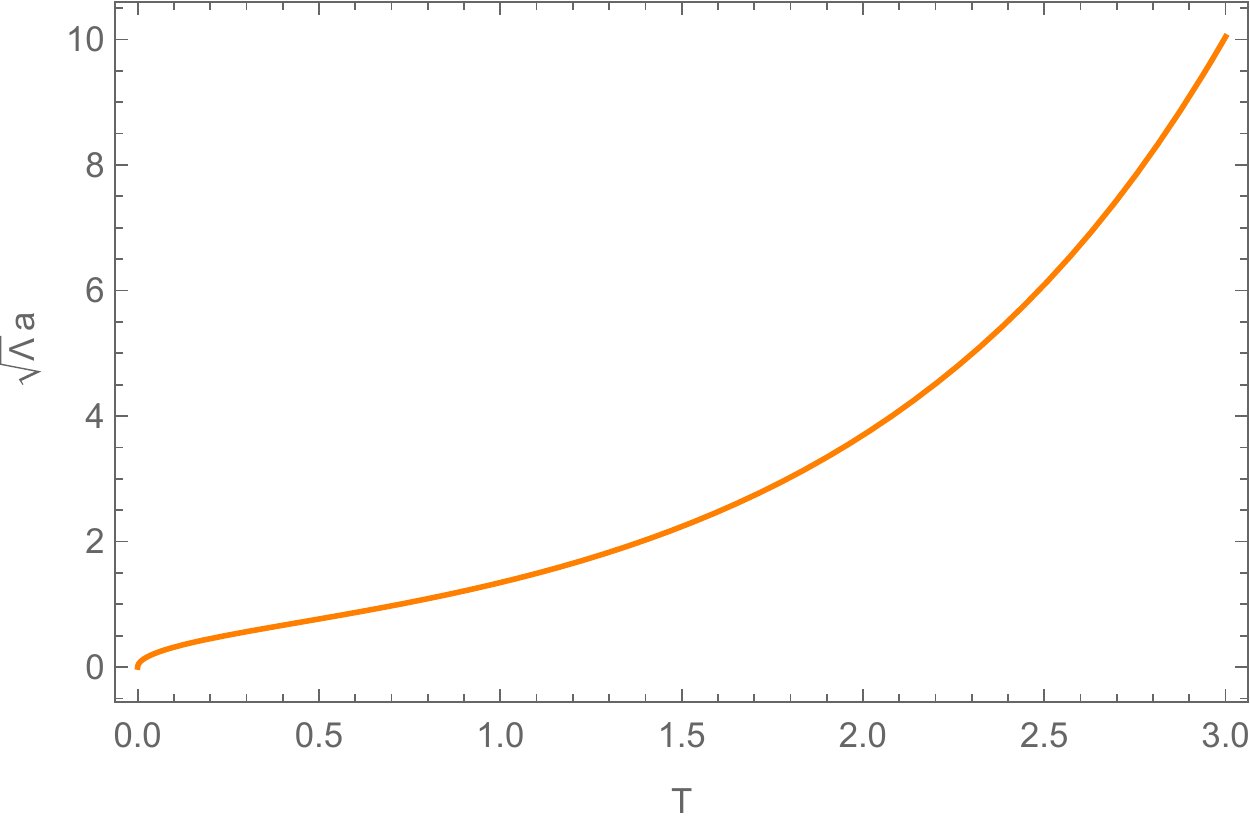}
     \includegraphics[width=5.5cm]{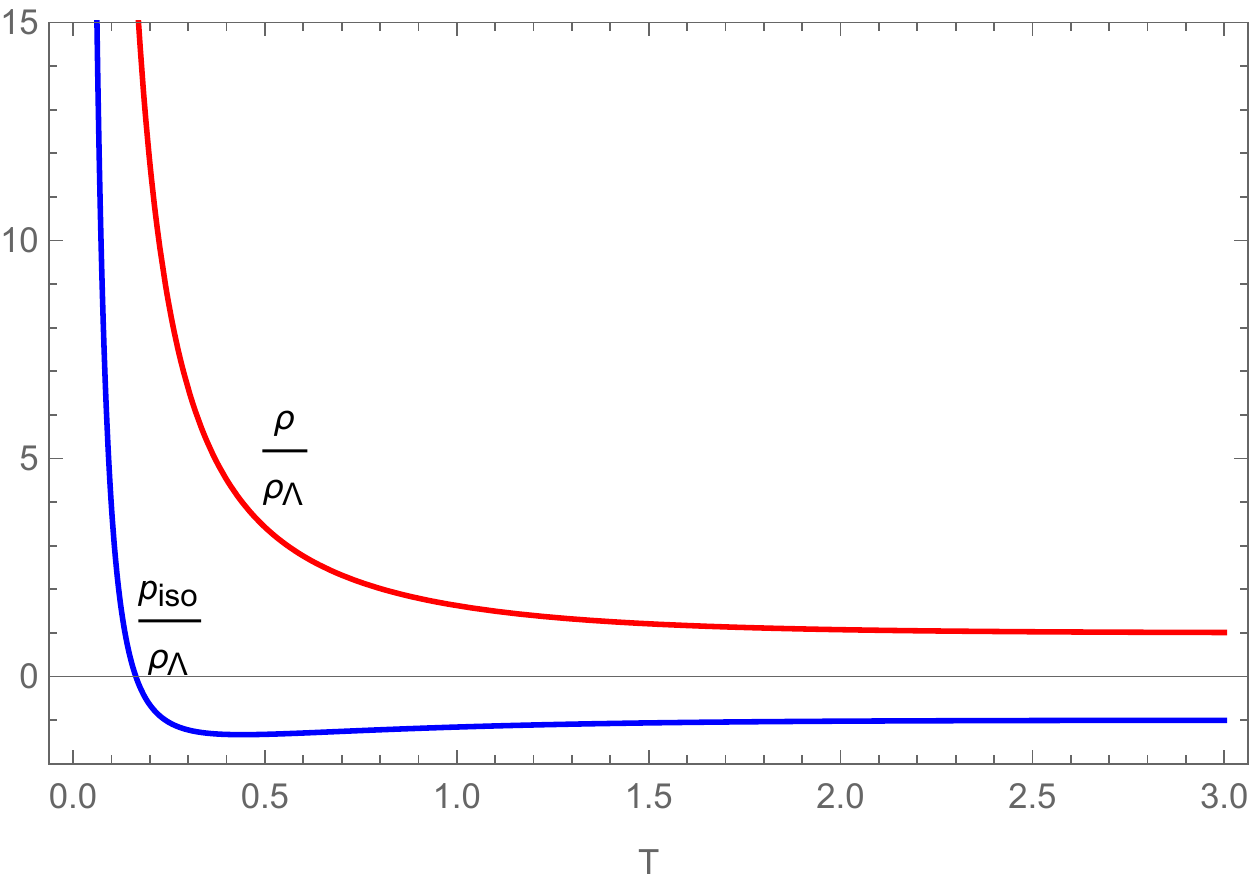}
     \includegraphics[width=5.5cm]{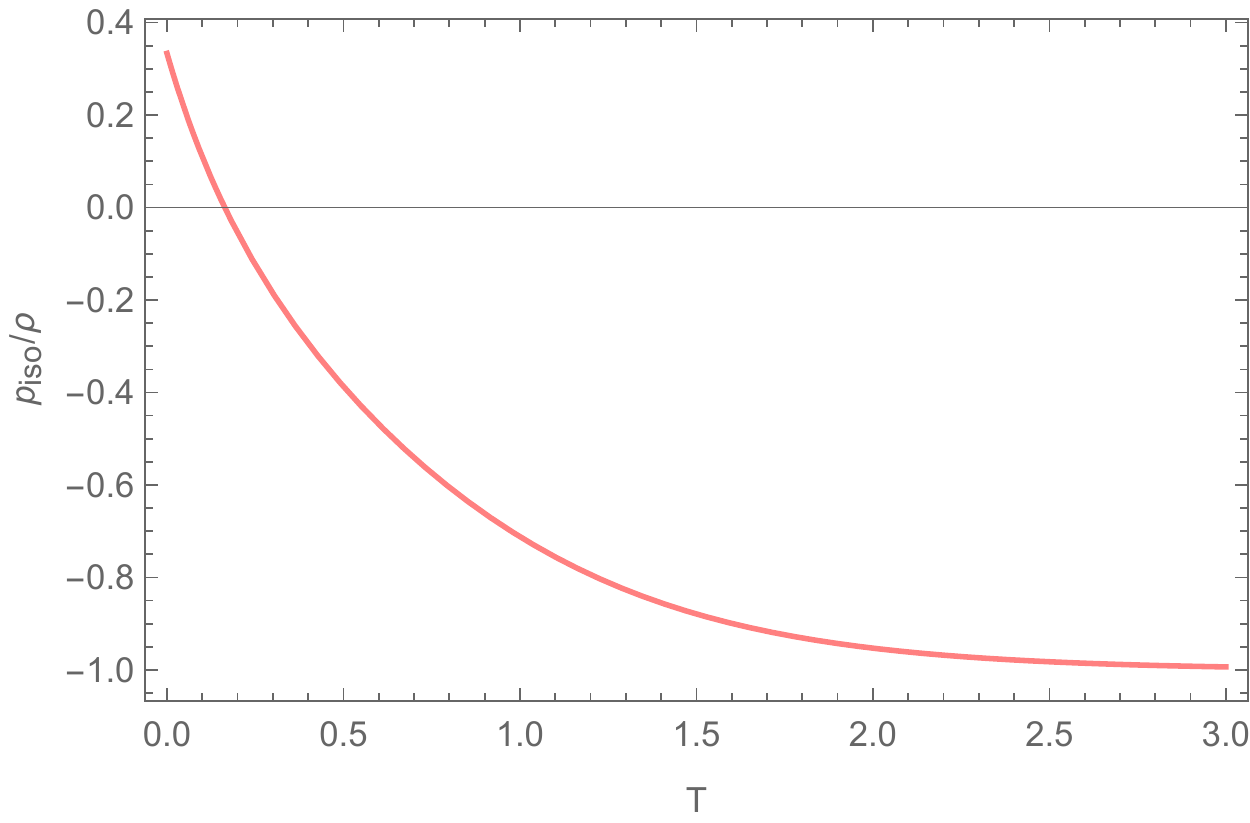}
     \caption{T evolution of the scale factor (specifically $\sqrt{\Lambda}a$), $\rho/\rho_\Lambda$, $p_{iso}/\rho_\Lambda$, and $p_{iso}/\rho$ for an example expanding system with $k=1/3$, $Y=1$. Notice that the behavior is radiation like at early times and vacuum energy like at late times. Choosing the minus sign in the time evolution Eq.~(\ref{NJaiso}) for a contracting system results in graphs which are mirrored across the $T$ axis. The behavior and these graphs are also qualitatively similar to the $k=0$ case with $Y>0$, such that specifically choosing $Y=1$ results in graphs which are extremely similar.}
     \label{NJisokp}
 \end{figure}

 However, there are two separate paths between vacuum like and infinite density configurations of Eq. (\ref{NJisobranch}) or Figure \ref{EOSbranchiso}, being  purely along the lower branch, or going through the zero density point and along the upper branch. 
 For positive $k,Y$ the evolution of the system seems to be purely along the lower branch. We show the time evolution of $a$, $\rho$, $p_{iso}$, $p_{iso}/\rho$ for an expanding universe with $k=1/3,~Y=1$ such that the start is at $T=0$ in Figure \ref{NJisokp}. Other values of positive $k,~Y$ have qualitatively similar behavior with the most obvious difference being the starting value of $T$, the graphs for contracting universes with corresponding $k,~Y$ look like mirror images about the $T$ axis. 
 
 For negative $k,Y$, we follow the other path which changes branches and goes through the zero density point. The evolution of the scale function $a$ looks similar to the positive $k,Y$ case, and at extremely small or extremely large times the energy-momentum functions are the same, but the behavior of the energy-momentum functions is different at intermediate times. When
 \begin{align}
     a=\sqrt{\frac{3k+Y}{-4\Lambda}}
 \end{align}
 the density and isotropic pressure go to zero. At this point, the system passes from the upper branch on the equation of state to the lower branch. Notice that this path traverses the lower branch to the left of the vacuum energy configuration, so the Null energy condition is violated at certain times.
  \begin{figure}
     \centering
     \includegraphics[width=5.5cm]{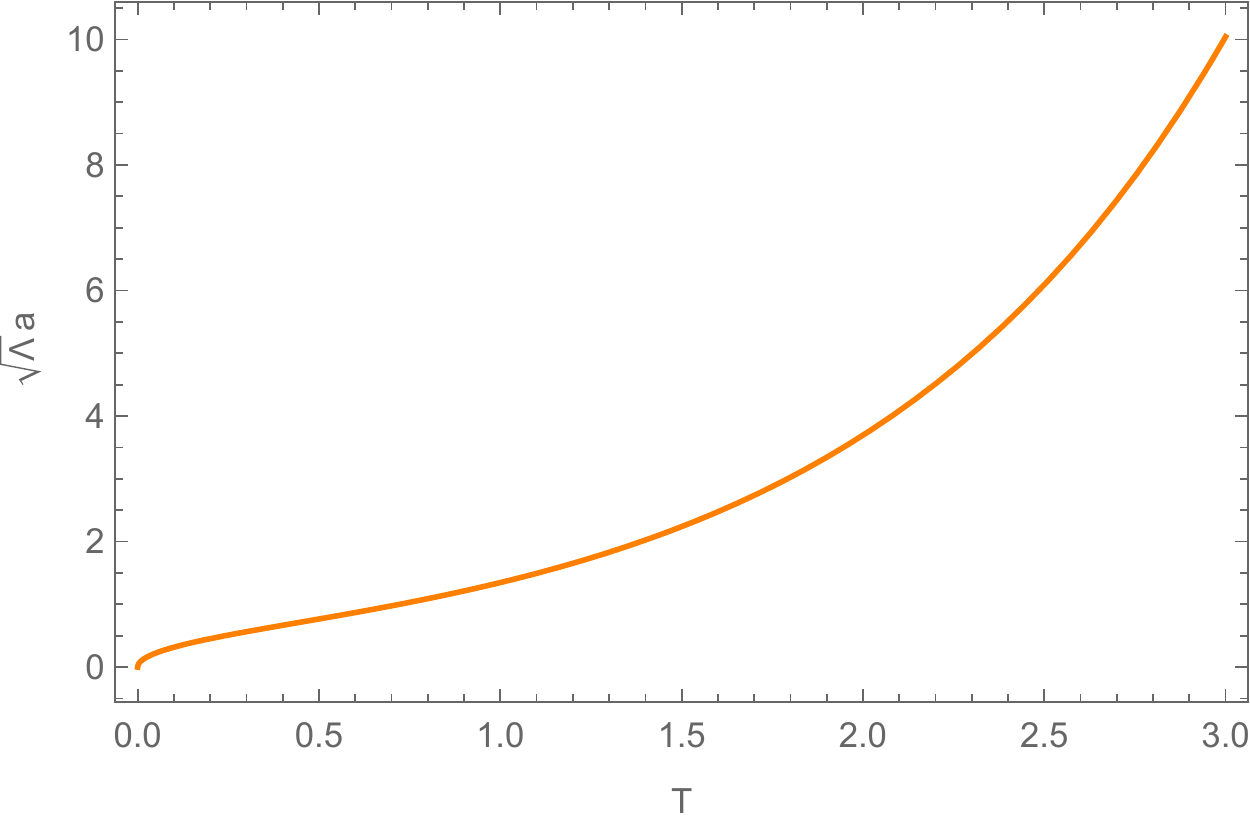}
     \includegraphics[width=5.5cm]{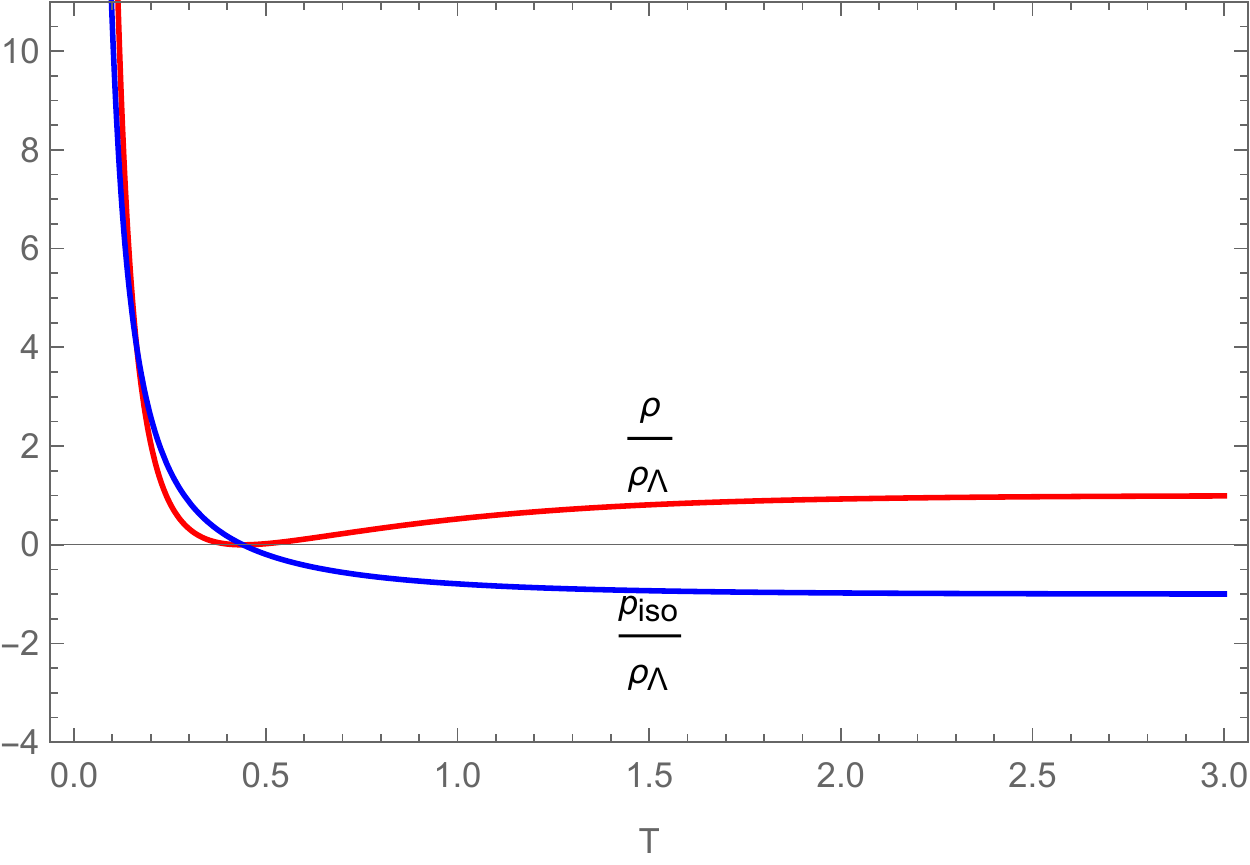}
     \includegraphics[width=5.5cm]{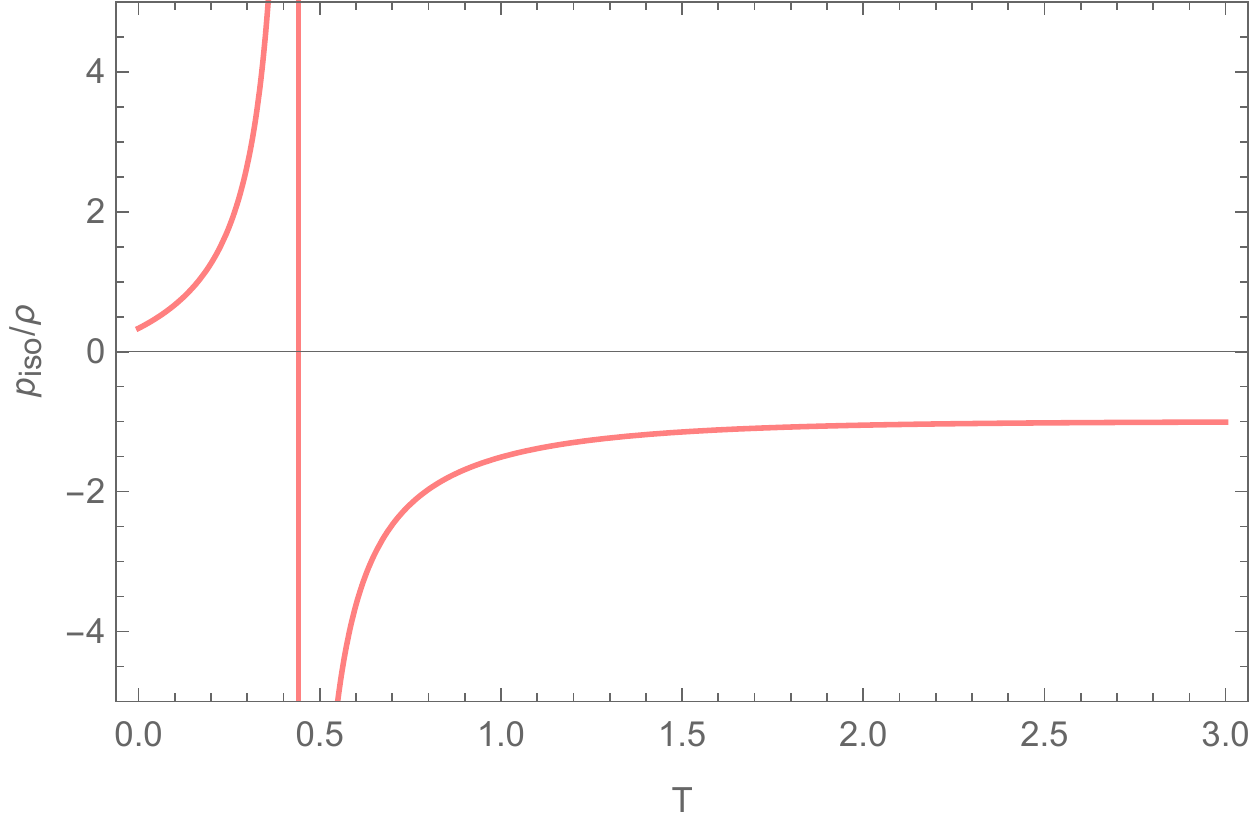}
     \caption{T evolution of the scale factor (specifically $\sqrt{\Lambda}a$), $\rho/\rho_\Lambda$, $p_{iso}/\rho_\Lambda$, and $p_{iso}/\rho$ for an example expanding system with $k=-1/3$, $Y=-1$. Choosing the minus sign in the time evolution Eq.~(\ref{NJaiso}) for a contracting system results in graphs which are mirrored across the $T$ axis. Notice that the evolution of the scale is similar to that for the case of positive $k,Y$ (see Figure \ref{NJisokp}) but the energy momentum tensor functions are rather different at intermediate times in that they pass from the upper branch of the equation of state onto the lower branch to the left of the vacuum energy point where the null energy condition is violated. As $T\rightarrow0$ and $T\rightarrow\infty$ we still approach radiation like and vacuum energy like configurations. }
     \label{NJisokm}
 \end{figure}
 
 \subsubsection{k=0 cases}
 If instead $k=0$, one should write the solution to Eq.(\ref{NJisoDEQ}) as
 \begin{align}
     a=\sqrt{\frac{e^{\pm\frac{ 2 (t-X)\sqrt{\Lambda}}{\sqrt{3}}}-Y}{4\Lambda}}.\label{NJaisokz}
 \end{align}
Here the constant $X$ does not correspond to $Z$ in Eq.~(\ref{NJaiso}), but $Y$ does in such a way that the density \label{NJaisokzz} can be computed by using Eq.~(\ref{NJisoden}) with $k=0$ and the appropriate expression for $a$. The energy-momentum tensor components directly in terms of $t$ are
\begin{align}
    \rho=\frac{\Lambda  e^{\frac{\pm 4 \sqrt{\Lambda }  (t-X)}{\sqrt{3}}}}{8 \pi  \left(e^{\frac{\pm 2
   \sqrt{\Lambda } (t-X)}{\sqrt{3}}}-Y\right)^2},\label{NJkzden}\\
   p_{iso}=-\frac{\Lambda  e^{\frac{\pm 2 \sqrt{\Lambda } (t-X)}{\sqrt{3}}} \left(3 e^{\frac{\pm 2
   \sqrt{\Lambda }  (t-X)}{\sqrt{3}}}-4 Y\right)}{24\pi \left(e^{\frac{\pm 2 \sqrt{\Lambda }
    (t-X)}{\sqrt{3}}}-Y\right)^2}.\label{NJkzp}
\end{align}
For $Y>0$, this system behaves in much the same way as when $k\ne0$; there is some singular point at which the scale function $a$ is zero and the density and pressure approach infinity in such a way that $p_{iso}=\rho/3$. At larger scale factors the system behaves like vacuum energy. Depending on the sign chosen in Eq.~(\ref{NJaisokz}) we either evolve towards or away from the singularity. Plotting  $\sqrt{\Lambda}a$, $\rho/\rho_\Lambda$, $p_{iso}/\rho_\Lambda$, and $p_{iso}/\rho$ for with the time variable $T=\sqrt{\Lambda/3}(t-X)$, choosing the $+$ sign for expansion and $Y=1$ such that $a=0$ at $T=0$ results in graphs which are qualitatively similar to those in Figure \ref{NJisokp}.

For $Y=0$, the matter functions Eqs.~(\ref{NJkzden},\ref{NJkzp}) describe vacuum energy at all times and Eq.~(\ref{NJaisokz}) is a simple exponential.

\begin{figure}
     \centering
     \includegraphics[width=5.5cm]{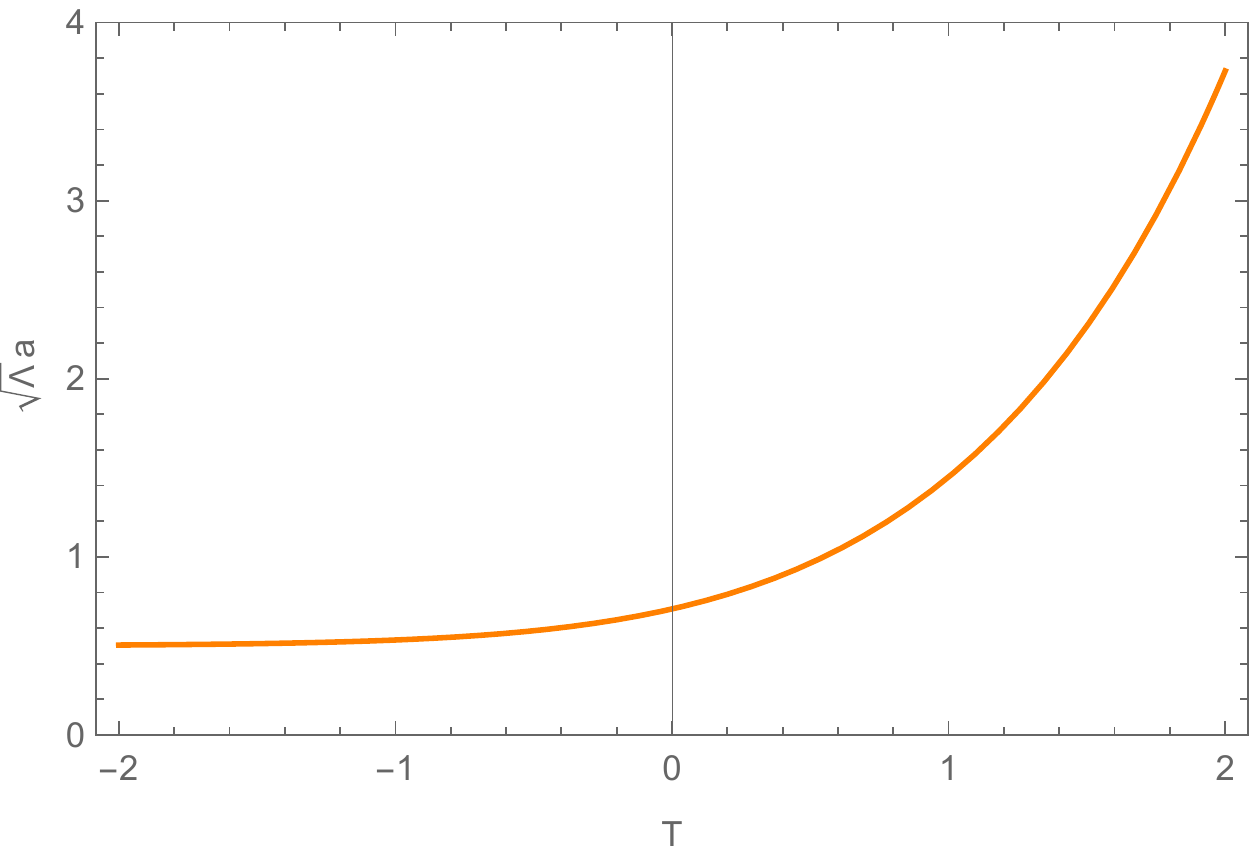}
     \includegraphics[width=5.5cm]{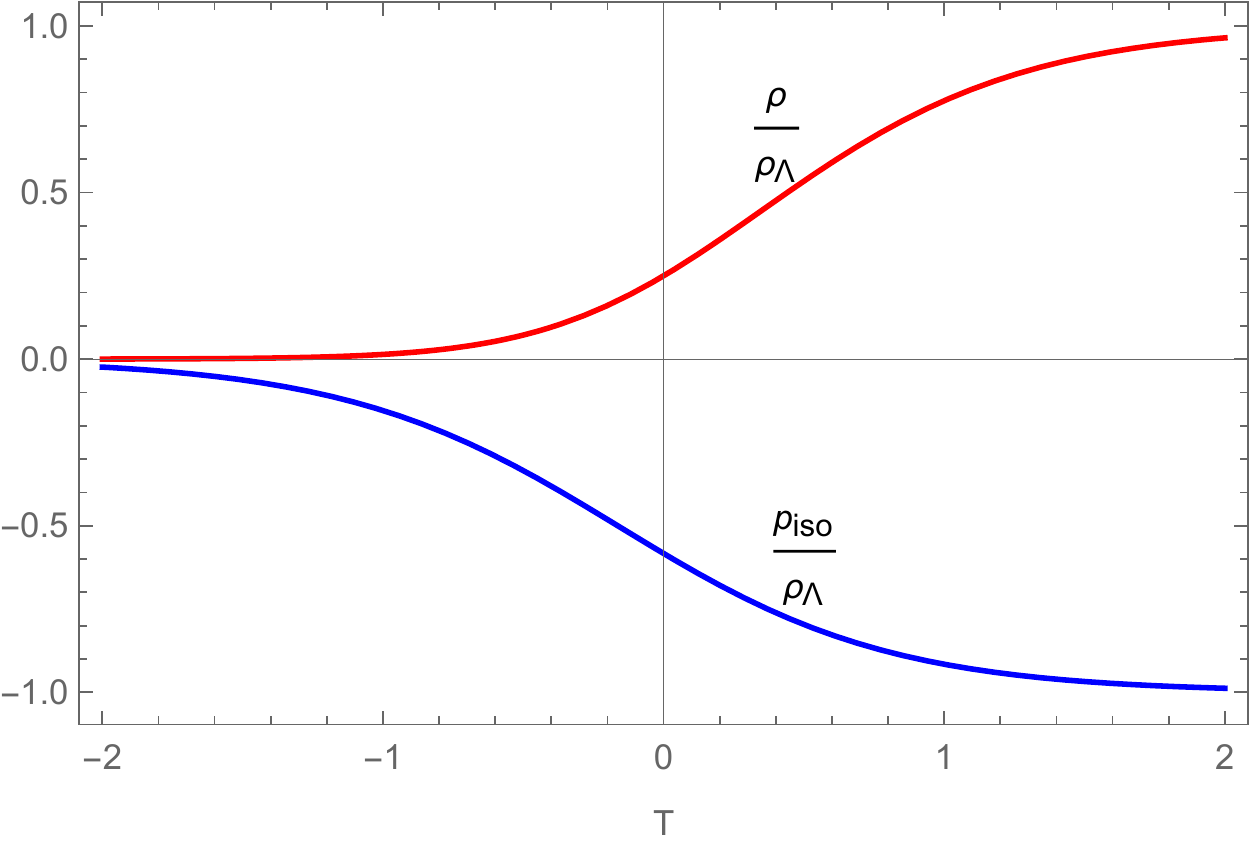}
     \includegraphics[width=5.5cm]{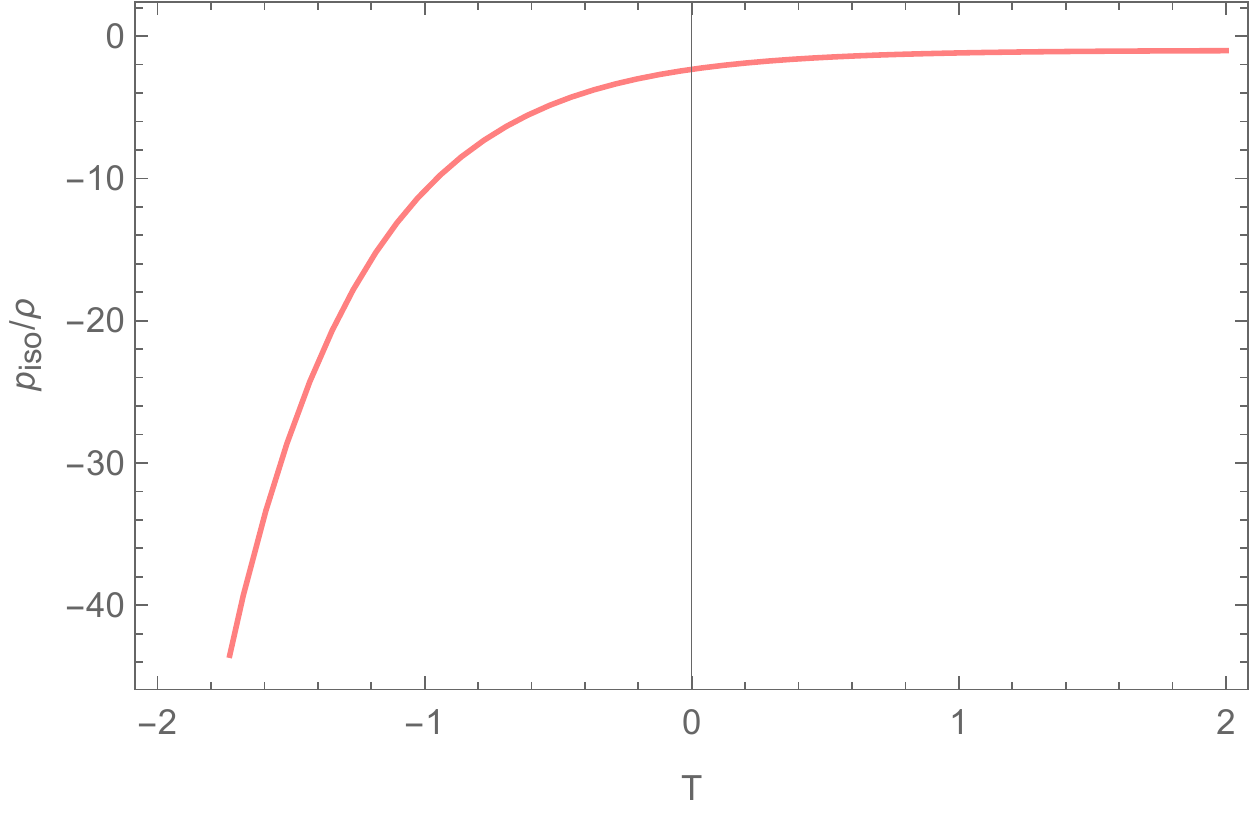}
     \caption{T evolution of the scale factor (specifically $\sqrt{\Lambda}a$), $\rho/\rho_\Lambda$, $p_{iso}/\rho_\Lambda$, and $p_{iso}/\rho$ for an example expanding system with $k=0$, $Y=-1$. Choosing the other sign in Eq.~(\ref{NJaisokz}) for a contracting system results in graphs which are mirrored across the $T$ axis. This is interesting since there is no singular point (the scale factor approaches a constant  and the energy-momentum tensor components approach zero in the distant past) and that the dark energy effectively ``turns on" in the vicinity of $T=0$, going from negligible to vacuum energy like. }
     \label{NJisoz}
 \end{figure}

For $Y<0$, the system behaves in a rather distinct way as there is no singular point. At one extreme in time, the scale factor approaches a constant $a\rightarrow\sqrt{|Y|/(4\Lambda)}$ and the energy/momentum terms approach 0, while at the other extreme, we have exponential behavior of $a$ and vacuum energy. The null energy condition is violated when we have negative $Y$ in that $p_{iso}+\rho<0$. The $k=0, Y<0$ system then is on the lower branch between the zero density point and vacuum energy point at all times. While the $k=0, Y<0$ violates the null energy condition, it is interesting that the dark energy effectively turns on, being negligible at $T<<0$ and vacuum energy like at $T>>0$. We show plots of the behavior of an example expanding $Y=-1,k=0$ case in Figure \ref{NJisoz}.

\section{Discussion and Conclusion}
\label{conclusion}
In this paper, we have systematically looked at cosmologies with Segre type [(11)(1,1)] matter, which is naturally constrained to behave like vacuum energy along a preferred axis. For systems with a long range uniformity of the direction of the preferred axis, we use a metric (\ref{unimet}) which describes either a Bianchi Type III or Type I based on the value of the transverse curvature parameter. For systems without long range uniformity of the preferred axis, we approximate the average behavior with a standard FLRW metric.  Some important results we have derived are as follows. 

We examine some simple equations of state in Section \ref{derivation section}. Here the Einstein equations for Eq. (\ref{unimet}) often have closed form solutions in terms of elementary functions. The case which obeys the vacuum energy equation of state $p_\perp=-\rho$ in particular may be of interest because it is similar to standard de Sitter solutions in some regards, but features nonzero contributions to the Weyl tensor because of the anisotropy.  The other simple cases, stringy $p_\perp=0$ and electromagnetic $p_\perp=\rho$, are not necessarily of great physical interest from the standpoint of dark energy, but serve to provide easier examples of the analysis techniques used for the vacuum energy and main equations of state. 

 We examine the uniform anisotropic case using metric (\ref{unimet}) and  our main equation of state (\ref{NJeos}) in Section \ref{NJsec}. We do not find an exact solution for the time dependence, but we demonstrate the existence systems which approach the standard inflationary vacuum energy solution at large times using perturbation theory, and we can retrodict the behavior to earlier times with numerical methods. We find in these numericaly derived systems the transverse scale function $b$ reaches a minimum at some point in the past, on the other side of which may be a situation where the transverse scale function is decreasing, reminiscent of big bounce universes in standard cosmology. The behavior at the minimum itself has a few possibilities. It can be a singular configuration approximating a very strong static electric field $p_\perp=\rho$. There also exist cases in which the Kretchman scalar is finite, despite the axial scale function $a$ going to zero; the behavior of the energy momentum tensor in these cases varies. We even find the possibility that at the critical point the energy-momentum tensor terms approach true vacuum $p_\perp=\rho=0$, but the Kretchman scalar is extremely high indicating large curvature in the Weyl sector.

In Section \ref{isoavg}, we argue that at large scales, systems with random ordering of the preferred spatial axis may approximate standard perfect fluid FLRW type cosmologies. Using this averaging procedure with the simple stringy, electromagnetic, and vacuum energy equations of state reproduced known spacetimes. For the isotropically averaged version of our central equation of state (\ref{NJeosiso}), we find that there is a closed form solution in terms of elementary functions, and that in many cases the universe evolves between configurations which approximate radiation at small scale factor and vacuum energy at large scale factor. We also found it was possible for such a universe to evolve between true and false vacuum configurations.

There are several possible future extensions to this work. First, one could look at other types of Segre [(11)(1,1)] matter with metric (\ref{unimet}) to examine their behavior to see if they have other interesting properties. Second, one could try to add additional components, such as standard ``dust" matter or radiation, to systems with generic anisotropic Segre [(11)(1,1)] dark energy, anisotropic dark energy which follows equation of state (\ref{NJeos}), or the isotropic average (\ref{NJeosiso}). For the isotropic average a standard FLRW metric (\ref{Fmet}) should be sufficient to consider this, while for the anisotropic dark energy it might be more appropriate to construct a model with metric (\ref{met}) or some other Bianchi cosmology. After additional components are added to the models, one could compute observational consequences. Third, one could attempt to find a Lagrangian density for which the energy momentum tensor eigenvalues obey the equation of state (\ref{NJeos}). Since nonlinear electrodynamics theories can naturally give Segre type [(11)(1,1)] energy momentum tensors, this might be a place to start. Alternatively, one could consider the equation of state (\ref{NJeosiso}) not as an isotropic averaging but as a fundamental equation of state for some other form of perfect fluid type matter, and try to define a Lagrangian density  for that perfect fluid.

\bibliographystyle{JHEP}
\bibliography{monster.bib}
\appendix
\section{Killing vectors, Geodesics, and Conserved Quantities}
\label{geoapdx}
Because the metric
\begin{align}
    ds^2=-dt^2+a(t)^2dz^2+b(t)^2\Big[\frac{dr^2}{1-k r^2/2 } +r^2 d\theta^2\Big] 
\end{align}
is independent of $z$ and $\theta$, there must be Killing vectors which can be written
\begin{align}
    K^\mu_z=(0,1,0,0),\\
    K^\mu_\theta=(0,0,0,1),
\end{align}
in $(t,z,r,\theta)$ coordinates.

In Cartesian type coordinates, with $x=r \cos(\theta)$, $y=r\sin(\theta)$ but the same $t$ and $z$, we obtain an alternate expression for the line element
\begin{align}
    ds^2=-dt^2+a(t)^2dz^2+b(t)^2\Big[\frac{2-ky^2}{2-k(x^2+y^2)}dx^2+\frac{2-kx^2}{2-k(x^2+y^2)}dy^2+2\frac{kxy}{2-k(x^2+y^2)}dxdy\Big]. \label{metcart}
\end{align}
The metric is easier in cylindrical type coordinates, but the Cartesian coordinates $(t,z,x,y)$ are useful since three linearly independent Killing vectors corresponding to translations point along the $z,x,y$ coordinates
\begin{subequations}
\begin{align}
    K^\mu_z=(0,1,0,0),\\
    K^\mu_x=(0,0,\sqrt{2-k(x^2+y^2)},0),\\
    K^\mu_y=(0,0,0,\sqrt{2-k(x^2+y^2}).
\end{align}
\label{kcart}
\end{subequations}
Notice that in the zero transverse curvature case $k=0$, metric (\ref{metcart}) becomes independent of the $x,y$ coordinates, diagonal, and that the Killing vector coefficients become constants (as opposed to depending on position for $K^\mu_x,~K^\mu_y$ for arbitrary $k$). These translational Killing vectors will be used in the next appendix for the Bianchi classification.

We call our affine parameter $\tau$ for simplify (although for null or spacelike geodesics $\tau$ cannot be identified with proper time) , such that $x^\mu=\big(t(\tau),z(\tau),r(\tau),\theta(\tau)\big)$. The normalization condition becomes
\begin{align}
    g_{\mu\nu}\frac{dx^\mu}{d\tau}\frac{dx^\nu}{d\tau}=n=\frac{b^2 \frac{dr}{d\tau}^2}{1-k r^2/2}+a^2\frac{dz}{d\tau}^2+r^2b^2\frac{d\theta}{d\tau}^2-\frac{dt}{d\tau}^2,\label{geonorm}
\end{align}
where $n=-1$ for timelike geodesics and $n=0$ for null geodesics. While it is possible to use the definition of the geodesic equations $\frac{d^2x^\mu}{d\tau^2}+\Gamma^\mu_{\rho \sigma}\frac{dx^\rho}{d\tau}\frac{dx^\sigma}{d\tau}=0$, it is more convenient in this case to derive the geodesic equations from the Euler-Lagrange equation $\frac{\partial n}{\partial x^\mu}-\frac{d}{d\tau}(\frac{\partial n}{\partial x^\mu_{,\tau}})=0$, which makes the existence of conserved quantities more clear. We have for instance
\begin{align}
    \frac{d}{d\tau}(r^2 b^2 \frac{d\theta}{d\tau})=0,\qquad \frac{d}{d\tau}(a^2 \frac{dz}{d\tau})=0,
\end{align}
so we can define conserved quantities
\begin{align}
    J_\theta=r^2 b^2 \frac{d\theta}{d\tau},\label{Jgeo}\\
    p_z=a^2 \frac{dz}{d\tau}\label{pgeo}.
\end{align}
These conserved quantities are related to the Killing vectors $K_\theta^\mu$  and $K_z^\mu$  such that $J_\theta=K^\mu_\theta u_\mu$ and $p_z=K^\mu_z u_\mu$. We can use the same property to define conserved quantities with respect to $K_x^\mu$ and $K_y^\mu$ which are not immediately obvious from the Euler-Lagrange equations, being
\begin{align}
    p_x=K^\mu_x u_\mu=b^2\left(\frac{2\frac{dr}{d\tau} \cos \theta}{\sqrt{2-kr^2}}-r\frac{d\theta}{d\tau} \sqrt{2-kr^2} \sin{\theta}\right),\\
     p_y=K^\mu_y u_\mu=b^2\left(\frac{2\frac{dr}{d\tau} \sin \theta}{\sqrt{2-kr^2}}-r\frac{d\theta}{d\tau} \sqrt{2-kr^2} \cos{\theta}\right).
\end{align}
When the transverse curvature $k=0$, $p_x$ and $p_y$ simplify to forms more reminiscent of the $p_z$ parameter, being 
\begin{align}
p_x(k\rightarrow0)=b^2\sqrt{2}(\frac{dr}{d\tau} \cos \theta-r\frac{d\theta}{d\tau}\sin{\theta})=\sqrt{2}b^2\frac{dx}{d\tau},\\
p_x(k\rightarrow0)=b^2\sqrt{2}(\frac{dr}{d\tau} \sin \theta+r\frac{d\theta}{d\tau}\cos{\theta})=\sqrt{2}b^2\frac{dy}{d\tau}.
\end{align}

Notice that the specification of the quantity $p_z$ and two of the quantities of $J_\theta,~p_x,$ or $p_y$ is sufficient to solve for $dz/d\tau,~dr/d\tau,~d\theta/d\tau$. One can then use the normalization condition to solve for $dt/d\tau$, i.e. if $p_x$ and $p_y$ are specified we obtain
\begin{align}
    \frac{dt}{d\tau}=\pm\sqrt{-n+\frac{p_z^2}{a^2}+\frac{(p_x \cos \theta+p_y \sin \theta)^2}{2b^2}+\frac{(p_y \cos \theta-p_x \sin \theta)^2}{(2-kr^2)b^2}}.
\end{align}

If we further constrain the spacetime to be Segre [(11)(1,1)] in the manner of our examples, by usage of Eq.~(\ref{unienforcer}), the only difference will be replacement of $a$ with $c\dot{b}$.

\section{Bianchi Classification of Metric \ref{met}}
\label{bianchi}
Because an explanation of the Bianchi classification method and proof of our classification was rather lengthy for the main body of the paper, as well as not being necessary to understand further results, it is detailed in this appendix.  
The classification of Bianchi spacetimes is based on Bianchi's earlier 1898 work \cite{1898MMFSI..11..267B} classifying three dimensional Lie algebras, the modern form of the classification scheme for spacetimes was presented in the 1960s \cite{1968JMP.....9..497E,Ellis:1968vb}, historical notes about the development of the classification can be found in \cite{Krasinski:2003zzb,Ellis:2006ba}, there is also discussion of the classification in \cite{Wald:1984rg}.

The procedure for determining the Bianchi classification is as follows:

1) find 3 linearly independent Killing vector fields corresponding to translations $K^\mu_A$

2) Take the Lie brackets
\begin{align}
    [J,L]^\gamma=J^\alpha \partial_\alpha L^\gamma-L^\alpha \partial_\alpha J^\gamma \label{liebracketdef}
\end{align} 
of each pair of Killing vector fields, write the result as 
\begin{align}
    [K_A,K_B]^\gamma=C^D_{AB}K_{D}^\gamma=\epsilon_{ABE}\overline{C}^{ED}K_{D}^\gamma,\label{bianciC}
\end{align}
 where capital Latin indices range over the set of Killing vector fields,  $\epsilon$ here is the Levi Civita symbol, $C$ is a set of numbers known as the structure constants, and $\overline{C}$ is a matrix.

3) Rewrite the matrix $\overline{C}$ in terms of a sum of a symmetric and an antisymmetric matrix 
\begin{align}
    \overline{C}=\overline{S}+\overline{A}.
\end{align} 

4) Find a transformation $P$ which diagonalizes the symmetric matrix $P^{-1}\overline{S}P=\hat{n}$, with $\hat{n}$ being diagonal, the apply it to the entire matrix 
\begin{align}
    P^{-1}\overline{C}P=P^{-1}\overline{S}P+P^{-1}\overline{A}P=\hat{n}+\hat{A}.
\end{align}

5) The transformed version of $\hat{A}$ should still be antisymmetric, allowing for it to be rewritten as 
\begin{align}
    \hat{A}^{AB}=\epsilon^{ABD} \hat{a}_{D}.\label{Ava}
\end{align}
The  $\hat{a}_D$ should have at most one nonzero component, having none if $\overline{C}$ was already symmetric and one otherwise. 

6) We now have 3 elements along the diagonal of $\hat{n}$ and the single nonzero component of $a$, it is these items which are actually tracked in the Bianchi classification. 

For the metric (\ref{met}), 

1) Three Killing vector fields corresponding to translations for the metric (\ref{met}) have $(t,z,r,\theta)$ components
\begin{subequations}
\begin{align}
    K_{z}^\mu=(0,1,0,0),\\
    K_y^\mu=(0,0,\sqrt{2-kr^2}\sin (\theta),\sqrt{2-kr^2}\cos (\theta)/r),\\
    K_x^\mu=(0,0,\sqrt{2-kr^2}\cos (\theta),-\sqrt{2-kr^2}\sin (\theta)/r). \label{kpolar}
\end{align}
\end{subequations}
These are the same as the Killing vectors introduced in (\ref{kcart}), but now expressed in the cylindrical type coordinates to simplify some of the following expressions.

2) The Lie brackets give components
\begin{align}
    [K_z,K_x]^\gamma=(0,0,0,0),\\
    [K_z,K_y]^\gamma=(0,0,0,0),\\
    [K_x,K_y]^\gamma=(0,0,0,-k),
\end{align}
from which we can deduce
\begin{align}
    C^x_{xy}=\frac{kr \sin(\theta)}{\sqrt{2-kr^2}},\\
    C^y_{xy}=\frac{-kr \cos(\theta)}{\sqrt{2-kr^2}},
\end{align}
 with all non-listed components are either identically zero or specified by the antisymmetry. The matrix $\overline{C}$ can be written as

\begin{align}
   \overline{C}= \left(
\begin{array}{ccc}
 0 & 0 & 0 \\
 0 & 0 & 0 \\
 \frac{k r \sin (\theta )}{\sqrt{2-k r^2}} & -\frac{k r \cos (\theta )}{\sqrt{2-k r^2}} & 0 \\
\end{array}
\right),
\end{align}
where the first index is the row and the second index is the column.

3) Breaking $\overline{C}$ into a sum of symmetric and antisymetric matrices, we get
\begin{align}
    \overline{C}=\overline{S}+\overline{A}=\left(
\begin{array}{ccc}
 0 & 0 & \frac{k r \sin (\theta )}{2 \sqrt{2-k r^2}} \\
 0 & 0 & -\frac{k r \cos (\theta )}{2 \sqrt{2-k r^2}} \\
 \frac{k r \sin (\theta )}{2 \sqrt{2-k r^2}} &- \frac{k r \cos (\theta )}{2 \sqrt{2-k r^2}} & 0
   \\
\end{array}
\right)+\left(
\begin{array}{ccc}
 0 & 0 & -\frac{k r \sin (\theta )}{2 \sqrt{2-k r^2}} \\
 0 & 0 & \frac{k r \cos (\theta )}{2 \sqrt{2-k r^2}} \\
 \frac{k r \sin (\theta )}{2 \sqrt{2-k r^2}} & -\frac{k r \cos (\theta )}{2 \sqrt{2-k r^2}} & 0
   \\
\end{array}
\right).
\end{align}

4) We diagonalize $S=P \hat{n} P^{-1}$, where $\hat{n}$ is a diagonal matrix of the eigenvalues and $P$ is a matrix with the columns being the eigenvectors\footnote{One is free to change the order of the columns/eigenvectors as well as the overall factor of each, this may be required in order to reach a canonical form with regard to the placement and sign of the $\hat{n}$ components or the possible nonzero component of $a$. } of $\overline{S}$, and apply the same transform to $\overline{A}$, we obtain in the new basis
\begin{align}
    \hat{C}=P^{-1}\overline{C}P=\hat{n}+\hat{A},
\end{align}
where the items can be explicitly written as
\begin{align}
    \hat{n}=\left(
\begin{array}{ccc}
 0 & 0 & 0 \\
 0 & \frac{k r}{2 \sqrt{2-k r^2}} & 0 \\
 0 & 0 & -\frac{k r}{2 \sqrt{2-k r^2}} \\
\end{array}
\right)
\end{align}
and
\begin{align}
    \hat{A}=\left(
\begin{array}{ccc}
 0 & 0 & 0 \\
 0 & 0 & \frac{k r}{2 \sqrt{2-k r^2}} \\
 0 & -\frac{k r}{2 \sqrt{2-k r^2}} & 0 \\
\end{array}
\right).
\end{align}

5) Decomposing $\hat{A}$ as in Eq. (\ref{Ava}) we obtain
\begin{align}
    \hat{a}=(\frac{k r}{2 \sqrt{2-k r^2}},0,0).
\end{align}

6) The diagonal elements of $\hat{n}$ and the nonzero component of $\hat{a}$ are
\begin{align}
    \hat{n}_1=0,\qquad\hat{n}_2=\frac{k r}{2 \sqrt{2-k r^2}},\qquad{n}_3=-\frac{k r}{2 \sqrt{2-k r^2}},\qquad \hat{a}_1=\frac{k r}{2 \sqrt{2-k r^2}}.
\end{align}
 Up to the overall constant factor, this has structure of
\begin{align}
  &(k\ne0)\qquad  \hat{n}_1=0,\hat{n}_2=1,\hat{n}_3=-1,\hat{a}_1=1\\
  &(k=0)\qquad  \hat{n}_1=\hat{n}_2=\hat{n}_3=\hat{a}_1=0
\end{align}
 which corresponds to Bianchi type $III$ when $k\ne0$ and corresponds to Bianchi type $I$ when $k=0$. It is important that the Bianchi type is independent of the specification of $a$ or $b$ as in Eq.~(\ref{unienforcer}), so metrics of the form (\ref{unimet}) have the same rules for Bianchi type as those of metric (\ref{met}).

\section{Constant $b=B$}
\label{constB}
If $b=B\ne f[t]$, then the metric (\ref{met}) describes a [(11)(1,1)] spacetime. The behavior is simpler but more restricted than the alternate case of using Eqs.~(\ref{unienforcer}) and (\ref{unimet}) which we use in the main body of the paper.
The only remaining nonzero orthonormal Riemann components from  are
\begin{subequations}
\begin{align}
    R_{\hat{0}\hat{1}\hat{0}\hat{1}}=\frac{-\ddot{a}}{a},\\
    R_{\hat{2}\hat{3}\hat{2}\hat{3}}=\frac{k}{2B^2}.
\end{align}
\label{ReimannBZ}
\end{subequations}
The $Q_{11}$ function, Ricci, and Kretchman scalars become
\begin{align}
    Q_{11}=\frac{k}{6B^2}+\frac{\ddot{a}}{3a},\\
    \mathcal{R}=\frac{k}{B^2}+2\frac{\ddot{a}}{a},\\
    \mathcal{K}=\frac{k^2}{B^4}+4\frac{\ddot{a}^2}{a^2}.
\end{align}
Finally
\begin{align}
    8\pi T^t_{~t}=8\pi T^z_{~z}=-\frac{k}{2 B^2},\label{rhoBZ}\\
     8\pi T^r_{~r}=8\pi T^\theta_{~\theta}=-\frac{\ddot{a}}{a}.\label{pTBZ}
\end{align}
Notice that, not only is the radial scale function $b=B$ restricted to not evolve with time, but the density is also restricted to not evolve with time. Imposing an equation of state of the form $p_\perp=f(\rho)$ therefore requires $p_\perp$ also not evolve with time. From Eq.~(\ref{pTBZ}) we then have
\begin{align}
   \frac{\ddot{a}}{a}=constant, 
\end{align}
but this removes all time dependence from the curvature or physical quantities listed here. 
 
 It is possible to have metric (\ref{met}) with constant $b$ describe a spacetime which satisfies the simple equations of state, for instance
 \begin{align}
     &p_\perp=0, && a=X t+Y,\\
     &p_\perp=-\rho, && a=X \cosh(\sqrt{\frac{k}{2B^2}}t)+Y \sinh(\sqrt{\frac{k}{2B^2}}t),\\
     &p_\perp=\rho, && a=X \cos(\sqrt{\frac{k}{2B^2}}t)+Y \sin(\sqrt{\frac{k}{2B^2}}t).
 \end{align}
 However, unlike using Eqs.~(\ref{unienforcer}), this $b=B$ case does not allow for describing the evolution of a system along the more complex equation of state (\ref{NJeos}).
\end{document}